\documentclass[useAMS,usenatbib]{mn2e}
\usepackage{graphicx}

\title[A multi-domain approach to asteroid families identification]
{A multi-domain approach to asteroid families identification}
\author[V. Carruba, R. C. Domingos, D. Nesvorn\'{y}, F. Roig, M. E. Huaman, 
and D. Souami] 
{V. Carruba$^{1}$\thanks{E-mail: vcarruba@feg.unesp.br},  
R. C. Domingos$^{2}$, D. Nesvorn\'{y}$^{3}$, F. Roig$^{4}$, 
M. E. Huaman$^{1}$, and D. Souami$^{5,6}$\\
$^{1}$UNESP, Univ. Estadual Paulista, Grupo de din\^{a}mica Orbital e
  Planetologia, Guaratinguet\'{a}, SP, 12516-410, Brazil \\
$^{2}$INPE, Instituto Nacional de Pesquisas Espaciais, S\~{a}o Jos\'{e} 
  dos Campos, SP, 12227-010, Brazil\\
$^{3}$SWRI, SouthWest Research Institute, 1050 Walnut Street, Suite 300, 
Boulder, CO 80302, USA\\
$^{4}$ON, Observat\'{o}rio Nacional, Rua General Jos\'{e} Cristino 77, 
Rio de Janeiro, RJ, Brazil\\
$^{5}$UPMC, Universit\'{e} Pierre et Marie Curie, 4 Place Jussieu, 75005, 
Paris, France\\
$^{6}$SYRTE, Observatoire de Paris, Syst\`{e}mes de R\'{e}f\'{e}rence 
Temps Espace, CNRS/UMR 8630, UPMC, Paris, France;
}

\begin{document}

\date{Accepted 2013 May 15.  Received 2013 May 14; in original 
form 2013 April 16th.}

\pagerange{\pageref{firstpage}--\pageref{lastpage}} \pubyear{2013}

\maketitle

\label{firstpage}

\begin{abstract}
It has been shown that large families are not limited to 
what found by hierarchical clustering methods (HCM) in the 
domain of proper elements (a,e,sin(i)), that seems to be biased 
to find compact, relatively young clusters, but that there exists 
an extended population of objects with similar taxonomy and geometric 
albedo, that can extend to much larger regions in proper elements and 
frequencies domains: the family ``halo''.  Numerical simulations can 
be used to provide estimates of the age of the family halo, that can 
then be compared with ages of the family obtained with other methods.  
Determining a good estimate of the possible orbital extension of a 
family halo is therefore quite important, if one is interested in 
determining its age and, possibly, the original ejection velocity field.  
Previous works have identified families halos by an analysis in proper 
elements domains, or by using Sloan Digital Sky 
Survey-Moving Object Catalog data, fourth release (SDSS-MOC4) 
multi-band photometry to 
infer the asteroid taxonomy, or by a combination of the two methods.   
The limited number of asteroids for which geometric albedo was known 
until recently discouraged in the past the extensive use of this 
additional parameter, which is however of great importance in 
identifying an asteroid taxonomy.  The new availability of geometric 
albedo data from the Wide-field Infrared Survey Explorer (WISE) 
mission for about 100,000 asteroids significantly increased the sample of 
objects for which such information, with some errors, is now known.   

In this work we proposed a new method 
to identify  families halos in a multi-domain space composed 
by proper elements, SDSS-MOC4 $(a^{*},i-z)$ colors, and WISE 
geometric albedo for the whole main belt (and the Hungaria and Cybele
orbital regions).  Assuming that most families were created by 
the breakup of an undifferentiated parent body, they are expected
to be  homogeneous in colors and albedo.
The new method is quite effective in determining objects 
belonging to a family halo, with low percentages of likely 
interlopers, and results that are quite consistent in term of 
taxonomy and geometric albedo of the halo members.

\end{abstract}

%

\section{Introduction}
\label{sec: intro}

Asteroid families are groups of asteroids that are supposed to have a common
origin in the collisional event that shattered the parent body.
They are usually determined by identifying clusters of objects close 
in proper elements domain $(a,e,sin(i))$.  The Hierarchical Clustering
Method (HCM hereafter) as described by Bendjoya and Zappal\'{a} 
(2002) operates by identifying all objects that are closer than a given 
distance (cutoff) with respect to at least one other member of a family. 
If an object is closer than this distance, it is associated to the 
dynamical family, and the procedure is repeated until no new family 
members are found.  The choice of this cutoff distance is then of paramount 
importance in determining the family.  For small values of the cutoff only 
the objects closest
in proper element domain are identified as family members: the family
``core''.  At larger cutoff one is able to identify objects that, while
still belonging to the collisional group, may have dynamically
evolved since the family formation and drifted
apart from the core: the family ``halo''~\footnote{The term halo
was first introduced to describe this population of
objects by Nesvorn\'{y} et al. (2006), and then
adopted by other authors such as Parker et al. (2008).}.  

\begin{figure}

  \centering
  \centering \includegraphics [width=0.45\textwidth]{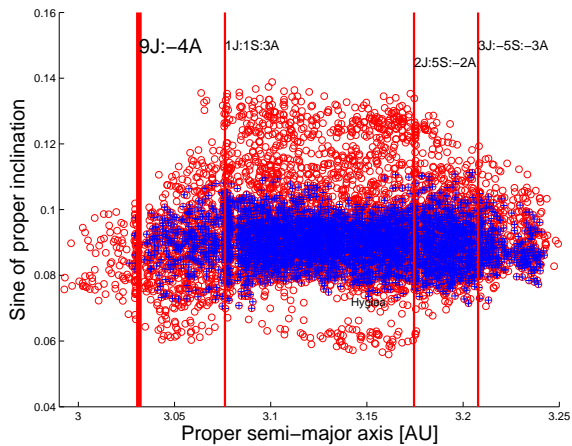}

\caption{An $(a,sin(i))$ projection of the Hygiea family core (blue crosses)
and halo (red circles), as obtained in Carruba (2013).  Vertical red lines
displays the location of the main mean-motion resonances in the region.}
\label{fig: hygiea_halo}
\end{figure}

To illustrate this issue, 
Fig.~\ref{fig: hygiea_halo} displays the family core (blue crosses)
and halo (red circles), as obtained in Carruba (2013) for the Hygiea
family, at cutoff of 66$~m/s$ and 76~$m/s$, respectively.  At cutoffs
larger than 76~$m/s$ the family merged with other dynamical groups in 
the region (such as the Veritas and Themis family) and with the 
local background, thus it was no longer identifiable as a separate entity.
Numerical simulations can be used to provide estimates of the age of the 
family halo, as done, for instance,
by Bro\v{z} and Morbidelli (2013) for the Eos halo, 
that can then be compared with ages of the family 
obtained with other methods.  Determining a good estimate of the 
possible orbital extension of a family halo is therefore quite 
important, if one is interested in determining its age and, possibly, 
the original ejection velocity field.   

One problem in obtaining a good determination of a family halo is 
however the presence of objects in the orbital region of the halo that 
might not be connected with the local family.   Assuming that most families 
were created by a breakup of an undifferentiated parent body, 
we would expect that most of its members should be homogeneous
in colors and albedo.  Objects that belongs to the dynamical families
but that differs in colors or albedo may possibly be asteroids of
the local background that just happened to lie in the orbital region 
of a given family: the interlopers.  An analysis of spectral 
properties of local asteroids may provide insights on the possible 
presence of such interlopers in groups found in proper elements domains, 
but such information is usually available only for 2\% or less of 
the main belt asteroids.   Yet including too many interlopers into the
family may change the perceived orbital structure of the group
and cause to obtain distorted estimates of its properties.
Recently, the Sloan Digital Sky 
Survey-Moving Object Catalog data, fourth release (SDSS-MOC4 
hereafter, Ivezic et al. 2002), provided multi-band photometry on 
a sample two order of magnitude larger than any available in any 
current spectroscopic catalogs (about 60000 numbered objects).   While 
for the purpose of deriving very reliable inferences about asteroid 
surface compositions, multi-band photometry is not as precise as 
spectroscopy.  Nesvorn\'{y} et al. (2005) showed that the 
SDSS-MOC is a useful data-set to study general statistical variations
of colors of main belt asteroids.  These authors used an automatic
algorithm of Principal Component Analysis (PCA) to analyze SDSS 
photometric data and to sort the objects into different taxonomical 
classes. In particular, PCA can be used to derive linear combinations of the 
five SDSS colors $(u, g, r, i, z)$, in order to maximize the separation 
between a number of different taxonomic classes in SDSS data.  Two 
large separated complexes were found in the PCA first two components: 
the C/X complex and the S complex, with various subgroups identified 
inside the main complexes.  A problem with this approach was however
the large errors that affected colors in the ultraviolet band u, and
that propagated into the computation of the principal components.
To avoid including the $u-$data, other authors (Ivezi\'{c} et al. 
2002, Parker et al. 2008) constructed a color-code diagram in a 
$(a^{*}, i-z)$ plane, where

\begin{equation}
a^{*}=C_1 *(g-r)+C_2 *(r-i)+C_3,
\label{eq: a_ast}
\end{equation}

\noindent and $C_1, C_2,$ and $C_3$ are numerical coefficients
that depend on the color values and on the number of observations
in the given database (Roig and Gil-Hutton 2006), and 
$g,r,i,$ and $z$, the other SDSS colors, had an accuracy 
of about 0.03 magnitudes, higher than the average errors in the $u-$band.
As in the plane of $(PC_1, PC_2)$, asteroids divided in the 
$(a^{*}, i-z)$ plane into three fairly distinct groups, the C-complex 
($a^{*} < 0$) the S-complex ($a^{*} > 0,~i-z > -0.13$) , 
and the V-type asteroids ($a^{*} > 0,~i-z < -0.13$, Parker et al. 
2008)~\footnote{The $a^{*}$ color is nothing but the first principal 
component $PC_1$ of the data distribution in the (g-r) vs (r-i) 
color-color diagram.}.

Asteroid taxonomy is however also defined by the geometric albedo $p_V$ 
(roughly, the ratio of reflected radiation from the surface to incident 
radiation upon it, at zero phase angle, (i.e., as seen from the light 
source), and from an idealized flat, fully reflecting, diffusive 
scattering (Lambertian) disk with the same cross-section).  C-type 
asteroids tend to have lower values of geometric albedo than S-type 
ones, and Tholen asteroid taxonomy (Tholen 1989) 
used values of $p_V$ to distinguish 
classes of asteroids inside the X-complex, such as the M-, E-, and 
P-types.   Until recently, however, only about two thousand 
asteroids had reliable values of geometric albedos (see Tedesco 
et al. 2002).   Initial 
results from the Wide-field Infrared Survey Explorer (WISE) 
(Wright et al. 2010), and the NEOWISE (Mainzer et al. 2011) enhancement 
to the WISE mission recently allowed to obtain diameters and 
geometric albedo values for more than 100,000 Main Belt asteroids 
(Masiero et al. 2011), increasing the sample of objects for which
albedo values were known by a factor 50.  Masiero et al. (2011) showed
that, with some exceptions, such as the Nysa-Polana group, asteroid
families typically show a characteristic albedo for all members, and 
that a strongly bimodal albedo distribution was observed in the 
inner, middle, and outer portions of the Main Belt.

Previous works, such as Bus and Binzel (2002a,b), Nesvorn\'{y} et al. 
(2005), found asteroid families in extended domains of proper elements and 
SDSS-MOC4 principal components data in order to minimize the number of 
possible interlopers, but such an analysis was not extended to asteroids' 
geometrical albedo.   Here we take full advantage of the newly available 
WISE data and we introduce a new hierarchical clustering method (HCM, 
see Bendjoya and Zappal\'{a} (2002) for details of the method in
proper elements space) 
in a multi-domain space composed by asteroids proper elements $(a,e,sin(i))$,  
SDSS-MOC4 colors $(a^{*},i-z)$, and WISE 
geometric albedo $(p_V)$, to identify halos associated with 
main belt asteroid families.   The great advantage of this approach is
that any group identified in these domains will most likely to
belong to the same taxonomical group, since its members have to
share not only similar values of proper elements, but also of
taxonomically related information such as $(a^{*},i-z)$, and $p_V$.
A shortcoming of this approach is related to the more limited number
of asteroids that have data in the three domains at the same time, 
when compared with the larger number of objects that have only proper
elements and frequencies, only SDSS-MOC4 principal components data,
only WISE albedo data, or a dual combination of these three quantities.
However, groups determined with this new
approach may serve as a first step in determining the real orbital extension
of the families cores and halos, with a precision that other methods already
in use in the literature may lack.

This work is so divided: in Sect.~\ref{sec: halo_metr} we discuss
the basics of our approach for finding halos in the multi-domain space.
In Sect.~\ref{sec: comp} we compare the efficiency of this
approach in finding low numbers of interlopers with the results of
other methods used for identifying asteroid families.  In 
Sects.~\ref{sec: inn_bel},~\ref{sec: cen_bel},~\ref{sec: out_bel} we apply
our method to all the currently known major families in the inner, central,
and outer main belt.  In Sects.~\ref{sec: cybele} and ~\ref{sec: hungaria} 
we discuss the case of the asteroids in the Cybele group and in the 
Hungaria region.  Finally, in 
Sect.~\ref{sec: concl} we present our conclusions.

\section{Methods}
\label{sec: halo_metr}

In this work we are trying to make best use of all the new data
on surface colors (SDSS-MOC4) and geometric albedo (WISE and NEOWISE) that 
is currently available to try to find the most possibly accurate determination
of all major main belt family halos.
For this purpose we determined the main belt asteroids 
with synthetic proper elements available at the AstDyS site 
http://hamilton.dm.unipi.it/cgi-bin/astdys/astibo, 
accessed on January $15^{th}$, 2013 (Kne\v{z}evi\'{c} and Milani 2003)
that also have SDSS-MOC4 and WISE albedo data, and errors 
in proper elements $(a,e,sin(i))$ less than what described 
as ``pathological'' in  Kne\v{z}evi\'{c} and Milani (2003), i.e., 
$\Delta a > 0.01~$AU, $\Delta e > 0.1$, and $\Delta sin(i) > 0.03$.
We computed the SDSS-MOC4 colors $(a^{*}, i-z)$ and theirs
errors, computed with standard propagation of uncertainty formulas 
under the assumption that the SDSS-MOC4 calibrated magnitude behave as 
uncorrelated variables. For our sample of 58955 asteroids with SDSS colors, we 
found values of the coefficients $C_1,C_2,$ and $C_3$ in Eq.~\ref{eq: a_ast} of 
0.93967, 0.34208 and -0.6324, respectively.  To avoid
including data affected by too large uncertainties, we eliminated
from our sample asteroids with errors in $a^{*}$ or $(i-z)$
larger than 0.1 magnitudes.  As a test of the validity of our
approach we also computed $PC_1, PC_2$ principal components
according to the approach of Novakovic et al. (2011), with their
errors, and also rejected objects with errors larger than 0.1.
While 68.1\% of the asteroids in the SDSS-MOC4 sample passed the 
conversion into $(a^{*}, i-z)$ colors and the rejection of noisy data,
only 42.08\% of the same asteroids had errors in  $PC_1, PC_2$ less 
than 0.1~\footnote{The large rejection of noisy data in this later 
approach is due to the inclusion of the magnitudes in the u filter, which 
are affected by larger errors that the magnitudes in the other filters. 
An alternative approach based on principal
components $PC_1, PC_2$ only in $g,r,i,$ and $z$ colors domain was also tried.
67.7\% of our data passed the conversion into this space
with errors less than 0.1.  Since the $(a^{*}, i-z)$ approach 
was slightly more efficient and it provided results that 
are easier to analyze in terms of taxonomies than the 
principal component approach, in this work we have decided to 
opt for the  $(a^{*}, i-z)$ method.}.  
Based on these results, we decided to work with 
$(a^{*}, i-z)$ colors rather than principal components.   
We also eliminated from our sample
asteroids with errors in $p_V$ larger than 0.05 if $p_V < 0.2$, 
and asteroids with errors in $p_V$ larger than 0.1 if $p_V > 0.2$.
The stringent constraint on errors in geometric albedo $p_V$ for
low albedo asteroids was required to better distinguish between
CX-complex asteroids ($p_V < 0.1$) and  S-complex asteroids ($p_V > 0.1$).
Since some objects in the inner main belt and Hungaria region have 
values of albedo in the WISE survey that are too high (up to 0.8-0.9) and are 
possibly an artifact of the method used to calculate absolute magnitude
(Masiero et al. 2011), we also eliminated all objects in these
two regions (essentially those with semi-major axis smaller than 
that of the center of the 3J:-1A mean-motion resonance, i.e., about 
2.5~AU) with $p_V > 0.5$ from our database.

\begin{figure}

  \centering
  \centering \includegraphics [width=0.45\textwidth]{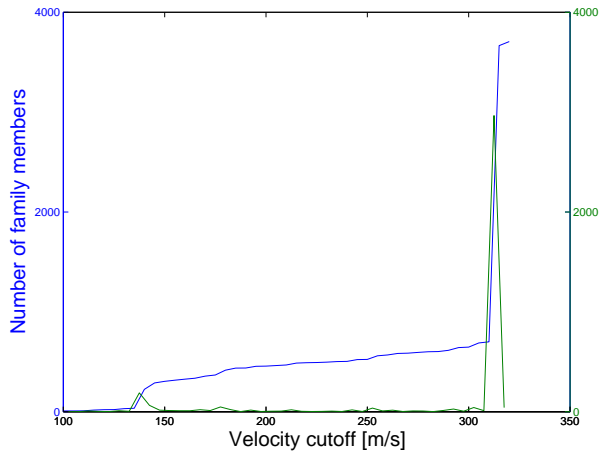}

\caption{The number of members (blue line) and new members (green line)
of the Themis group as a function of the velocity cutoff $d_{md}$.}
\label{fig: fam_count}
\end{figure}

We then defined a distance metrics between two asteroids in a 
multi-domain space as

\begin{equation}
d_{md} = \sqrt{d^2+C_{SPV}[(\Delta {a^{*}})^2+ (\Delta (i-z))^2 + 
(\Delta {p}_V)^2]},
\label{eq: multi_dom_metr}
\end{equation}

\noindent where, $\Delta a^{*} = a^{*}_2- a^{*}_1$ and similar relations
hold for $\Delta (i-z)$ and $\Delta {p}_V$.  
Following the approach of Bus and Binzel (2002a,b) for
a similar distance metric of proper elements and SDSS-MOC principal
components (see also Nesvorn\'y et al. 2005,  
Carruba and Michtchenko 2007), $C_{SPV}$ is a 
weighting factor 
equal to $10^6$  (other choices in a range between $10^4$ to $10^8$ have been
tested without significantly changing the robustness of the results), and
$d$ is the standard distance metrics in proper element domain 
defined in Zappal\'{a} et al. (1995) as:

\begin{equation}
d = na \sqrt{k_1 (\frac{\Delta a}{a})^2 +k_2(\Delta e)^2+k_3(\Delta sin(i))^2}, 
\label{eq: stand_metr}
\end{equation}

\noindent 
where $n$ is the asteroid mean motion; $\Delta x$ the difference in proper
$a, e,$ and $sin(i)$; and $k_1, k_2, k_3$ are weighting factors, defined as
$k_1$ = 5/4, $k_2$ = 2, $k_3$ = 2 in Zappal\'{a} et al. (1990, 1995).  
As first halo members, we selected 
asteroids that belong to the asteroids family, whose spectral type is 
compatible with that of the other members according to Moth\'{e}-Diniz 
et al. (2005), Nesvorn\'{y} et al. (2006), Carruba (2009a,b, 2010b) and other 
authors, and that, of course, also have acceptable 
SDSS-MOC4 and WISE/NEOWISE data.  For families
not treated by these authors, we consulted the list of asteroid families
available at the AstDyS site, and the Nesvorn\'{y} (2012) 
HCM Asteroid Families V2.0, on the Planetary Data System, 
available at http://sbn.psi.edu/pds/resource/nesvornyfam.html, 
accessed on March $13^{th}$~2013.  We then 
obtained dynamical groups using Eq.~\ref{eq: multi_dom_metr}, for a value
of cutoff $d_{md}$ a bit less than the value for which the family halo 
merges with the local background (and other families in the region).
As an example of this procedure, we choose the case of the Themis
family.  Fig.~\ref{fig: fam_count} displays the total number of 
members of this
group (blue line) and the number of new members of the group (green
line, as new members we mean the number of objects that became part
of the group at that given velocity cutoff), as a function of the
velocity cutoff $d_{md}$.   For $d_{md} = 315$~m/s the Themis halo
merged with other local groups, such as the Hygiea family, so we choose in this
case to work with a halo defined at $d_{md} = 310$~m/s.

\begin{figure*}

  \begin{minipage}[c]{0.5\textwidth}
    \centering \includegraphics[width=3.0in]{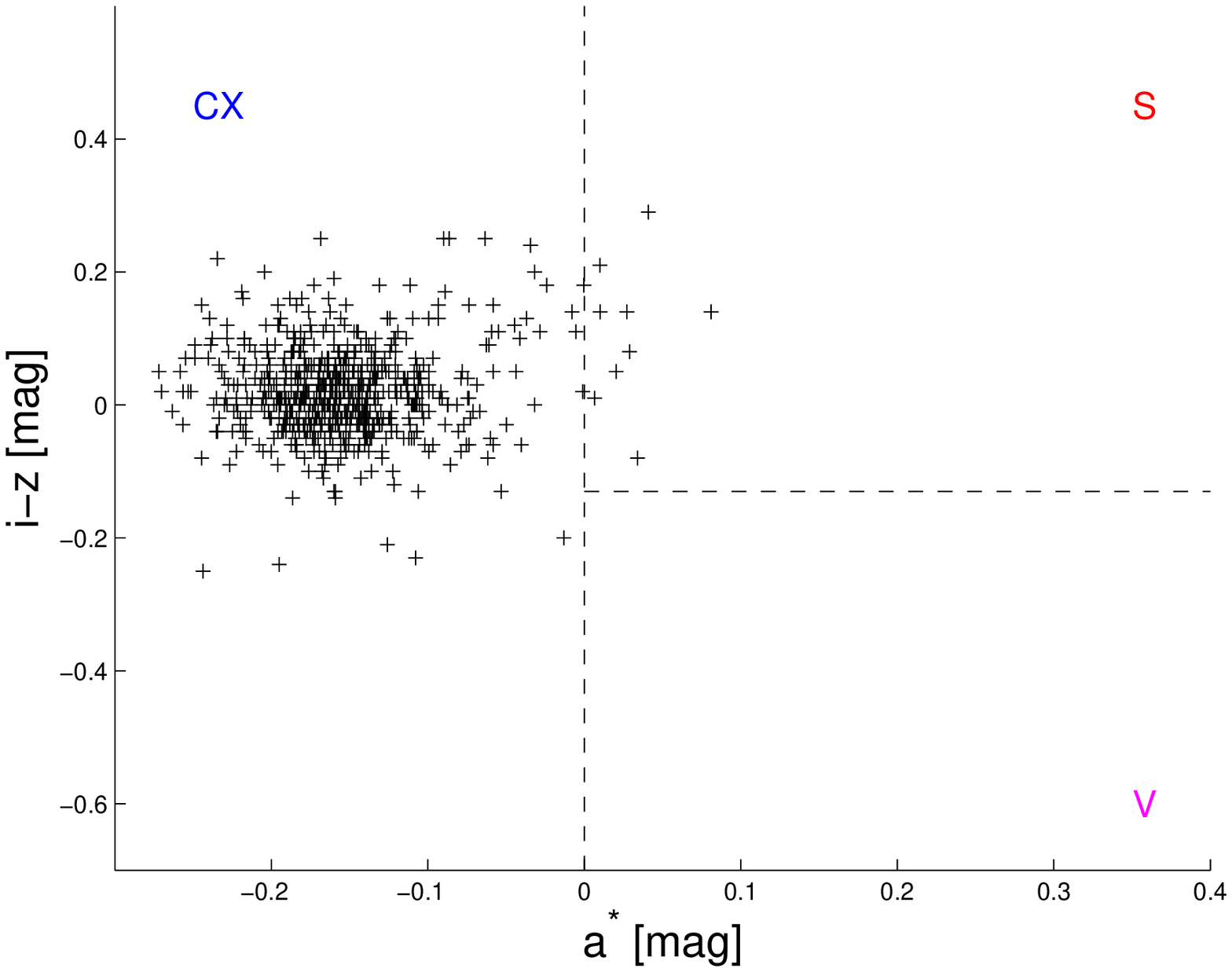}
  \end{minipage}%
  \begin{minipage}[c]{0.5\textwidth}
    \centering \includegraphics[width=3.0in]{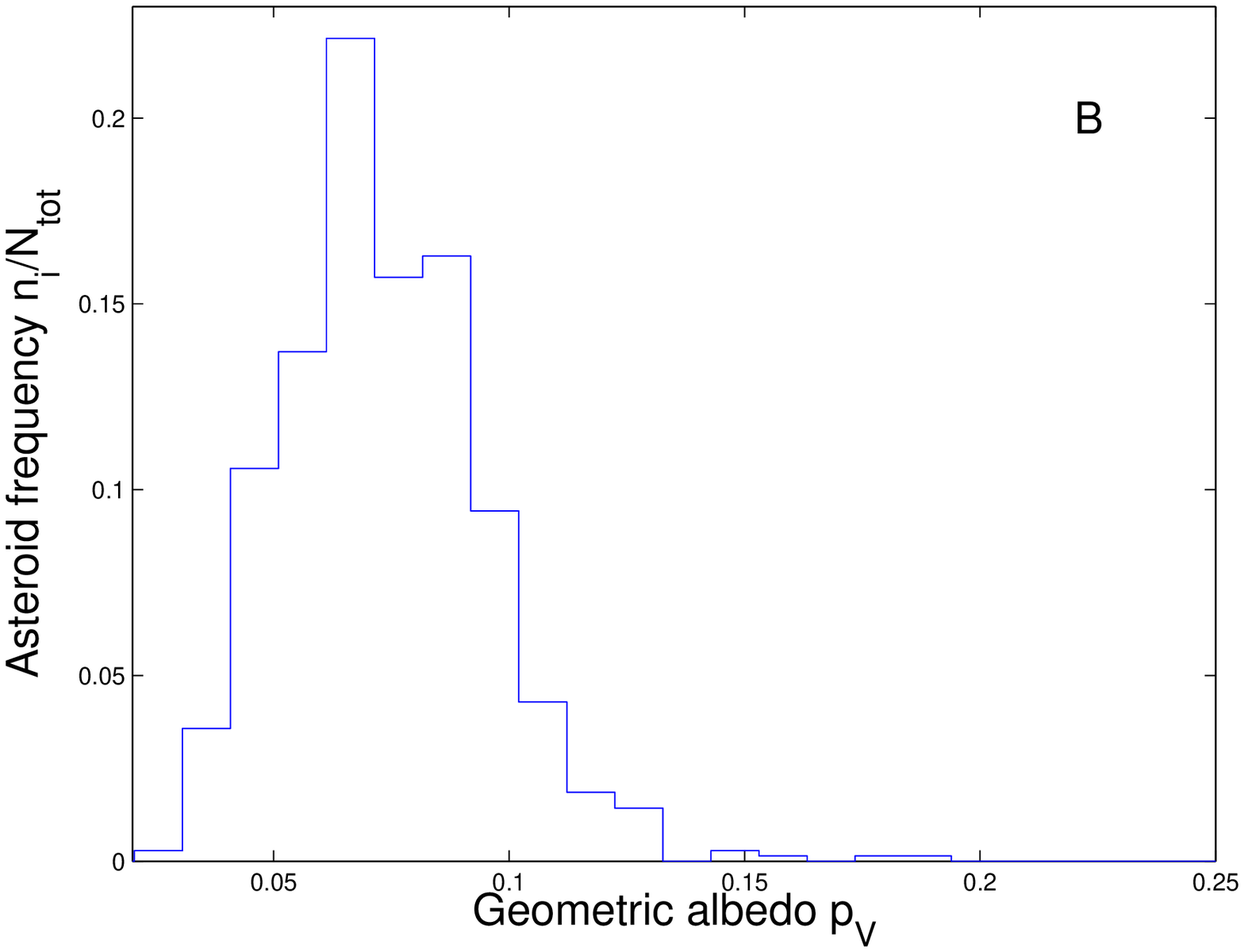}
  \end{minipage}

\caption{An $(a^{*},i-z)$ projection (panel A), 
and a histogram of the relative distribution of $p_V$ 
values (panel B) of members of the Themis halo.}
\label{fig: aiz_PV}
\end{figure*}

An advantage of the method here proposed 
is that it should automatically select 
asteroids close in proper elements, SDSS-MOC4, and WISE albedos, so 
reducing the number of interlopers usually found in dynamical group 
encountered in proper elements (or frequencies) domains only.
This can be verified by an analysis of the SDSS-MOC4 and WISE albedo
data of the group so obtained.  Again, for the case of the Themis
family,  Fig.~\ref{fig: aiz_PV} shows a projection in the $(a^{*},i-z)$
plane (panel A) and a histogram of the relative distribution of $p_V$ 
values (panel B) of members of the Themis halo obtained with this 
method.   The Themis family is made mostly by asteroids with CX-complex 
taxonomy, that in the $(a^{*},i-z)$ plane appear on the left of
the vertical dotted line, but 9 asteroids have colors incompatible
with such classification and should be considered as interlopers.
This is confirmed by an analysis of $p_V$ values, where most 
of the albedos are below 0.1, the threshold for CX-complex asteroids,
but there is a tail of objects with higher albedos.  The percentage of 
possible interlopers found with this method, 1.30\%, is indeed quite inferior
to the $\simeq 10$\% statistically expected in dynamical families obtained
only in proper elements domains (Migliorini et al. 1995).
We will discuss how this new approach fares when compared with other
methods already known in the literature in the next section.

\section{Comparison with HCM in other domains}
\label{sec: comp}

A natural question that may arise is why studying families halos in 
a domain of proper elements, SDSS-MOC4 colors, and WISE
albedos.  How do the results obtained with this approach compare
to those obtained with more traditional methods, such as the HCM
in proper elements domain, or  
in a domain of proper elements and SDSS-MOC $a^{*}$ and $i-z$ colors?
To answer this question we obtained asteroid families halos 
for several groups with 
the standard distance metrics in proper element domain 
$d$ of Zappal\'{a} et al. (1995), with a metrics 
in proper elements and SDSS-MOC4 colors
domains, given by:

\begin{equation}
d_{md} = \sqrt{d^2+C_{SPV}[(\Delta {a^{*}})^2+ (\Delta (i-z))^2]},
\label{eq: bus_metr}
\end{equation}

\noindent where $C_{SPV}$ , as discussed in Sect.~\ref{sec: halo_metr}, 
 is a weighting factor equal to $10^6$, and with a newly defined
distance metric in proper elements and WISE geometric albedo $p_V$ domain,
given by:

\begin{equation}
d_{md} = \sqrt{d^2+C_{SPV}(\Delta {p}_V)^2}.
\label{eq: pv_metr}
\end{equation}

\noindent We determined families halos with the standard metrics of
Zappal\'{a} (1995), and Eqs.~\ref{eq: bus_metr},~\ref{eq: pv_metr}, 
and ~\ref{eq: multi_dom_metr} for several large families in the main 
belt .  Table~\ref{table: halo_metrics} summarizes our results 
for the Hygiea, Koronis, and Eos
family halos, where we report the value of the cutoff at which 
the family was found, the number of halo members, the percentage of
SDSS-MOC4 and geometric albedo likely interlopers (see 
Sect.~\ref{sec: halo_metr} for a definition of the concept of SDSS-MOC4
and geometric albedo likely interlopers), for the four methods
that we used (we will refer to these methods as 
metrics $D$, $DS$, $DPV$, and $DSPV$).  
The last column, which reports
the sum of the percentage of SDSS-MOC4 and geometric albedo likely 
interlopers, gives a measure of the efficiency of the method in finding
likely interlopers: the lower this index, the better the method is
working in avoiding taxonomically uncorrelated asteroids to the family 
halo.  Among the several large families halos that we analyzed,
we choose to display the results for the Hygiea, Koronis, and Eos
groups because these are families for which the new multi-domain
method showed one of the best, medium, and worse results in term
of not finding interlopers when compared with the other methods, respectively.

\begin{table*}
\begin{center}
\caption{{\bf Efficiency of distance metrics in several domains
in finding asteroid families halo members.}}
\label{table: halo_metrics}
\vspace{0.5cm}
\begin{tabular}{|c|c|c|c|c|c|}
\hline
 &              &                   &                      &      &    \\
Family & $d_{md}$ cutoff value & Number of & Percentage of SDSS-MOC4 & 
Percentage of $p_V$ &  Metric efficiency \\ 
Name & [$m/s$] &  members &  likely interlopers  & likely interlopers &     \\
     &            &                   &                      &      &    \\
\hline
Hygiea halo  &     &                   &                      &      &    \\
\hline
     &            &                   &                      &      &    \\
Metric $D$      &  80 & 6152 &  7.58 & 15.59 & 23.17 \\
Metric $DS$     & 225 &  728 &  5.63 & 14.02 & 19.65 \\
Metric $DPV$    & 140 & 2977 &  8.63 &  6.45 & 15.08 \\
Metric $DSPV$   & 290 &  426 &  1.41 &  8.45 &  9.86 \\ 
     &            &                   &                      &      &    \\
\hline
Koronis halo  &     &                   &                      &      &    \\
\hline
     &            &                   &                      &      &    \\
Metric $D$      &  65 & 6958 & 15.71 & 14.36 & 30.07 \\
Metric $DS$     & 230 & 1054 & 13.54 & 13.11 & 26.65 \\
Metric $DPV$    &  80 & 1366 & 14.32 & 21.74 & 36.06 \\
Metric $DSPV$   & 215 &  200 & 16.00 &  7.00 & 23.00 \\ 
     &            &                   &                      &      &    \\
\hline 
Eos halo  &     &                   &                      &      &    \\
\hline
     &            &                   &                      &      &    \\
Metric $D$      &  40 & 5322 & 52.70 & 17.35 & 70.05 \\
Metric $DS$     & 135 & 1886 & 55.99 & 11.08 & 67.07 \\
Metric $DPV$    &  80 &  846 & 57.01 & 18.94 & 75.95 \\
Metric $DSPV$   & 165 &  738 & 51.36 & 17.89 & 69.25 \\ 
             &    &                   &                      &      &    \\
\hline
\end{tabular}
\end{center}
\end{table*}

The Hygiea family case was the one for which the new method had the best
results among the families analyzed, with an overall efficiency of only 
9.86\%.  In the case of the Koronis family the efficiency was  
lower (23.00\%), but the new method still provided the best results 
when compared with other approaches.
The Eos family was a very peculiar case: most of the family members
are K-type, an S-complex type whose $a^{*}$ values are very close to zero,
the limiting value separating CX-complex asteroids and S-complex ones
The family is surrounded by CX-complex asteroids, and 
an analysis only based on distance metric inevitably recognizes
as family members many background objects not necessarily connected to the
family. Only in the case of this family, we found an efficiency
of the new method slightly inferior to the results of the $DS$ metric 
(69.25\% with respect to 67.07\%).  Overall, the new approach was at its best
a factor of two more efficient in eliminating interlopers than other
methods, and at its worse provided comparable results to 
what obtained in the domain of proper elements and SLOAN colors.

Having concluded that the method described by Eq.~\ref{eq: multi_dom_metr}
is the most efficient in term of low numbers of interlopers, we are now
ready to start our analysis of the main belt.  We will do this by 
investigating asteroids family halos in the inner main belt.

\section{Inner Main Belt}
\label{sec: inn_bel}

The inner main belt is dynamically limited in semi-major axis by the 
3J:-1A mean-motion resonance at high $a$ (Zappal\'{a} et al. 1995).
The 7J:-2A mean-motion resonance is sometimes used by some authors
as the boundary between the inner main belt at high inclination
and the region of the Hungaria asteroids.   In this work we will
just use the upper limit in $a$ given by the 3J:-1A mean-motion resonance.
The linear secular resonance ${\nu}_6$ separates the low-inclined 
asteroid region from the highly inclined area, dominated by the Phocaea family
(see Carruba 2009b, 2010a for a discussion of the local families 
groups and dynamics).  The Phocaea family is located in a stable island
limited by the 7J:-2A and 3J:-1A in semi-major axis, and by the 
${\nu}_6$ and ${\nu}_5$ secular resonances in inclination (Kne\v{z}evi\'{c}
and Milani 2003).  We found 2366 objects that have proper elements and 
frequencies, SDSS-MOC $(a^{*},i-z)$ colors, WISE geometric albedo data 
in the inner main belt, and reasonable errors, according to the criteria 
defined in Sect.~\ref{sec: halo_metr}.  We will start our analysis 
by studying the case of the Belgica family.

\subsection{The Belgica family}
\label{sec: belgica}

The Belgica group was a clump associated with the former Flora family and
identified by Moth\'{e}-Diniz et al. (2005) as a small and sparse group
of only 41 members at a cutoff in proper element domain of 57.5~$m/s$.
Here we found that the halo of the Belgica family merges with that of 
the Baptistina group already at a cutoff of 100~$m/s$.  We will therefore 
treat the Belgica family together with the Baptistina cluster.

\subsection{The Baptistina family}
\label{sec: baptistina}

The Baptistina family, as the Belgica group, was studied by Moth\'{e}-Diniz
et al. (2005) and was part of the former Flora family.  It is located
in a very complex dynamical region (Michtchenko et al. 2010), being 
crossed by powerful mean-motion resonances such as the 7J:-2A
and interacting with secular resonances such as the $z_2 = 2(g-g_6)+(s-s_6)$.  
It has been obtained in the $(n,g,s)$ frequency domain by Carruba 
and Michtchenko (2009) to study possible diffusion of its members 
in $s-$type resonances such as the ${\nu}_{17}+{\nu}_4+{\nu}_5-2{\nu}_6$.
Here we identified a 56 members CX halo at a cutoff of $250~m/s$.
The taxonomical structure of the halo is indeed very complex and puzzling.
The majority of the Baptistina halo members have SDSS-MOC4 data compatible
with a CX-complex taxonomy, with only 2 members (3.6\% of the total) that 
are possible interlopers.  The albedo data is however very puzzling, since
47 members (83.9\% of the total) have values of $p_V > 0.1$, not usually 
associated with dark CX-complex asteroids. Baptistina family members
seem to behave as the members of the Hungaria group, a CX-complex
family, characterized by large values of albedos (see 
Sect.~\ref{sec: hungaria_fam}).  Understanding the properties
of the Baptistina halo will require a much more in depth analysis than
what we performed in this work.

\subsection{The Vesta family}
\label{sec: vesta}

The Vesta family is unique in the main belt, since it is made mostly by V-type
asteroids, that are associated with a basaltic composition, typical of
differentiated objects with a crust.  Of the many possible differentiated
or partially differentiated asteroids that may have existed in the primordial
main belt, (4) Vesta is the largest remnant for which a basaltic crust
is still present and was observed by a space mission (Russel et al. 2012).
Many V-type objects are observed outside the limits of the traditional HCM 
family (Carruba et al. 2005, Nesvorn\'{y} et al. 2008), 
making this family a test-bed for the application of methods on halo 
determinations.  

We determined a 161 members halo at a cutoff of $d_{md}~=~275~m/s$.  46 
halo members (28.6\% of the total, a considerable fraction of the halo) 
are possible SDSS-MOC4 interlopers, and 26 asteroids (16.2\% of the total)
are possible albedo interlopers.  Among the asteroids with $a^{*} > 0$, 
58.6\% are in a region of the ($a^{*},i-z$) plane associated
with V-type objects, according to the criteria defined in 
Sect.~\ref{sec: intro}, and can be considered as possible V-type 
candidates.

How efficient is the new method in identifying V-type asteroids outside
the Vesta family as members of the halo?  Among the V-type asteroids not 
connected to the traditional HCM Vesta family listed in Carruba et al. 
(2005), only four objects are present in our multi-domain
catalog: (3849), (3869), (4188), and (4434).  Of these, two (50\% of the 
total), (3869) and (4188), were part of the Vesta halo as found by our method.  
Having such a limited sample of objects in our catalog and in the halo, 
we are not able to achieve any conclusions on the validity of the method 
for the Vesta halo.  The Sidwell, Somekawa, Henninghaack, and 
Ausonia AstDyS family merge with the Vesta halo at a cutoff of less than 
$320~m/s$.

\subsection{The Erigone family}
\label{sec: erigone}

The Erigone family was identified by Nesvorn\'{y} et al. (2006) as 
a CX-complex group at a cutoff of $80~m/s$.  Their analysis appears
to be confirmed by this work: we found a 57 member CX-complex
group at a cutoff $d_{md}$ of 400$~m/s$.   No SDSS-MOC4 interlopers 
were found in the halo, and only one object (1.8\% of the total) 
was (barely) a possible albedo interloper.  The Maartes AstDyS family
merges with the Erigone halo at a cutoff of $d_{md} = 165 ~m/s$.

\subsection{The Massalia family}
\label{sec: massalia}

The Massalia family was identified by Nesvorn\'{y} et al. (2006) as 
an S-complex family at a cutoff of $50~m/s$.  In this work we identified
19 S-complex members at a cutoff $d_{md}$ of 250$~m/s$.  Seven SDSS-MOC4 
interlopers (36.8\% of the total) were found in the halo, and
10 objects (52.6\% of the total) were possible albedo interlopers. 
Incidentally, (20) Massalia itself is a C-type asteroid, and quite
likely an interloper in its own family. 

\subsection{The Nysa/Mildred/Polana family}
\label{sec: Nysa/Mildred/Polana}

The Nysa/Polana family was studied by Moth\'{e}-Diniz et al. (2005) that
confirmed previous results about the dual structure of the family,
made by an S-type member group around (878) Mildred, and an F-type
group around (142) Polana.
In this work we find a CX-complex halo of 147 members at a cutoff $d_{md}$ 
of 280$~m/s$.  The halo is dominated by the CX-complex Polana group,
that is also made by the largest bodies in the area (Moth\'{e}-Diniz 
et al. 2005):  we found no possible SDSS-MOC4 interlopers, and
1 (0.7\%  of the total) albedo interloper.  A smaller halo associated
with the Mildred family merges with the larger Polana halo at about
150$~m/s$, and the Clarissa Planetary Data System is englobed at 
a cutoff of 345~$m/s$.

\subsection{The Euterpe family}
\label{sec: euterpe}

The Euterpe family is a low-inclination 
group listed by the Planetary Data System.  In this
work we identified an S-complex halo at a cutoff $d_{md}$ 
of 335$~m/s$.  Two objects (22.2\% of the total) were possible  
SDSS-MOC4 interlopers, and one asteroid (11.1\% of the total) is a possible
albedo interloper.

\subsection{The Lucienne family}
\label{sec: lucienne}

The Lucienne family is a relatively high-inclination group listed by the 
Planetary Data System. Unfortunately we could not find any member of this
group in our multi-domain sample of asteroids for the inner main belt, so
no conclusions are possible to achieve on this cluster.

\subsection{The Phocaea family}
\label{sec: phocaea}

The Phocaea family has been studied by Kne\v{z}evi\'{c} and Milani (2003) and 
by Carruba (2009b).  It is located in a stable island bounded by the ${\nu}_6$
and ${\nu}_5$ secular resonances in inclination and the 7J:-2A and
3J:-1A mean-motion resonances in semi-major axis.  Despite the peculiar
dynamical configuration, Carruba (2009b) concluded that it was likely that 
the Phocaea family was a real S-complex collisional family, with an estimated
age of about 2.2 Byr.  In this work we found an 80 members S-complex 
halo at a cutoff $d_{md} > 800~m/s$, which is the value for which all 
asteroids in the stable island in our database were found connected to 
the Phocaea family.  27 objects (33.8\% of the total) were possible
SDSS-MOC4 interlopers, and 16 asteroids (20.0\% of the total) 
had values of $p_V < 0.1$.  Overall, we confirm the analysis of Carruba 
(2009b) on the possible reality of the Phocaea family as an S-complex
collisional group.

\subsection{The inner main belt: an overview}
\label{sec: inner_sum}

Our results for the inner main belt are summarized 
in Table~\ref{table: halo_inner}, where we give
information on the first halo member used to determine the family halo,
the cutoff value used to identify the halo, the number of bodies in the halo,
the spectral complex to which the majority of halo members belongs,
and the number of possible interlopers, according to SDSS-MOC4 and geometric
albedo considerations.

\begin{table*}
\begin{center}
\caption{{\bf Asteroid families halos in the inner main belt.}}
\label{table: halo_inner}
\vspace{0.5cm}
\begin{tabular}{|c|c|c|c|c|c|}
\hline
                 &                &    &                      &         & \\
First halo & $d_{md}$ cutoff value &   Number of & Spectral &Number of SDSS-MOC4  & Number of $p_V$ \\ 
 member    &   [$m/s$] & members  &   Complex  & likely interlopers &  likely interlopers \\
                 &                   &   &                   &         &  \\
\hline
                 &      &     &     &    &   \\
(298) Baptistina: (4691)&250  &  56 & CX  & 2 & 47\\
(4) Vesta: (2011)    & 275  & 161 & S(V)& 46& 26\\
(163) Erigone: (9566)  & 400  &  57 & CX  & 0  &  1\\
(20) Massalia: (10102) & 250 &  19 &  S  & 7  &  10\\
(44) Nysa/Mildred/Polana: (1768)&280 &147&CX&0&  1\\
(27) Euterpe: (5444) & 335   &   9 &  S  & 2 &  1\\
(25) Phocaea: (3322)  & $>800$ &80 & S  & 27 & 16\\
                 &      &     &     &    &   \\
\hline
\end{tabular}
\end{center}
\end{table*}

\begin{figure*}

  \begin{minipage}[c]{0.5\textwidth}
    \centering \includegraphics[width=3.0in]{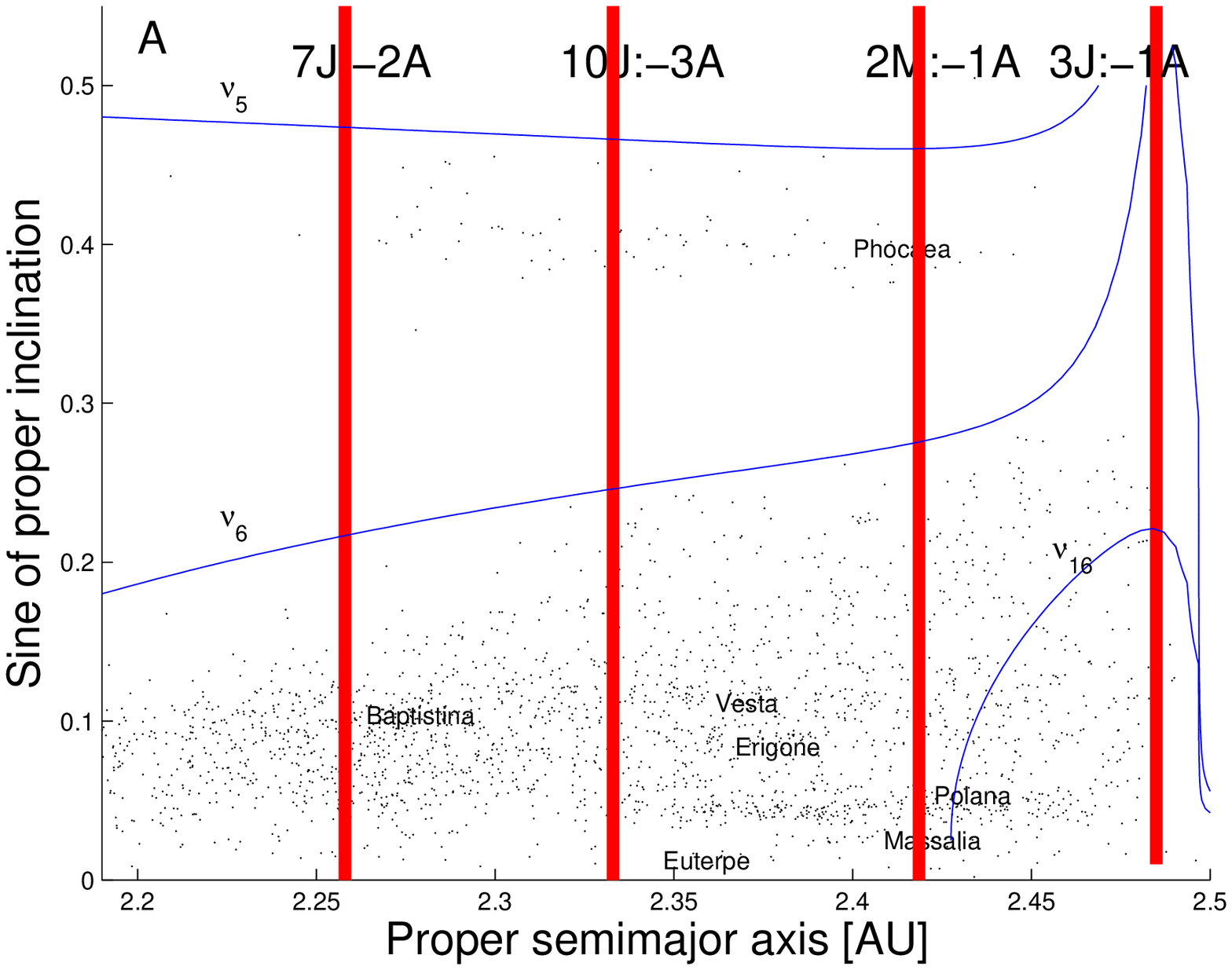}
  \end{minipage}%
  \begin{minipage}[c]{0.5\textwidth}
    \centering \includegraphics[width=3.0in]{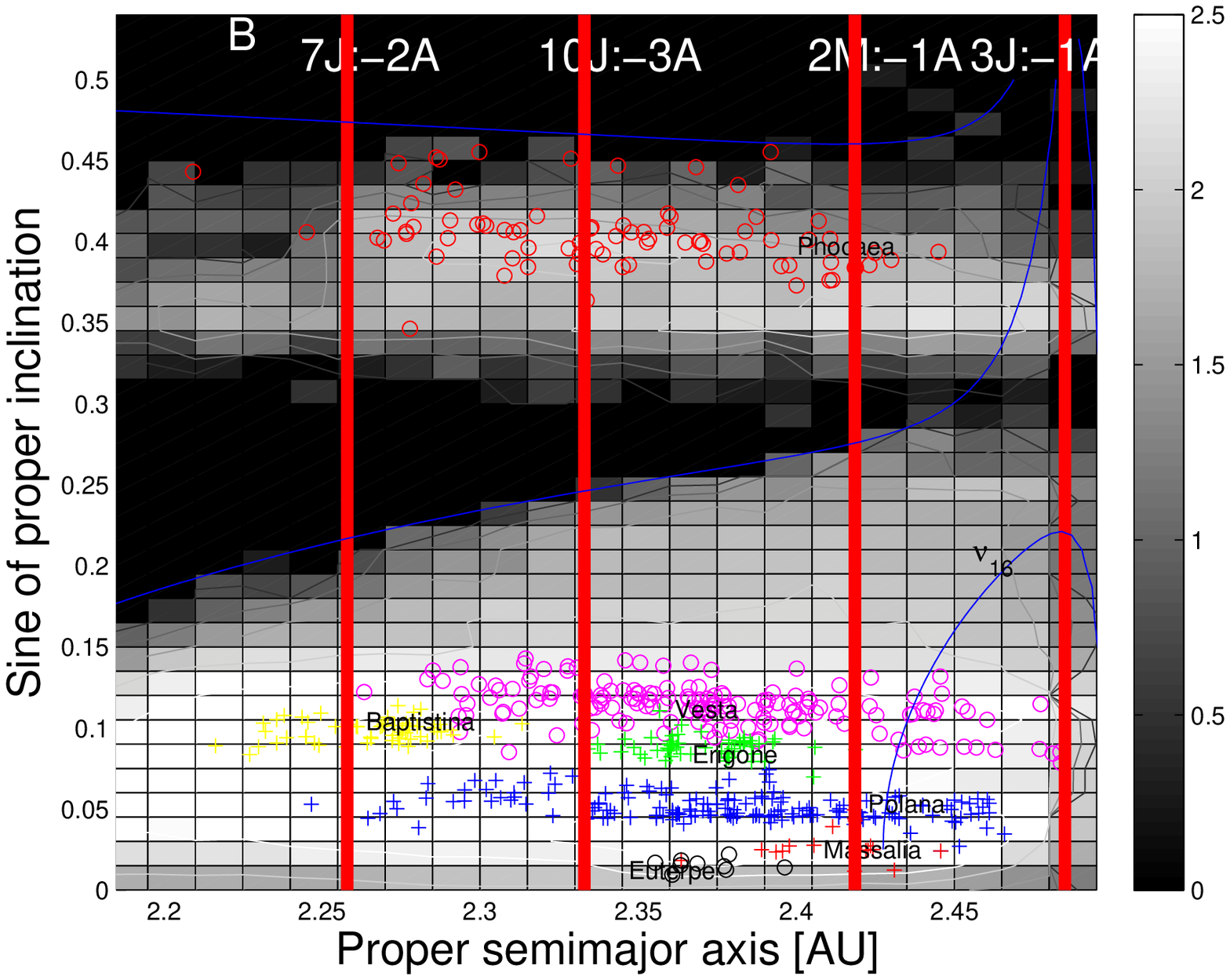}
  \end{minipage}

\caption{Panel A: An $(a,sin(i))$ projection of inner main belt 
asteroids in our multi-variate sample.  Panel B: contour plot
of the number density of asteroids in the proper element sample.  
Superimposed, we display the orbital location of asteroid 
families in the CX-complex (plus signs), and in the S-complex 
(circles).}
\label{fig: inner_el}
\end{figure*}

Fig.~\ref{fig: inner_el}, panel A, displays an $(a,sin(i))$ projection of
asteroids in our multi-variate sample in the central main belt.  Vertical red
lines identify the orbital position of the main mean-motion resonances
in the area.  Blue lines show the location of the main linear secular 
resonances, using the second order and fourth-degree secular perturbation
theory of Milani and Kne\v{z}evi\'{c} (1994)
to compute the proper frequencies $g$ and $s$ for the grid of $(a,e)$
and $(a,\sin(i))$ values shown in Fig.~\ref{fig: inner_el}, panel A, and the
values of angles $ \Omega, \omega, M$, and eccentricity of (25) 
Phocaea, the highly inclined
asteroid associated with the largest family in the region (Carruba 2009b).
The orbital position in the $(a,\sin(i))$ plane
of the first numbered asteroid in all the inner main belt 
groups is also identified in Fig.~\ref{fig: inner_el}, panel A.  In panel B 
of the same figure we display a density map of the inner
main belt, according to the approach described in Carruba and Michtchenko 
(2009).  Density maps display regions characterized by strong mean-motion
or secular resonances by a relatively low number of asteroids per unit bin.
To quantitatively determine the local density of asteroids,  we
computed the $log_{10}$ of the number of all asteroids with proper elements 
per unit square in a 22 by 67 grid in $a$ (starting
at $a$~= 2.18~AU, with a step of 0.015 AU) and $sin(i)$ (starting at 
0, with a step of 0.015).  Superimposed to the density map, we also 
show the orbital projection
of the halos found in this work shown as plus signs for CX-complex
families, and circles for S-complex families.   
The other symbols are the same as in 
Fig.~\ref{fig: inner_el}, panel A.

\begin{figure*}

  \begin{minipage}[c]{0.5\textwidth}
    \centering \includegraphics[width=3.0in]{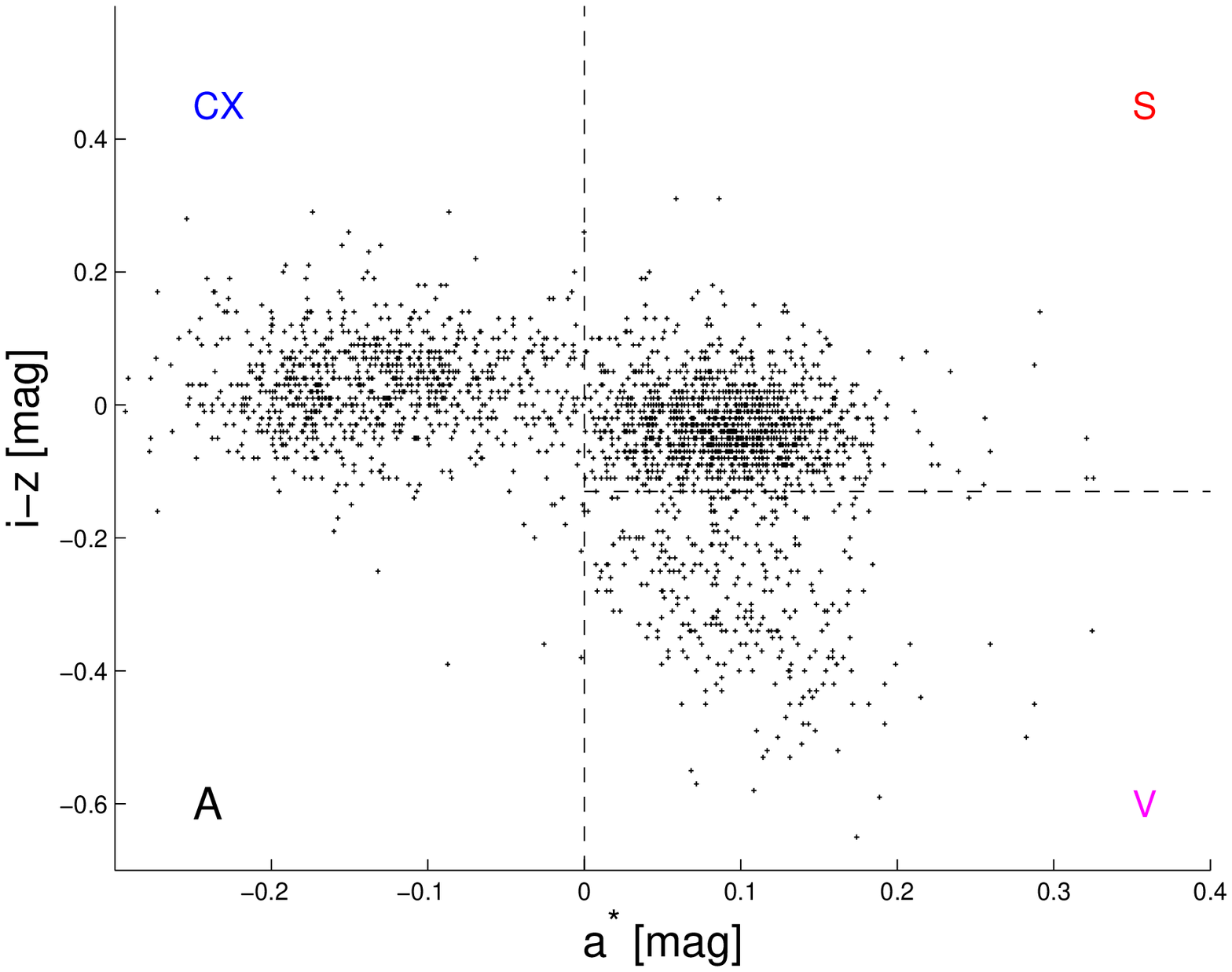}
  \end{minipage}%
  \begin{minipage}[c]{0.5\textwidth}
    \centering \includegraphics[width=3.0in]{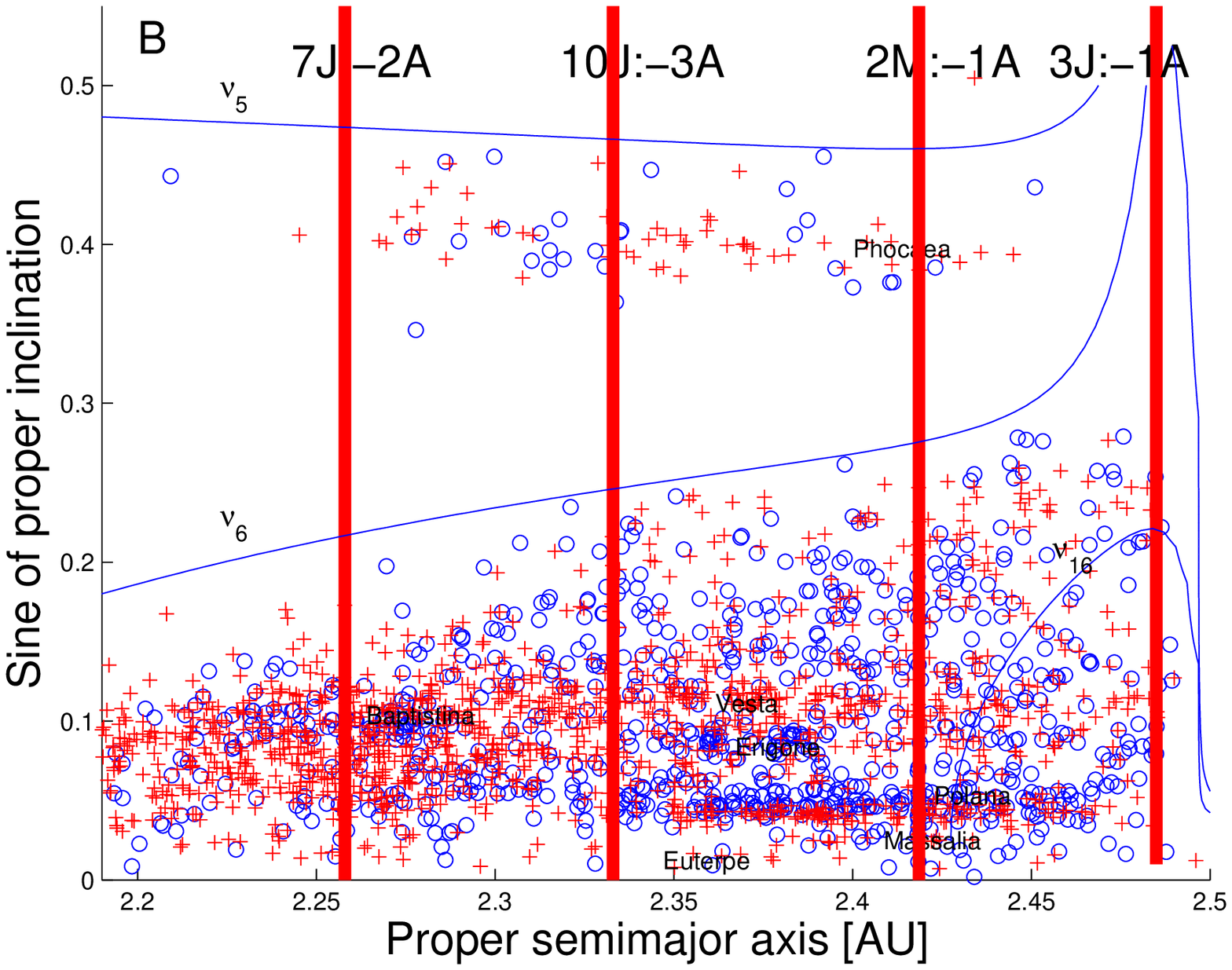}
  \end{minipage}

\caption{Panel A: an $(a^{*},i-z)$ projection of inner main belt 
asteroids in the in our multi-domain sample.  Panel B: an 
$(a,sin(i))$ projection of the same asteroids,
where objects in the CX complex are shown as blue circles, 
and asteroids in the S-complex are identified as red plus signs.}
\label{fig: inner_PC}
\end{figure*}

Fig.~\ref{fig: inner_PC} displays a projection 
in the $(a^{*},i-z)$ plane of all asteroids in our 
multi-domain sample (panel A), and an $(a,sin(i))$ projection of 
the same asteroids, (panel B), where objects in the CX complex are 
shown as blue circles, and asteroids in the S-complex are identified 
as red plus signs.
The inner main belt is slightly dominated by S-complex
asteroids, but with a significant minority of CX-complex bodies.
V-type asteroids are mostly concentrated in the Vesta family, but 
with a significant population outside the dynamical group
(see also Carruba et al. 2005).

\begin{figure*}

  \begin{minipage}[c]{0.5\textwidth}
    \centering \includegraphics[width=3.0in]{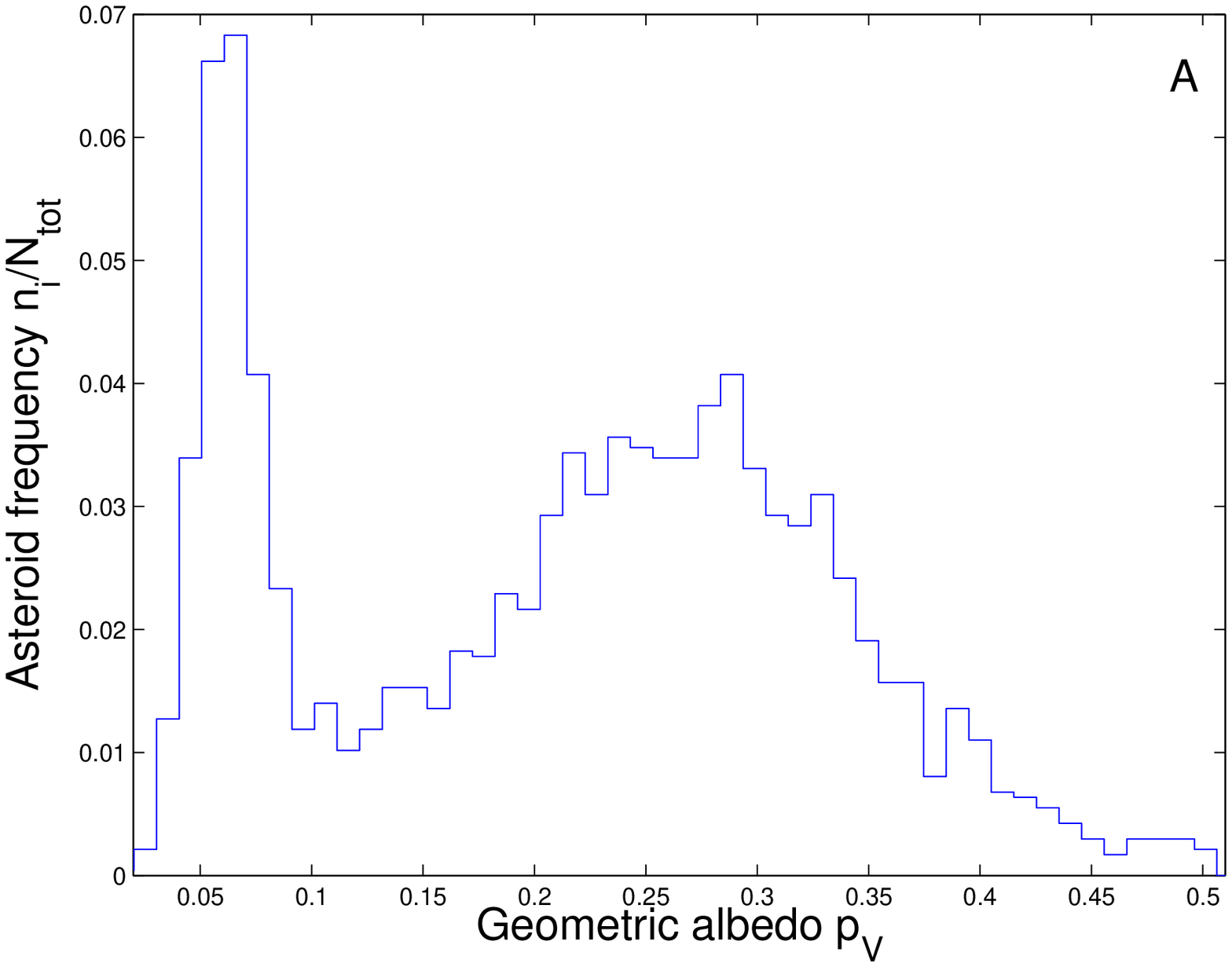}
  \end{minipage}%
  \begin{minipage}[c]{0.5\textwidth}
    \centering \includegraphics[width=3.0in]{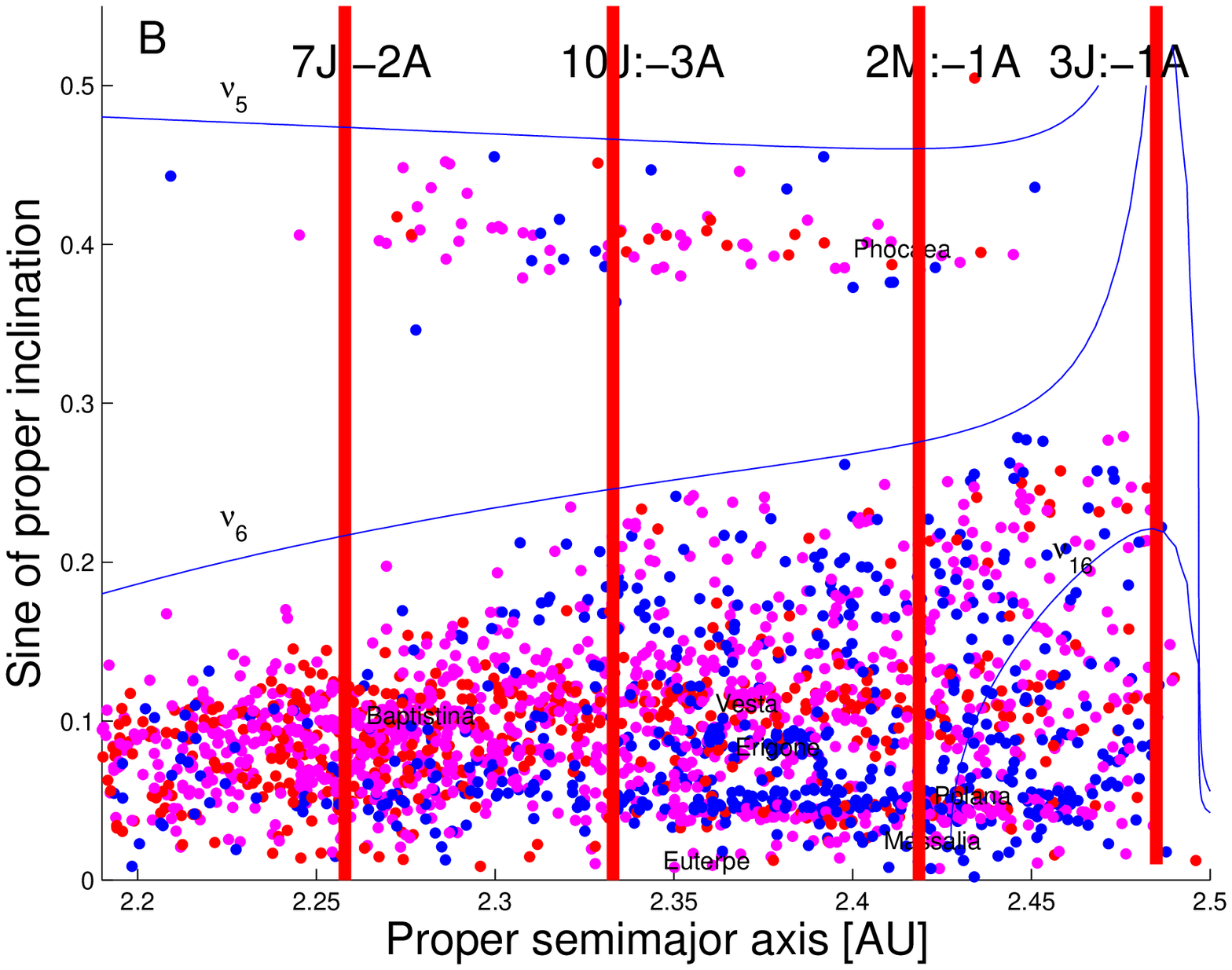}
  \end{minipage}

\caption{Panel A: a histogram of number frequency values 
$n_i/N_{Tot}$ as a function of geometric albedo $p_V$ for 
inner main belt asteroids in our multi-domain sample.  Panel B: an
$(a,sin(i))$ projection of the same asteroids,
where blue full dots are associated with asteroids with $p_V < 0.1$,
red full dots display asteroids with $0.1 < p_V < 0.3$, and magenta
full dots show asteroids with $p_V > 0.3$.}
\label{fig: inner_pv}
\end{figure*}

This is confirmed by an analysis of WISE
$p_V$ geometrical albedo data, a histogram of which is presented
in Fig.~\ref{fig: inner_pv}, panel A.  Fig.~\ref{fig: inner_pv}, panel B,
displays an $(a,sin(i))$ projection of the same asteroids,
where blue full dots are associated with asteroids with $p_V < 0.1$,
red full dots display asteroids with $0.1 < p_V < 0.3$, and magenta
full dots show asteroids with $p_V > 0.3$.  The majority of asteroids
in the inner main belt is made by high albedo objects, associated
with S-complex taxonomies, but with a significant minority of CX-complex
bodies.

\section{Central Main Belt}
\label{sec: cen_bel}

The central main belt is dynamically limited in semi-major axis by the 
3J:-1A and 5J;-2A mean-motion resonances  (Zappal\'{a} et al. 1995).   
The linear secular resonance ${\nu}_6$ separates the low-inclined 
asteroid region from the highly inclined area, dominated by the 
Hansa and Pallas families (Carruba 2010b).  As discussed in Carruba (2010b),
in the highly inclined region the local web of linear secular resonances
and mean-motion resonances divided the region into six separated stable islands,
each hosting one or more major families, and that can be considered
as a stable archipelago.  Of particular interest in this region 
is the Tina family, whose members are all in anti-aligned states of
the ${\nu}_6$ linear secular resonance (Carruba and Morbidelli 2011).
We found 3693 objects that have proper elements and frequencies, 
SDSS-MOC4 $a^*$ and $i-z$ colors, WISE geometric albedo data,
and satisfy our error analysis criteria, in the central
main belt, and we will start our analysis by studying the case of the 
Hestia family.

\subsection{The Hestia family}
\label{sec: hestia}

The Hestia family was identified in Nesvorn\'{y} et al. (2005) as a 154 members
group with S-taxonomy at a cutoff in proper element domain of 80~$m/s$.
Here we obtained a CX-halo of 26 members at a cutoff $d_{md} =~360~m/s$. 
Eleven objects (42.3\% of the total) were SDSS-MOC4 interlopers, and 
twelve (46.2\% of the total) had $p_V < 0.1$.  

\subsection{The Astraea family}
\label{sec: astraea}

The Astraea family is listed at the AstDyS.  We identified a small CX-complex
halo of four members at a cutoff of $320~m/s$, with no interlopers.

\subsection{The Aeolia family}
\label{sec: aeolia}

The Aeolia family was identified in Nesvorn\'{y} et al. (2005) as a group
of 28 members at a cutoff of 20~$m/s$ with no identifiable dominant 
taxonomy.  Here we obtained
a halo of 14 members at a cutoff $d_{md} =~320~m/s$, all with 
CX-complex-taxonomies.  Two objects 
(14.3\% of the total) have values of $p_V > 0.1$.

\subsection{The Chloris family}
\label{sec: chloris}

The Chloris family was a group of 135 members identified in 
Nesvorn\'{y} et al. (2005) at a cutoff in proper element domain of
120~$m/s$.  Most of the members of this group belonged to the C-class.
In this work we found a halo of 35 members at a cutoff of 340~$m/s$.
One object (2.9\% of the total) was a possible SDSS-MOC4 interloper, and
eigth objects (22.9\% of the total) had values of $p_V > 0.1$.

\subsection{The Misa family}
\label{sec: misa}

The Misa family was a large C-class group of 119 asteroids identified at 
a cutoff of 80~$m/s$ in proper element domain by Nesvorn\'{y} et al. (2005).
Here we found a halo of 33 members at a cutoff $d_{md} =~355~m/s$
all belonging to the CX-complex.
One object (3.0\% of the total) was a possible SDSS-MOC4 
and albedo interloper.  The Leonidas AstDyS group merges with this 
family at a cutoff of less than 150$~m/s$.

\subsection{The Brangane family}
\label{sec: brangane}

The Brangane group was a 30 members S-type cluster identified in 
proper element domain by Nesvorn\'{y} et al. (2005) at a cutoff of 30~$m/s$.
In this work we identified an S-complex halo of just 3 members at 
a cutoff $d_{md} =~355~m/s$. One member (33.3\% of the total) was a 
possible SDSS-MOC4 interloper, and, as observed for some other 
S-complex families, all objects had $p_V < 0.1$, 

\subsection{The Bower family}
\label{sec: bower}

The Bower family was a 82 cluster identified by Nesvorn\'{y} et al. (2005)
at a cutoff of 100~$m/s$ with no dominant taxonomical information.  In this
work we identified a 27 members halo at a cutoff 
$d_{md} =~260~m/s$.  Most of the members belonged to the CX-complex, 
but 6 (22.2\% of the total) were possible  SDSS-MOC4 
interlopers, and 7 objects (25.9\% of the total) had $p_V > 0.1$.

\subsection{The Cameron family}
\label{sec: cameron}

The Cameron group was identified at a cutoff of 60~$m/s$ by Nesvorn\'{y} et 
al. (2005).  It was a 162 group made mostly by S-type asteroids.  The halo
that we identified in this work had 3 members at $d_{md} =~310~m/s$.
Contrary to what was found by Nesvorn\'{y} et 
al. (2005), all members have 
SDSS-MOC4 colors compatible with a CX-complex-taxonomy, but 
two objects (66.7\% of the total) had $p_V > 0.1$.  The Innes AstDyS 
group merges with this family at cutoff lower than 150~$m/s$.

\subsection{The Rafita family}
\label{sec: rafita}

The Rafita family was an S-complex group identified by 
Nesvorn\'{y} et al. (2005) in the $(a,e,sin(i))$ proper elements 
domain at a cutoff of 
100~$m/s$.  Unfortunately we could not identify any member of this family 
in our multi-domain sample of proper elements, SDSS-MOC4 colors, and 
geometric albedos.  Therefore, we could not analyze this family halo.

\subsection{The Eunomia family}
\label{sec: eunomia}

The Eunomia family is the largest family in the central main belt.
Moth\'{e}-Diniz et al. (2005) analyzed the spectra of 43 members of
this family, most of which belonging to the S-complex, but with a 
large taxonomical diversity that suggested surface inhomogeneities
or the action of space weathering.  The presence of T- and X-class
asteroids, classes these compatible with iron meteorites, 
suggested the possibility that the formation of the Eunomia family may 
have been the result of the catastrophic break-up of a differentiated
(or partially differentiated) parent body.  The identification of
three V-type asteroids in the orbital proximity of the Eunomia
family provided further hints for this possibility.  Carruba et al. 
(2007) showed that it is possible to migrate from the 
Eunomia dynamical family to the current orbital location of (21238) 1995 WV7,
the largest of the V-type asteroids in the Eunomia region,
via the interplay of the Yarkovsky effect and the 
${\nu}_5 -{\nu}_6+{\nu}_{16}$ nonlinear secular resonance, on 
time-scales of at least 2.6 Gyr.

In this work we identified a halo with 52
members, at a cutoff $d_{md}$ of 90 $m/s$.  As found in Moth\'{e}-Diniz et 
al. (2005), the Eunomia family halo is quite diverse, with a predominance
of objects belonging to the S-complex, but with a fairly large minority of
C- and X-complex asteroids.  We found 7 SDSS-MOC4 interlopers and
7 asteroids with $p_V < 0.1$, which yields a percentage of 13.5\% 
likely interlopers.  The Planetary Data System 
group of Schulhof merges with the Eunomia family at a cutoff of
$195~m/s$. 

\subsection{The Iannini family}
\label{sec: iannini}

The Iannini family was studied in Nesvorn\'{y} et al. (2005), where it was 
identified in proper element domain at a cutoff of 30~$m/s$.  The group
was listed as an S-type, but here we found a 93 members halo dominated 
by CX-complex asteroids, at a cutoff $d_{md}$ of 305 $m/s$.   The discrepancy
with Nesvorn\'{y} et al. (2005) spectral classification may possibly be caused
by the low number (18) of objects found in this family at the time.
There were no SDSS-MOC4 interloper, and 6 asteroids (6.5\% of 
the total) had $p_V > 0.1$.

\subsection{The Gefion family}
\label{sec: gefion}

The Gefion family, previously identified as the Ceres family (Zappal\'{a}
et al. 1995) and also as the Minerva/Gefion family 
(Moth\'{e}-Diniz et al. 2005),
was identified in Moth\'{e}-Diniz et al. (2005) as a fairly homogeneous
family, with members mostly belonging to the S-complex.
Because of its orbital proximity to (1) Ceres, it was studied in 
Carruba et al. (2003) as a test case for chaotic diffusion caused
by close encounters with massive asteroids.  The Gefion family halo 
was identified at a cutoff $d_{md}$ of 210 $m/s$, with 146 members.   
Moth\'{e}-Diniz et al.
(2005) found that the local background of this family is mostly dominated
by distinguished C-type asteroids.  Indeed our halo is contaminated
by a minority of bodies belonging to the C-complex:  we found
43 SDSS-MOC4 interlopers and
33 asteroids with $p_V < 0.1$, which yields a percentage of 29.5\% and
22.6\% likely interlopers, respectively.  The Minerva AstDys group merges
with this family halo at cutoff lower than 150~$m/s$.

\subsection{The Adeona family}
\label{sec: adeona}

The Adeona family was analyzed by Moth\'{e}-Diniz et al. (2005) that 
found it to be a very homogeneous family, made mostly in its entirety
by asteroids belonging to the CX-complex.  Because of its orbital
proximity to (1) Ceres, it was also studied in Carruba et al. (2003) 
to understand the long-term effects of diffusion caused
by close encounters with massive asteroids.  The Adeona family halo 
has been identified in this work 
at a cutoff $d_{md}$ of 295 $m/s$, with 149 members.  We found one SDSS-MOC4
interloper (0.7\% of the total), and four 
objects with geometric albedo (barely) larger
than 0.1, which corresponds to a percentual of possible interlopers of 2.7\%.
This high uniformity of the Adeona albedo confirms the results 
found in Moth\'{e}-Diniz et al. (2005).  

\subsection{The Maria and Renate families}
\label{sec: maria}

The Maria family was analyzed together with the Renate family in 
Moth\'{e}-Diniz et al. (2005), and both families had a majority of
members with known taxonomies belonging to the S-complex, indistinguishable
from the local background.  Zappal\'{a} et al. (1997) analyzed this family
and found that the spectra of 10 family members were compatible 
with those of near-Earth asteroids (433) Eros and (1036) Ganymede, conclusion
not supported by the work of Moth\'{e}-Diniz et al. (2005).
The Maria family halo has been identified in this work at a cutoff of 
240 $m/s$ with 135 members.  We found 21 objects that can be classified
as SDSS-MOC4 interlopers, five of which barely in the area of the CX-complex, 
and 10 asteroids with $p_V < 0.1$, which yields a percentual of 
possible interlopers of 15.6\% and 7.4\%, respectively.  The Renate family,
considered together with the Maria family in Moth\'{e}-Diniz et al. (2005), and
also classified as an S-complex group in that work,  
merges with the Maria family at a cutoff of
225~$m/s$.  For the purpose of halo analysis, the two families
can be considered as an unique group.

\subsection{The Padua family}
\label{sec: padua}

This family, previously associated to the asteroid (110) Lydia, is made
mostly by X-type asteroids indistinguishable from the local background,
according to Moth\'{e}-Diniz et al. (2005).  The family is very
important from a dynamical point of view, since it is the second 
family, after the Agnia, to have most of its members in a nonlinear
secular resonance configuration.  More than 75\% of its members,
according to Carruba (2009a), are currently in a $z_1$ librating state.
Conservation of the $K_2^{'} = \sqrt{2-e^2} (2\cos{i})$ quantity 
associated with this secular 
allowed to set limits on the original ejection velocity field, that was in 
agreement with result obtained with an alternative Monte Carlo model
that included Yarkovsky and YORP semi-major axis mobility.
The current spread of values in the $(\sigma, g-g_6+s-s_6)$ plane,
where $\sigma$ is the resonant argument of the $z_1$ resonance allowed
to set a lower limit on the age of the family of 25 Myr, that 
was then used to set an upper limit on the effect of low-energy collisions.
The Padua halo was identified at a cutoff of 130 $m/s$, with 31 members,
and no interlopers.  The Zdenekhorsky AstDyS group merges with the Padua 
halo at cutoff lower than $100~m/s$.

\subsection{The Juno family}
\label{sec: juno}

The Juno family was identified in Nesvorn\'{y} et al. (2005) as a
74 members S-type group.  Here we identified a halo of 61 members
at a cutoff of 275~$m/s$, that, contrary to what published in 
Nesvorn\'{y} et al. (2005), is made mostly by CX-complex bodies, with (3) 
Juno itself, an Sk object, a possible interloper.
There were no SDSS-MOC4 interlopers, and 4 asteroids
(6.6\% of the total) had values of $p_V > 0.1$.

\subsection{The Dora family}
\label{sec: dora}

The Dora family was classified by Moth\'{e}-Diniz et al. (2005) as 
a very homogeneous C-complex family, with the majority of
members belonging to the Ch class, and five objects in the C and B 
classes.  The family was very differentiated from the local background, 
made mostly by asteroids belonging to the S-complex.  The Dora halo
was identified at a cutoff of 265~$m/s$, with 108 members.  Only 
2 members were possible SDSS-MOC4 interlopers and had
$p_V > 0.1$ (1.9\%), confirming the very homogeneous nature of
this family, as found in Moth\'{e}-Diniz et al. (2005).

\subsection{The Merxia and Nemesis family}
\label{sec: merxia}

The Merxia family was found to be made mostly by S-complex asteroids
in Moth\'{e}-Diniz et al. (2005), and was dominated by the two largest 
bodies, (808) Merxia, and (1327) Namaqua, the second of which was most likely
an interloper because of its low albedo.  The family is crossed by
the 3J:-1S:-1A three-body mean-motion resonance, which divides it into 
two lobes and cause a depletion in the number of members at the center
of the family, and it was well differentiated from the local background,
dominated by CX-complex objects, according to Moth\'{e}-Diniz et al. 
(2005).  Nesvorn\'{y} et al. (2005) also identified in the region the 
Nemesis family, but its halo merges with that of the Merxia family at a 
cutoff of $\simeq200~m/s$, and we therefore 
decided to treat the two families as a single case.
We found a CX-halo at a cutoff of 250~$m/s$, with 19 members, 
5 of which could be SDSS-MOC4 interlopers and 9 of which have $p_V < 0.1$.
The large percentual of possible interlopers (26.3\% and 42.1\%) 
may be caused by the fact that, possibly, there is no Merxia halo, 
and the family is small and limited to the S-complex core found
in Moth\'{e}-Diniz et al. (2005)

\subsection{The Agnia family}
\label{sec: agnia}

The Agnia family, previously identified as the Liberatrix family,
was the first group to be found having the 
majority of its members in $z_1$ librating states (Vokrouhlick\'{y} et 
al. 2006b).  Conserved quantities of the $z_1$ resonance and 
spread in the $(\sigma, g-g_6+s-s_6)$ plane, as discussed
for the case of the Padua family, were introduced in that work 
to obtain constraints on the family original ejection velocity
field and age.  The family, first analyzed by Bus (1999), appears
compatible with an S-complex taxonomy in Moth\'{e}-Diniz et al. (2005),
while the local background is dominated by CX-complex bodies.
Here we determined a halo at a cutoff of 190~$m/s$, with 12 members.
As for the Merxia family halo, we found a large number of possible
interlopers: 4 SDSS-MOC4 CX-complex members, and 4 $p_V < 0.1$ asteroids
(33.3\% of the total), which may suggest
that the actual Agnia family is small and with a limited halo.

\subsection{The Astrid family}
\label{sec: astrid}

The Astrid family was identified in Bus (1999) and 
Moth\'{e}-Diniz et al. (2005) as a very
tight clump, with most members belonging to the C-complex.  No 
asteroid in the local background had taxonomical information
at the time of Moth\'{e}-Diniz et al. (2005) analysis.
Here we found a very robust and isolated group, with a halo
that was separated from the local background for cutoffs as large
as 435~$m/s$, with 6 members, and no interlopers, confirming that
this is a very homogeneous and robust group.

\subsection{The Hoffmeister family}
\label{sec: hoffmeister}

The Hoffmeister family was found to be a very compact and spectrally
homogeneous CX-group in Moth\'{e}-Diniz et al. (2005).  Here we determined
a halo with 62 members at a cutoff of 210~$m/s$.  No interlopers were 
detected, so confirming previous analysis of this group.

\subsection{The Lavrov family}
\label{sec: lavrov}

The Lavrov group, previously known as the Henan clump, is a small group
formed mostly by L-type asteroids, that are also typical of the local 
background (Moth\'{e}-Diniz et al. 2005).  We identified a 
halo of 8 members at a cutoff of 200~$m/s$.
We identified only 1 possible SDSS-MOC interloper and 2 asteroids with
$p_V < 0.1$ (12.5\% and 25.0\% of the total, respectively), 
which confirms that this should be a fairly compact and 
robust L-class group.

\subsection{The 1995 SU37 family}
\label{sec: su37}

The 1995 SU37 group is listed at the Planetary Data System.  We identified
a small S-complex halo of four members, at a cutoff of $105~m/s$, with 
no interlopers.

\subsection{The Watsonia family}
\label{sec: watsonia}

The Watsonia family is listed at the AstDyS.  We identified a CX-complex halo
of ten members at a cutoff of $425~m/s$.  Three objects (30.0\% of the total)
were possible SDSS-MOC4 interlopers, and 5 objects (50\% of the total) 
had $p_V > 0.1$.

\subsection{The Ino, Atalante, and Anacostia families}
\label{sec: atalante}

The Ino, Atalante, and Anacostia families are listed at the AstDyS.  
Unfortunately we could not find any of their members in our 
multi-domain database. No information 
is therefore available for this family in this work.

\subsection{The Gersuind family}
\label{sec: gersuind}

With the Gersuind family we start the analysis of the highly inclined
$\sin{i} > 0.3$ asteroid groups in the central main belt, that were 
the subject of the study of Carruba (2010b).  The Gersuind family
was studied in Gil-Hutton (2006).  While having most of its members
at $\sin{i} > 0.3$, it lies at lower inclinations than the center 
of the ${\nu}_6$ resonance, and it is not therefore considered a proper
high-inclination family by other authors, such as Machuca and Carruba (2011).
The few objects with SDSS-MOC 3 data in the family obtained by
Carruba (2010b) were compatible with an S-complex taxonomy. 
Here we found a halo at 310~$m/s$ with 7 members, the majority of 
which were compatible with an S-complex taxonomy. Three objects
(42.9\% of the total) were SDSS-MOC4 interlopers, and 2 objects
(28.6\% of the total) had albedos smaller than 0.1, confirming the
analysis of Carruba (2010b).  The Planetary Data System Emilkowalski 
group merges with this family at cutoff lower than 100$~m/s$.

\subsection{The Myriostos family}
\label{sec: myriostos}

The Myriostos family is listed at the AstDyS.  We identified a 7 members
CX-complex halo at a cutoff of 580~$m/s$, with two SDSS-MOC4 
interlopers (28.6\% of the total).   Six objects (85.7\% of the total) 
have $p_V > 0.1$. 

\subsection{The Kunitaka family}
\label{sec: kunitaka}

The Kunitaka family is listed at the AstDyS. We could not find any of 
its members in our multi-domain database, so no information is available
on this group in this work.

\subsection{The Hansa family}
\label{sec: hansa}

The Hansa family is the largest dynamical group among high inclination
families in the central main belt.  The Hansa family was originally 
proposed by Hergenrother et al. (1996), studied in Gil-Hutton (2006),
and re-analyzed in Carruba (2010b), that found a large group compatible
with an S-type taxonomy.   The family is located in a stable island limited
in inclination by the ${\nu}_6$ and ${\nu}_5$ linear secular resonances.
This is confirmed by our analysis: we found
a 20 members halo at a cutoff of $> 535~m/s$ (at cutoff
as large as 1000$~m/s$ the family does not yet connect with the 
local background), with all but one member (95\% of
the total) with an S-complex taxonomy.  We did not 
detected asteroids with $p_V < 0.1$.  The 2001 YB113 AstDyS group
merges with this family halo for cutoff less than $150~m/s$.

\subsection{The Brucato family}
\label{sec: brucato}

The Brucato family was first identified as a clump in proper elements domain
and as a family in the $(n,g,g+s)$ proper frequencies domain in Carruba 
(2010b)~\footnote{See Carruba and Michtchenko (2007) for a more
in depth discussion of frequency families.}.  Novakovi\'{c} et 
al. (2011) then re-obtained this group
in proper element domains as a family, using a larger sample of 
asteroid proper elements.   The family is located in a stable island 
limited in inclination by the ${\nu}_5$ and ${\nu}_{16}$ 
linear secular resonances.
The group was made mostly by CX-complex
asteroids, and this is confirmed by the current analysis: we identify
a family halo at a cutoff of $950~m/s$ with 32 members, all belonging
to the CX-complex.  Two albedo interlopers 
(6.3\% of the total) were identified in this family halo.  The 1998 DN2, 
1999 PM1, 1998 LF3, and 2004 EW7 AstDyS groups merge with this family at 
cutoffs lower than 150$~m/s$.

\subsection{The Dennispalm family}
\label{sec: dennispalm}

The Dennispalm family is listed at the AstDys.  We could not find any of
its members in the multi-domain database, so no information is available
on this group in this work.

\subsection{The Barcelona family}
\label{sec: barcelona}

The Barcelona family was first identified as a clump in Gil-Hutton (2006).
Carruba (2010b) identified the group as a dynamical family, and 
this was confirmed by the later work of Novakovi\'{c} et al. (2011).
The Barcelona family was made mostly by Sq asteroids.  Very few objects
were present in our multi-domain sample of asteroids at this inclinations:
we identified an S-complex halo of only one member at a cutoff of 
730~$m/s$. 

\subsection{The Tina family}
\label{sec: tina}

The Tina family, first identified in Carruba (2010b), is unique 
in the Solar System because all of its members are in ${\nu}_6$ 
anti-aligned librating states, making it the only family currently
known to lie in a stable island of a linear secular resonance.
Carruba and Morbidelli (2011) studied its dynamics and obtained
estimates of the family age and possible survival time before the family
members escape from the stable island (both events
happened and will happen on timescales of 150 Myr).
(1222) Tina itself belongs to the X-complex.  The one halo
object that we identified at a cutoff of 890~$m/s$ is
compatible with such taxonomy.  The very limited number of 
objects with known taxonomy does not, however,
allow to determine if the Tina's group is a real family or a conglomerate
of asteroids happening to be lying in the local stable island, yet.

\subsection{The Gallia family}
\label{sec: gallia}

The Gallia family was first identified as a clump in Gil-Hutton (2006), 
and was re-obtained as a family in Carruba (2010b), and Novakovi\'{c} 
et al. (2011).  It is located in a stable island 
limited in inclination by the ${\nu}_5$ and ${\nu}_{16}$ 
linear secular resonances.
Its taxonomy was compatible with an S-complex composition,
according to the analysis of Carruba (2010b).   Here we identified a halo 
of just three members at a cutoff of $> 410~m/s$.  All members were compatible
with an S-complex taxonomy.

\subsection{The Pallas family}
\label{sec: pallas}

Williams (1992) first proposed the Pallas family, that was later re-analyzed 
by Gil-Hutton (2006), Carruba (2010b), and and Novakovi\'{c} et al. (2011).
Most of the Pallas family members have B-type taxonomies, but C-type objects
are also observed in the orbital region.  In this work we identified a halo
of 8 members at a cutoff of $> 920~m/s$.  No SDSS-MOC4 interloper was found,
but, as observed for Hungaria family members, all 8 asteroids have 
large values of $p_V$, in principle incompatible with a B- or C-type taxonomy. 

\subsection{The central main belt: an overview}
\label{sec: central_sum}

Our results for the central main belt are summarized 
in Table~\ref{table: halo_central}, that has the same format as 
Table~\ref{table: halo_inner}.

\begin{table*}
\begin{center}
\caption{{\bf Asteroid families halos in the central main belt.}}
\label{table: halo_central}
\vspace{0.5cm}
\begin{tabular}{|c|c|c|c|c|c|}
\hline
                 &                &    &                      &         & \\
First halo & $d_{md}$ cutoff value &   Number of & Spectral &Number of SDSS-MOC4  & Number of $p_V$ \\ 
 member    &   [$m/s$] & members  &   Complex  & likely interlopers &  likely interlopers \\
                 &                   &   &                   &         &  \\
\hline
                 &                   &   &                   &         &  \\
(46) Hestia: (7321)  & 360 &  26 & CX & 11 & 12\\
(5) Astraea: (4018) & 320 &  4  & CX & 0 &   0\\
(396) Aeolia: (76144) & 320 &  14 & CX & 0 &   2\\
(410) Chloris: (9545) & 340 &  35 & CX & 1 &   8\\
(569) Misa: (2289)    & 355 &  33 & CX & 1 &   1\\
(606) Brangane: (56748) & 355&  3 &  S & 1 &   3\\
(1639) Bower: (26703)  &  260 & 27 & CX & 6 &   7\\
(2980) Cameron: (4067) &  310 &  3 & CX & 0 &   2\\
(15) Eunomia: (630)  & 90  & 52 &  S &  7 &  7\\
(4652) Iannini: (143366) &305  & 93& CX & 0 &   6\\
(1272) Gefion: (2373)   & 210  & 146 & S& 43&  33\\
(145) Adeona: (1783)   & 295  & 149 & CX & 1 & 4\\
(170) Maria/Renate: (4104)& 240 & 135 &  S &21 & 10\\
(363) Padua: (2560)    & 130  & 31 & CX & 0 & 0\\
(3) Juno: (22216)    & 275  & 61 & CX & 0 &  4\\
(668) Dora: (1734)     & 265  & 108 & CX & 2 & 2\\
(808) Merxia/Nemesis: (3439)& 250 &19&CX & 5 &  8\\
(847) Agnia: (1020)    & 190  &  12 &  S & 4 &  4\\
(1128) Astrid: (2169)   & 435  &   6 & CX & 0 &  0\\
(1726) Hoffmeister: (1726)& 210&  62 & CX & 0 &  0\\
(2354) Lavrov: (2354)   & 200  &   8 &  S & 1 &  2\\
(18466) 1995 SU37: (95534)& 105 &   4 &  S & 0 &  0\\
(729) Watsonia: (5492) & 425  &  10 & CX & 3 &  5\\
(686) Gersuind: (14627)& 310  &   7 &  S & 3 &  2\\
(10000) Myriotos: (101897) & 580 &  7 & CX & 2 &  6\\
(480) Hansa: (13617)   & $> 535$& 20 & S & 1 &  0\\
(4203) Brucato: (4203)  & 950 & 32& CX & 0 &  2\\
(945) Barcelona: (11028)& 730 &   1 &  S & 0 &  0\\
(1222) Tina: (16257)    &  890 &   1 & CX & 0 &  0\\
(148) Gallia: (40853)  & $> 410$& 3 &  S & 0 &  0\\
(2) Pallas: (24793)  & $> 920$& 8 & CX & 0 &  8\\
                 &      &     &     &    \\
\hline
\end{tabular}
\end{center}
\end{table*}

\begin{figure*}

  \begin{minipage}[c]{0.5\textwidth}
    \centering \includegraphics[width=3.0in]{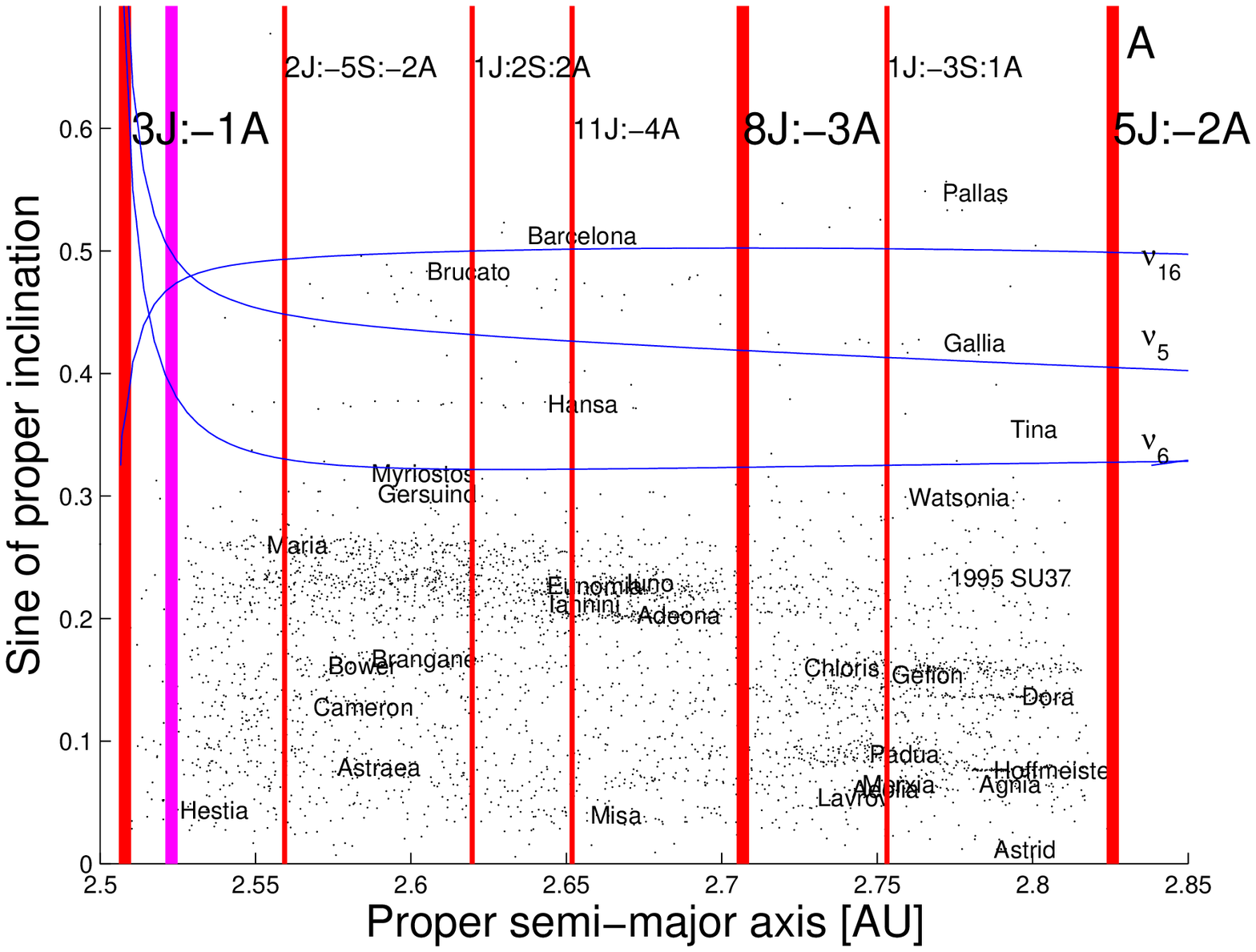}
  \end{minipage}%
  \begin{minipage}[c]{0.5\textwidth}
    \centering \includegraphics[width=3.0in]{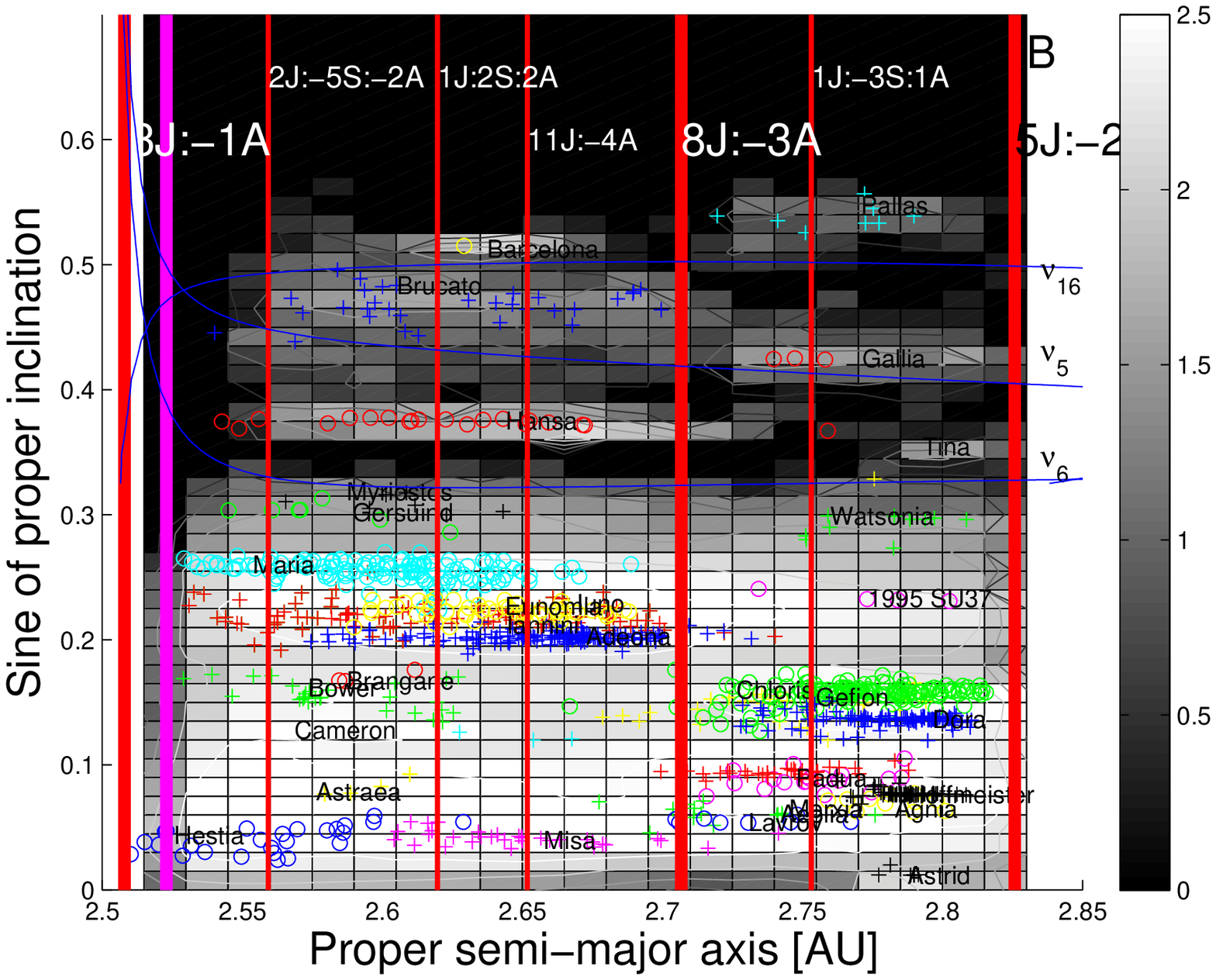}
  \end{minipage}

\caption{Panel A: An $(a,sin(i))$ projection of central main belt 
asteroids in our multi-variate sample.  Panel B: contour plot
of the number density of asteroids in the proper element sample.  
Superimposed, we display the orbital location of asteroids
of families in the CX-complex (plus signs) or in the S-complex 
(circles).}
\label{fig: central_el}
\end{figure*}

Fig.~\ref{fig: central_el}, panel A, displays an $(a,sin(i))$ projection of
asteroids in our multi-variate sample in the central main belt.  
 Blue lines show the location of the main linear secular 
resonances, using the second order and fourth-degree secular perturbation
theory of Milani and Kne\v{z}evi\'{c} (1994)
to compute the proper frequencies $g$ and $s$ for the grid of $(a,e)$
and $(a,\sin(i))$ values shown in Fig.~\ref{fig: outer_el}, panel A, and the
values of angles and eccentricity of (480) Hansa, the highly inclined
asteroid associated with the largest family in the region (Carruba 2010b).   
Other symbols have the same meaning as in Fig.~\ref{fig: inner_el}, 
panel A.  In panel B of the same figure we display a density map of the central
main belt.  We computed the $log_{10}$ of the number of all asteroids with 
proper elements per unit square in a 22 by 67 grid in $a$ (starting
at $a$~= 2.500~AU, with a step of 0.015 AU) and $sin(i)$ (starting at 
0, with a step of 0.015).  Superimposed to the density map, we also 
show the orbital projection
of the halos found in this work shown as red plus signs for CX-complex
families and as blue circles for S-complex families.   The other
symbols have the same as in Fig.~\ref{fig: outer_el}, panel B.
The reader may notice that regions with higher number density of asteroids
are associated with the families halos found in this work.  The halos that 
we determined do not have multiple membership, i.e., an asteroid in a 
given family halo is not present in another family halo. 

\begin{figure*}

  \begin{minipage}[c]{0.5\textwidth}
    \centering \includegraphics[width=3.0in]{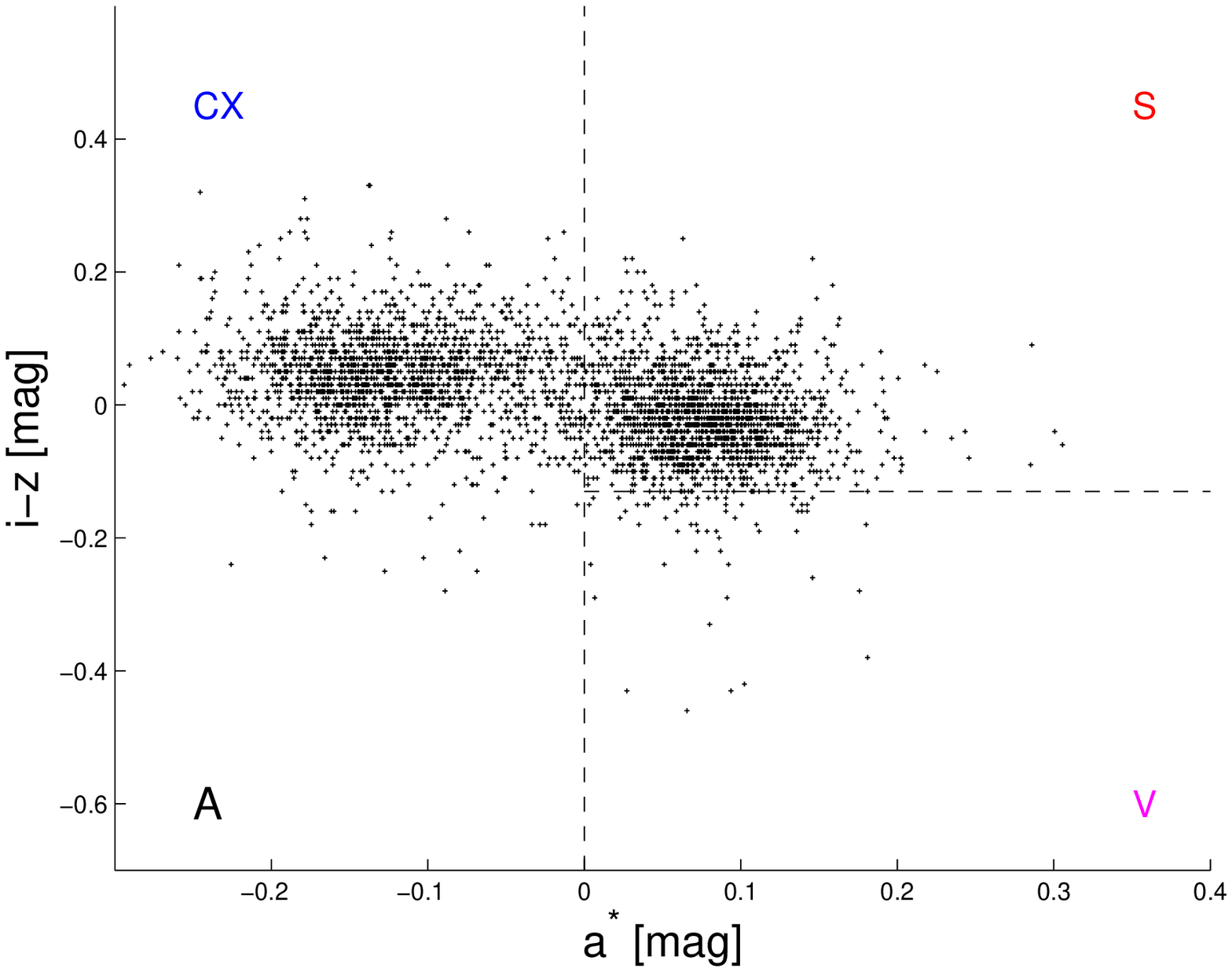}
  \end{minipage}%
  \begin{minipage}[c]{0.5\textwidth}
    \centering \includegraphics[width=3.0in]{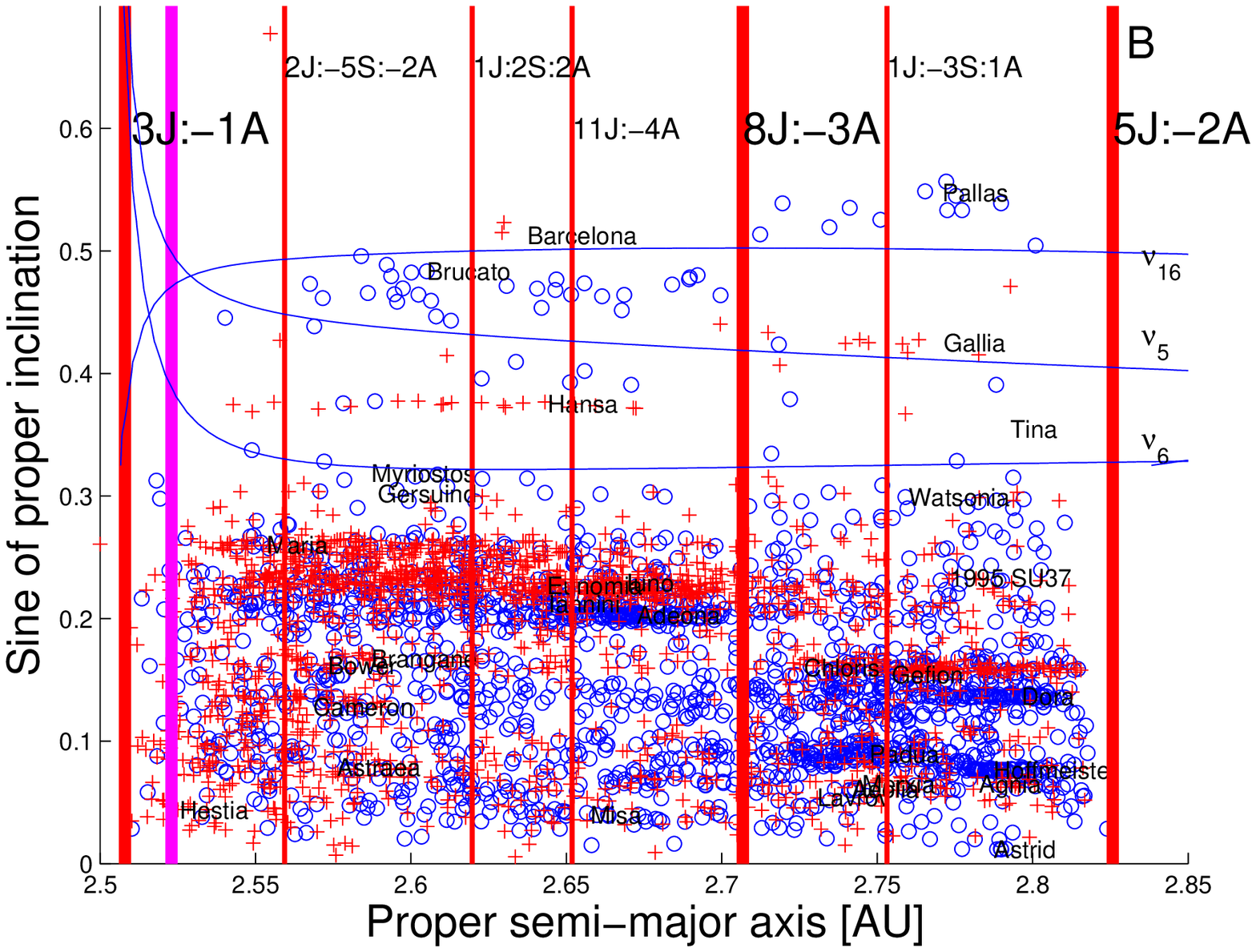}
  \end{minipage}

\caption{Panel A: an $(a^{*},i-z)$ projection of central main belt 
asteroids in our multi-domain sample.  Panel B: an
$(a,sin(i))$ projection of the same asteroids,
where objects in the CX complex are shown as blue circles, and
asteroids in the S-complex are identified as red plus signs.}
\label{fig: central_PC}
\end{figure*}

Fig.~\ref{fig: central_PC} displays a projection 
in the $(a^{*},i-z)$ plane of all asteroids in our 
multi-domain sample (panel A), and an $(a,sin(i))$ projection of 
the same asteroids, (panel B).  We refer the reader to the caption
of Fig.~\ref{fig: inner_pv} for a more detailed description of this figure 
symbols.  The central main belt is a transitional region,
where CX-complex and S-complex asteroids are rather mixed.  
A greater proportion of S-type asteroids can be found at lower semi-major
axis (and vice-versa), but overall, no group dominates the local
taxonomy. 

\begin{figure*}

  \begin{minipage}[c]{0.5\textwidth}
    \centering \includegraphics[width=3.0in]{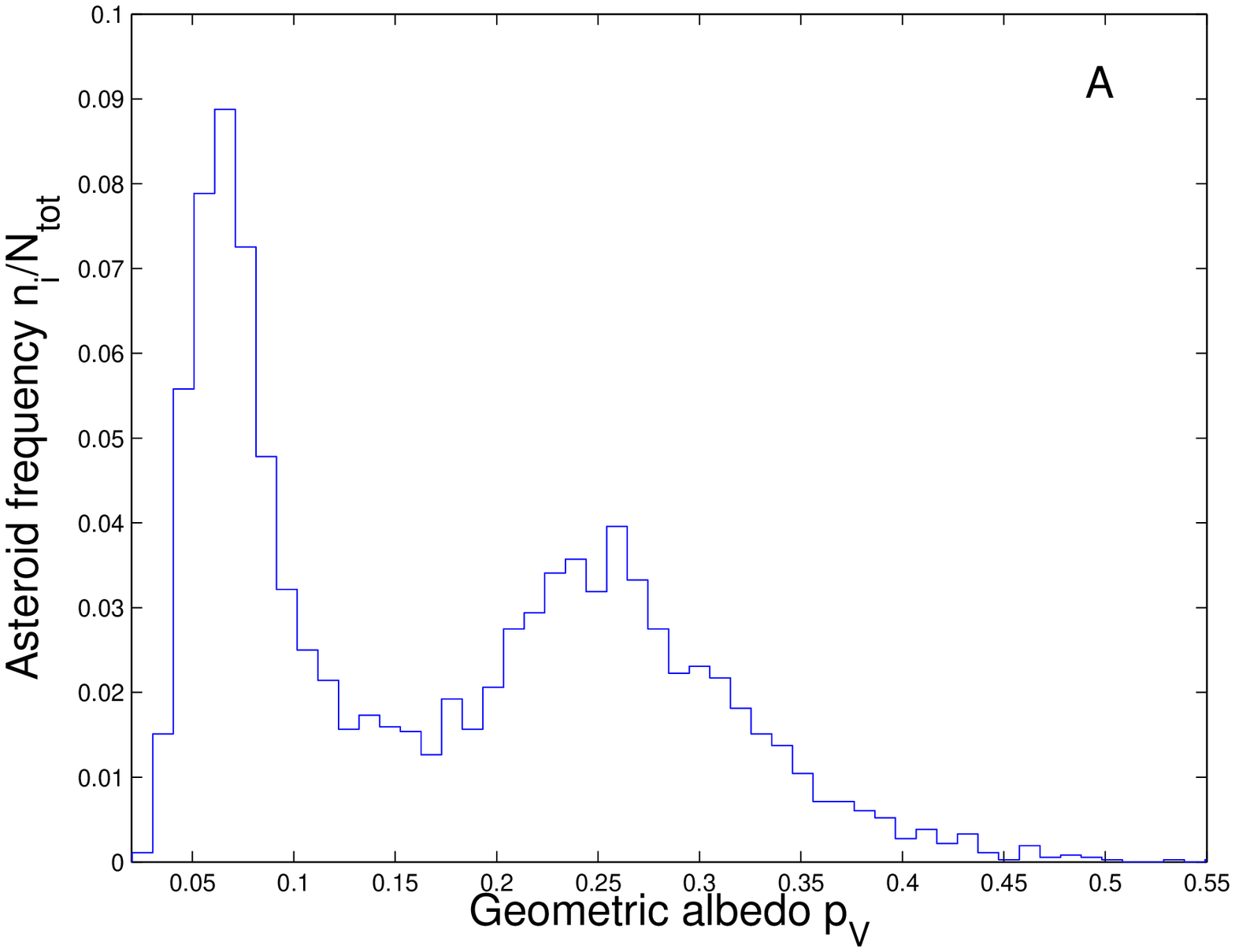}
  \end{minipage}%
  \begin{minipage}[c]{0.5\textwidth}
    \centering \includegraphics[width=3.0in]{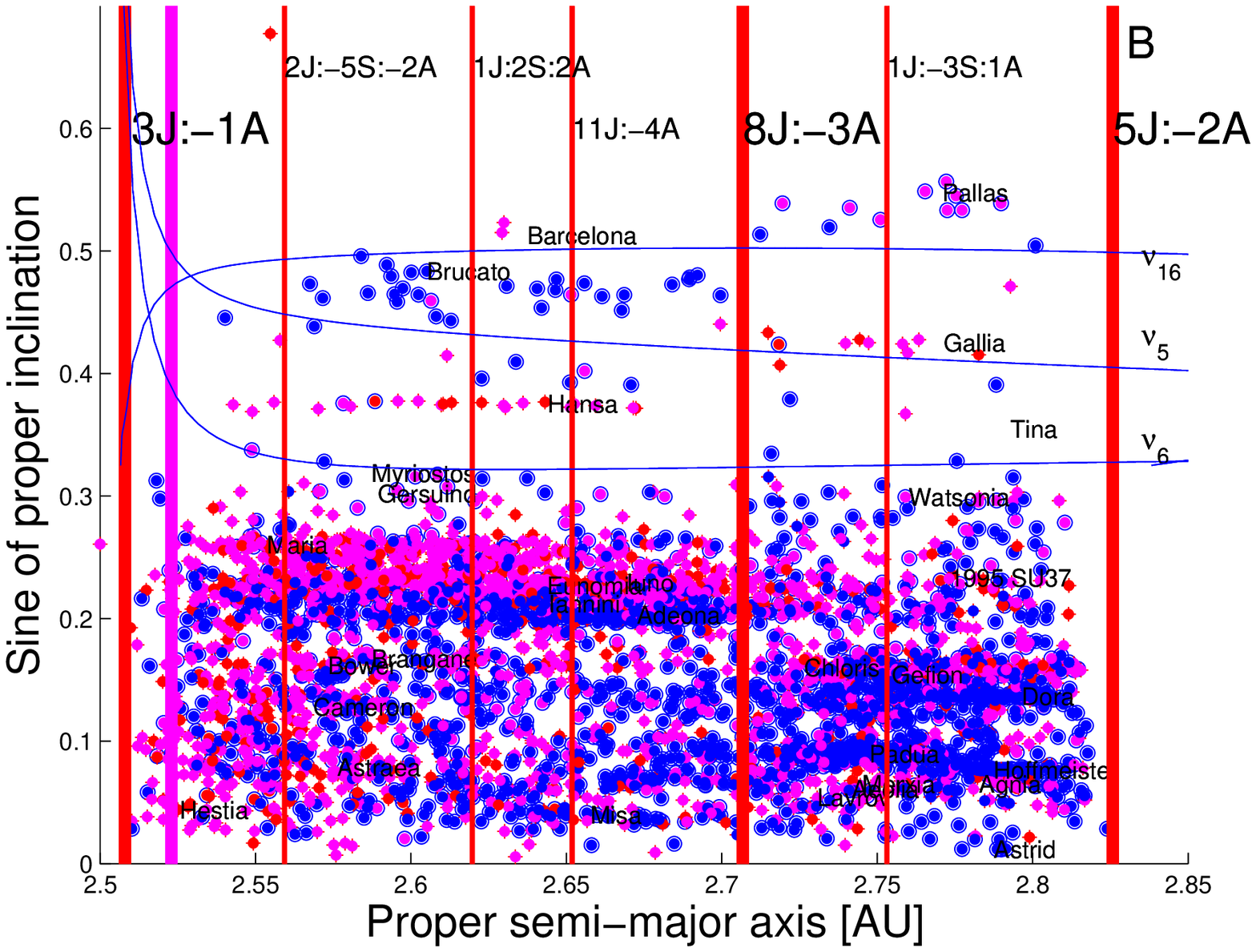}
  \end{minipage}

\caption{Panel A: a histogram of number frequency values 
$n_i/N_{Tot}$ as a function of geometric albedo $p_V$ for 
central main belt asteroids in our multi-domain sample.  Panel B: an 
$(a,sin(i))$ projection of the same asteroids,
where blue full dots are associated with asteroids with $p_V < 0.1$,
red full dots display asteroids with $0.1 < p_V < 0.3$, and magenta
full dots show asteroids with $p_V > 0.3$.}
\label{fig: central_pv}
\end{figure*}

This is confirmed by an analysis of WISE
$p_V$ geometrical albedo data, a histogram of which is presented
in Fig.~\ref{fig: central_pv}, panel A.  Fig.~\ref{fig: central_pv}, panel B,
displays an $(a,sin(i))$ projection of the same asteroids, 
where we used the same color code as in Fig.~\ref{fig: inner_pv}, panel B.

One can notice the clearly bi-modal distribution of asteroid albedos, and
the mixture of objects with low albedo and high albedo that dominates
the central main belt.  Having analyzed the families of the central
main belt, we are now ready to move on to the outer main belt.

\section{Outer Main Belt}
\label{sec: out_bel}

The outer main belt is dynamically limited in semi-major axis by the 
5J:-2A and 2J:-1A mean-motion resonances  (Zappal\'{a} et al. 1995).   
The linear secular resonance ${\nu}_6$ separates the low-inclined 
asteroid region from the highly inclined area, dominated by the 
Euphrosyne family (Machuca and Carruba 2011).  We found 5385 objects 
that have proper elements and frequencies, SDSS-MOC4 colors, WISE 
geometric albedo data, and satisfy our error analysis criteria 
in the outer main belt.  We start our analysis of families halos 
by investigating the Koronis family.

\subsection{The Koronis family}
\label{sec: koronis}

The Koronis family is one of the most interesting families in the main belt.
Bottke et al. (2002) explained its shape in eccentricity as originating
by the interaction of asteroids evolving via Yarkovsky effect into secular
resonances such as the $2{\nu}_5-3{\nu}_6$.   Carruba and Michtchenko (2007)
showed that its upper boundary in eccentricity is limited by the 
${\nu}_6+2{\nu}_5-2{\nu}_7+{\nu}_{16}$ nonlinear secular resonance.
The interesting subgroup of the Karin cluster, identified by 
Nesvorn\'{y} et al. (2002a) opened new perspectives in the 
understanding of recent breakups among asteroids.
Moth\'{e}-Diniz et al. (2005) identified this family as predominantly belonging
to the S-complex, with a few interlopers belonging to the C- and D- types,
that predominate in the background.   The presence of a few K-type 
asteroids in the family was somehow puzzling, and justified by 
Moth\'{e}-Diniz et al. (2005) as a possible remnant of a pre-existing
family.

In this work we identified a halo with 200
members, at a cutoff $d_{md}$ of 215 $m/s$.  32 of the halo members have 
SDSS-MOC4 colors not compatible with an S-complex taxonomy,
which yields to a 16.0\% fraction of possible interlopers in the halo,
somewhat in agreement with the finding of Moth\'{e}-Diniz et al. (2005).
About 14 asteroids (7.0\% of the total) have values of $p_V$ 
smaller than 0.1, which is the limit
for S-complex asteroid albedos, and that may be related with the presence
of the interloper population found in Moth\'{e}-Diniz et al. (2005).

\subsection{The Lau family}
\label{sec: lau}

The Lau cluster is listed at the Planetary Data System.   In this work 
we identified a 10 objects CX halo at a cutoff of 490~$m/s$.  No interloper
was found in this group.

\subsection{The (1993) FY12 family}
\label{sec: 1993 FY12}

The cluster around (18405) (1994 FY12) was identified for the first time
in Nesvorn\'{y} et al. (2005) at a cutoff of $50~m/s$ as an 11 
asteroids X group.  Here we find a halo at a cutoff $d_{md}$ of 355 $m/s$
of just 2 members, one of which (50.0\% of the total) 
have $p_V > 0.1$.  The possible X-type composition of this small group 
appears to be confirmed by our analysis.

\subsection{The Fingella family}
\label{sec: fingella}

The Fingella family is listed at the Planetary Data System.   In this work 
we identified a 16 objects CX halo at a cutoff of 480~$m/s$.  No interloper
was found in this group.

\subsection{The Naema family}
\label{sec: naema}

The Naema group is another family discussed in Nesvorn\'{y} et al. (2005), 
where it was visible at a cutoff of 40~$m/s$ as a 64 C-type group.   Here 
we found a halo of 43 members at a cutoff of 390~$m/s$, with all 
members belonging to the CX-complex.  No interlopers were found
in this group.

\subsection{The Brasilia family}
\label{sec: brasilia}

The Brasilia family was identified as a small clump of 96 members in
Moth\'{e}-Diniz et al. (2005), of which only 4 had known spectral type,
all belonging to the CX-complex.  We identified a small halo of 24 objects,
most belonging to the CX-complex, that merges with the local background 
at a cutoff of 440 $m/s$.  While all but 3 halo members (12.5\% of the total) 
have SDSS-MOC4 colors compatible with a taxonomy in the 
CX-complex, the vast majority of these asteroids (17, 70.8\% of the total) 
has large albedos ($p_V > 0.1$), not usually
associated with a CX-composition.   The reason for the large albedo values
of these objects remains unexplained.

\subsection{The Themis family}
\label{sec: themis}

The Themis family is a rather large, homogeneous group, made mostly
by asteroids of types belonging to the C/X-complex.  Moth\'{e}-Diniz et al.
(2005) discuss that the spectral type of the family members is not 
distinguishable from that of asteroids in the background, that is
dominated by C and X-type objects.   Tanga et al. (1999) suggests that 
this family originated from the break-up of a large ($\simeq$ 370 ~km in 
diameter) C-type parent body.  In this work we identified a halo with 700
members, at a cutoff $d_{md}$ of 310 $m/s$.  9 of the halo members have 
SDSS-MOC4 colors not compatible with a C/X-complex taxonomy,
which yields a 1.3\% fraction of possible interlopers in the halo.
50 asteroids have values of $p_V$ larger than 0.1, which is the limit
for CX-complex asteroid albedos, which yields a percentage of possible
albedo interlopers of 7.1\%.  The AstDys Ashkova group merges with 
the Themis halo at a cutoff of 205$~m/s$.

\subsection{The Eos family}
\label{sec: eos}

Vokrouhlick\'{y} et al. (2006b) studied in detail the dynamical evolution
of this asteroid family, also pioneering the use of Monte Carlo 
methods for asteroid families chronology.   The Eos family is crossed
by the powerful 9J:-4A mean-motion resonance, and interacts with 
the $z_1$, $2{\nu}_5-3{\nu}_6$ and $2{\nu}_5-2{\nu}_6+{\nu}_{16}$
secular resonances (see also Carruba and Michtchenko 2007).
More recently, Bro\v{z} and Morbidelli (2013) identified  
Eos halo members based on SDSS-MOC4 data, and obtained an 
estimate of the halo age, that is in
agreement with what found by Vokrouhlick\'{y} et al. (2006b).
In this work we identified a halo at a cutoff of 165 $m/s$, with 738 members.
In agreement with what was found by Moth\'{e}-Diniz et al. (2005), the
taxonomy of the family is quite inhomogeneous, but well distinguished
from the local background that is dominated by C and X asteroids.
Most members of the Eos family have a characteristic K-type taxonomy,
and we found that many members of the halo members have $(a^{*},i-z)$ 
colors compatible with an S-complex taxonomy. But, as discussed
in Sect.~\ref{sec: comp}, the Eos family lies at the separation
between CX-complex and S-complex asteroids in the $(a^{*},i-z)$.
Our simple criterion for identifying S-complex asteroids, $a^{*} > 0$,
does not apply well to the case of the Eos family.
379 asteroids (51.4\% of the total) have values of $a^{*} < 0$ and
may be considered interlopers. 132 Eos halo 
asteroids have values of $p_V < 0.1$, which yields that 17.9\% 
of the halo asteroids may actually be 
albedo interlopers.  The quite diverse mineralogy
of bodies in the Eos family area provides challenges that should
be confronted with more advanced tools than the one used in this
paper, in our opinion.  The Telramund cluster, that 
was identified as an S-type 70 members group at 
a cutoff of 60~$m/s$ by Nesvorn\'{y} et al. (2005), merges
at very low cutoff with the Eos family, so we consider this group 
as a substructure of the larger family.

\subsection{The Hygiea family}
\label{sec: hygiea}

(10) Hygiea is the fourth most massive asteroid of the main belt, and
the family associated with this body has been studied in Zappal\'{a} et 
al. (1995), Moth\'{e}-Diniz et al. (2005), and, more 
recently by Carruba (2013).  It is made mostly by bodies 
belonging to the CX-complex, and, as 
the Themis family, is not easily distinguishable from the local background
of objects.    We identified a halo with 426 objects at a cutoff of
290 $m/s$.  The slightly 
higher number of objects that we found in the Hygiea halo
with respect to Carruba (2013) migth be due to the larger sample of
bodies in our data-set.  Of the 426 possible halo members, 6 (1.4\%) 
may be interlopers based on SDSS-MOC4 colors analysis, 
and 36 have $p_V > 0.1$, that is incompatible with CX-complex asteroids.
Overall, we found a maximum of 8.5\% of objects that are possibly 
not correlated with the Hygiea halo.  The AstDys Filipenko group merges
with this family halo at a cutoff of $125~m/s$.

\subsection{The Emma family}
\label{sec: emma}

The Emma family was identified by Nesvorn\'{y} et al. (2005) as a 76 
members group
at a cutoff of 40~$m/s$ in proper elements domain.  No information on its
taxonomy was given in that article.   In this work we found
a family halo of 43 members at a cutoff $d_{md}$ of 270 $m/s$.  No interlopers
were identified in this family halo.

\subsection{The Veritas family}
\label{sec: veritas}

The Veritas family is a relatively small group, made mostly by CX-complex
asteroids.  Milani and Farinella (1994) used for the first time chaotic
chronology on this family to determine its relatively young age, later
confirmed by other works.
We identified a halo of 148 members at a cutoff of 240 $m/s$.  2 objects
have colors not compatible with a CX-complex taxonomy, and 9 had
values of $p_V > 0.1$.  Overall, up to 6.1\% of the halo members encountered 
may be interlopers of the Veritas halo.

\subsection{The Lixiaohua family}
\label{sec: lixiaohua}

The Lixiaohua family was identified by Nesvorn\'{y} et al. (2005) as 
a 97 CX-complex group at a cutoff of 50~$m/s$.   It was the subject 
of a dynamical study by Novakovi\'{c} et al. (2010) that extensively 
studied the local dynamics and the diffusion in the $(e_p,i_P)$ plane
using Monte Carlo modeling.   In this work we identified a 69 members
CX halo at a cutoff of $d_{md}~=~255~m/s$.   Only one object (1.4\% of the total)
was a possible SDSS-MOC4 interloper, and all asteroids in the halo
had $p_V < 0.1$.  The AstDyS Gantrisch group merges with this family
at cutoffs lower than 50~$m/s$.

\subsection{The Aegle family}
\label{sec: aegle}

The Aegle family is an AstDyS group.  We identified a 21 CX-complex
group at a cutoff of $290~m/s$.  Two objects (9.5\% of the total) were
possible SDSS-MOC4 interlopers, and all members had low albedos.

\subsection{The Meliboea family}
\label{sec: meliboea}

The Meliboea family was discussed by Zappal\'{a} et al. (1995) and 
Moth\'{e}-Diniz et al. (2005).  It is a small group, mainly composed
by asteroids belonging to the CX-complex.  It is a rather inclined
family ($i_p \simeq 15^{\circ}$), characterized by the presence of 
several weak mean-motion and secular resonances.   We identified a halo with 
73 members at a cutoff of 270 $m/s$.  As found by Moth\'{e}-Diniz et al.
(2005), the Meliboea family is fairly homogeneous, with only three 
members of the halo with colors not compatible with a CX-complex
taxonomy.  Only one halo member has $p_V > 0.1$, which yields
a percentage of up to 4.1\% possible interlopers.   The AstDyS 
group of Inarradas merges with this family halo at a cutoff of
140~$m/s$, while the Traversa cluster is englobed at a cutoff
of 205~$m/s$.

\subsection{The Klumpkea/Tirela family}
\label{sec: Klumpkea}

The Klumpkea family was identified in 
Machuca and Carruba (2011) and corresponds to the old Tirela family of 
Nesvorn\'{y} et al. (2005).  Nesvorn\'{y} et al. (2005) listed the Tirela
family as a D-group.
Here we found a halo at a cutoff of 290 $m/s$ 
with 21 members, 2 of which (9.5\% of the total) have colors
(barely) in the S-complex area.  All members of the halo have $p_V < 0.1$,
which makes this family compatible with a C-complex taxonomy.  The 
AstDyS Zhvanetskij cluster is annexed by this halo at a cutoff
of $165~m/s$, the Ursula family is englobed at $200~m/s$, and 
the Pannonia group merges at a cutoff of $245~m/s$.

\subsection{The Theobalda family}
\label{sec: theobalda}

The Theobalda family is also an AstDyS group.  We identified
a 34 members CX-complex halo, with two (5.9\% of the total)
possible albedo interlopers.

\subsection{The Kartvelia family}
\label{sec: kartvelia}

The Kartvelia family is listed at the AstDyS.  We found a 26 members
CX-complex halo at a cutoff of $280~m/s$, with just one (3.8\% of the total)
possible SDSS-MOC4 interloper.

\subsection{The Alauda family region}
\label{sec: alauda}

The orbital region of the Alauda family has been most recently analyzed
by Machuca and Carruba (2011) that found several small groups, among
which the Alauda and Luthera families, in the area.  In this work
we identified a 158 CX-complex group at a cutoff of 420 $m/s$.
Two members, 1.3\% of the total,  have colors in the S-complex region,
and 19 objects, 12.0\% of the total, have $p_V > 0.1$.  The AstDys Higson
cluster merges with this halo at a cutoff of $235~m/s$, the 
AstDyS Moravia and Snelling groups merge at a cutoff 
of $240~m/s$, while the AstDyS Vassar cluster is annexed at $255~m/s$.

\subsection{The Euphrosyne family}
\label{sec: euphrosyne}

Machuca and Carruba (2011) most recently 
analyzed the orbital region of this highly 
inclined asteroid family.  This family is characterized by its
interaction with linear secular resonances.  In particular, 
13 of its members are in ${\nu}_6$ anti-aligned
librating states, one in a ${\nu}_5$ anti-aligned
librating state (242435), and one in a ${\nu}_5$ aligned
librating state (2009 UL136), according to Machuca and Carruba (2011).
The long-term effect of close encounters of asteroids with 
absolute magnitude $H < 13.5 $ with (31) Euphrosyne was recently
studied in Carruba et al. (2013).
As discussed in Machuca and Carruba (2011), the Euphrosyne family is
separated by the near regions of (69032) and Alauda in 
inclination by areas with very low asteroid densities.  It is a region
with a relatively small population of objects, separated among them
by large distances, which explains why the 75 members 
halo that we identified in this work is encountered at the high 
cutoff value of 575 $m/s$.  All halo members have colors 
in the S-complex area, and three members have $p_V > 0.1$.
As for the case of the Klumpkea family, this is a group highly compatible
with a C-complex taxonomy, with a 4.0\% of possible interlopers.

\subsection{The outer main belt: an overview}
\label{sec: outer_sum}

Having obtained estimates for the halos of the main families in the 
outer main belt, we are now ready to outline our results. 
As done for previous asteroid regions, our results are summarized 
in Table~\ref{table: halo_outer}.

\begin{table*}
\begin{center}
\caption{{\bf Asteroid families halos in the outer main belt.}}
\label{table: halo_outer}
\vspace{0.5cm}
\begin{tabular}{|c|c|c|c|c|c|}
\hline
                 &                   &        &              &         & \\
First halo & $d_{md}$ cutoff value & Number of & Spectral & Number of SDSS-MOC4  & Number of $p_V$ \\ 
 member    &   [$m/s$] & members &    likely interlopers &  Complex & likely interlopers \\
                 &                   &         &             &         &  \\
\hline
                 &                   &         &             &         &  \\
(158) Koronis: (761)   & 215  & 200 &  S & 32  &  14\\
(18405) 1993 FY12: (29959)& 355 &   2 & CX &  0  &  1 \\
(10811) Lau: (51707)     & 490  &  10 & CX &  0  &  0 \\
(709) Fingella: (1337) & 480  &  16 & CX &  0  &  0 \\
(845) Naema: (21257)   & 390  &  43 & CX &  0  &  0 \\
(293) Brasilia: (3985) & 440  &  24 & CX &  3  &  17\\
(24) Themis: (981)    & 310  & 700 & CX &  9  &  50\\
(221) Eos: (320)       & 165  & 738 &  S & 379 & 132\\
(10) Hygiea: (867)    & 290  & 426 & CX &  6  &  36\\
(283) Emma: (3369)     & 270  &  43 & CX &  0  &   0\\
(490) Veritas: (5592)  & 240  & 148 & CX &  2  &   9\\
(3556) Lixiaohua: (18477)& 255 &  69 & CX &  1  &   0\\
(96) Aegle: (29579)     & 290  & 21  & CX & 2   &  0\\
(137) Meliboea: (1165) & 270  & 73 & CX &  3  &   1\\
(1040) Klumpkea/Tirela: (18399)& 290 & 21 &CX&2  &   0\\
(778) Theobalda:(3432)  &  310 & 34  & CX & 0   &  2\\
(781) Kartivelia:(11911)&  280 & 26  & CX & 1   &  0\\
(702) Alauda/Luthera: (11911)& 420 & 158&CX&  2&  19\\
(31) Euphrosyne: (16712) & 575 &75 & CX &  0  &   3\\
                 &                   &         &             &         & \\
\hline
\end{tabular}
\end{center}
\end{table*}

\begin{figure*}

  \begin{minipage}[c]{0.5\textwidth}
    \centering \includegraphics[width=3.0in]{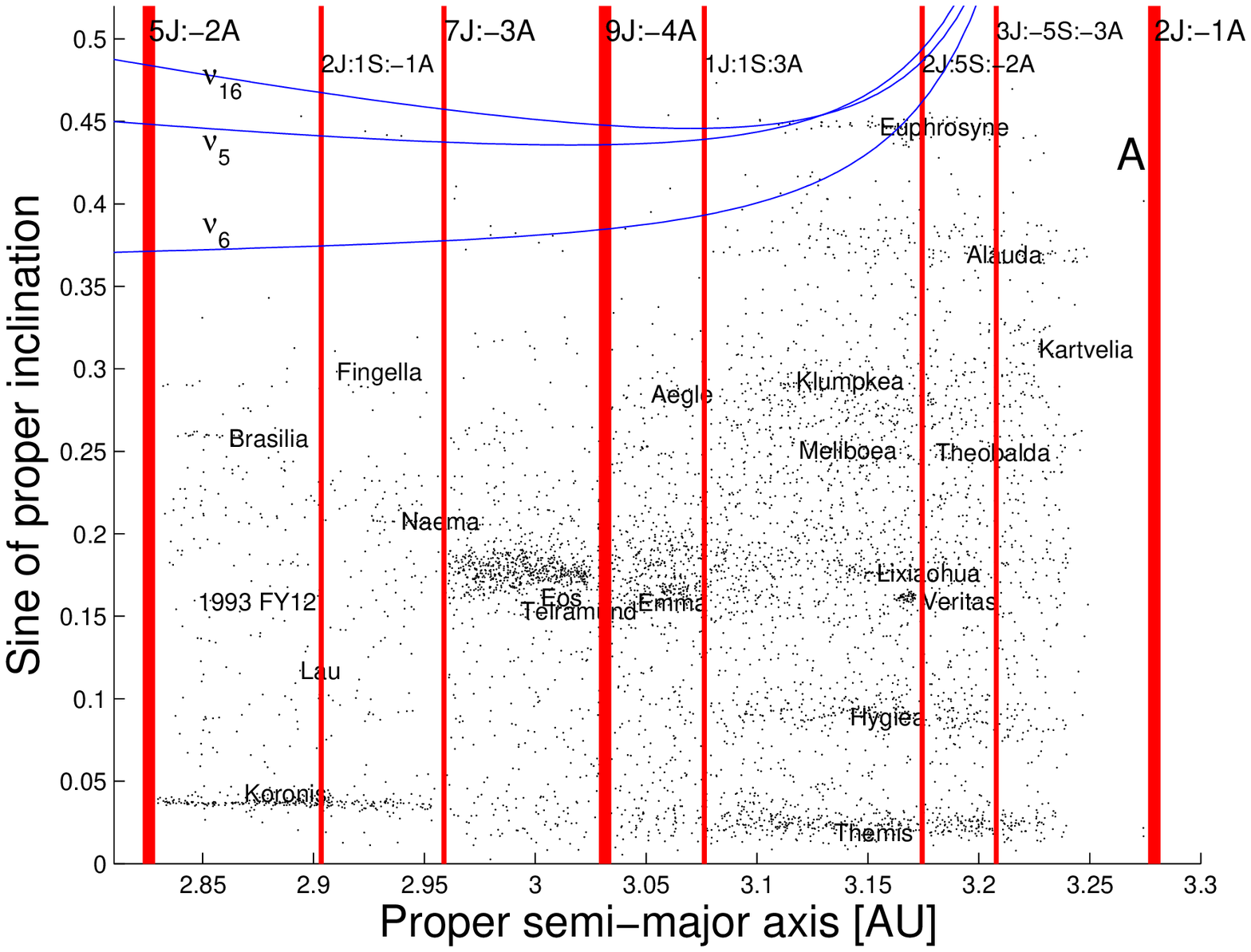}
  \end{minipage}%
  \begin{minipage}[c]{0.5\textwidth}
    \centering \includegraphics[width=3.0in]{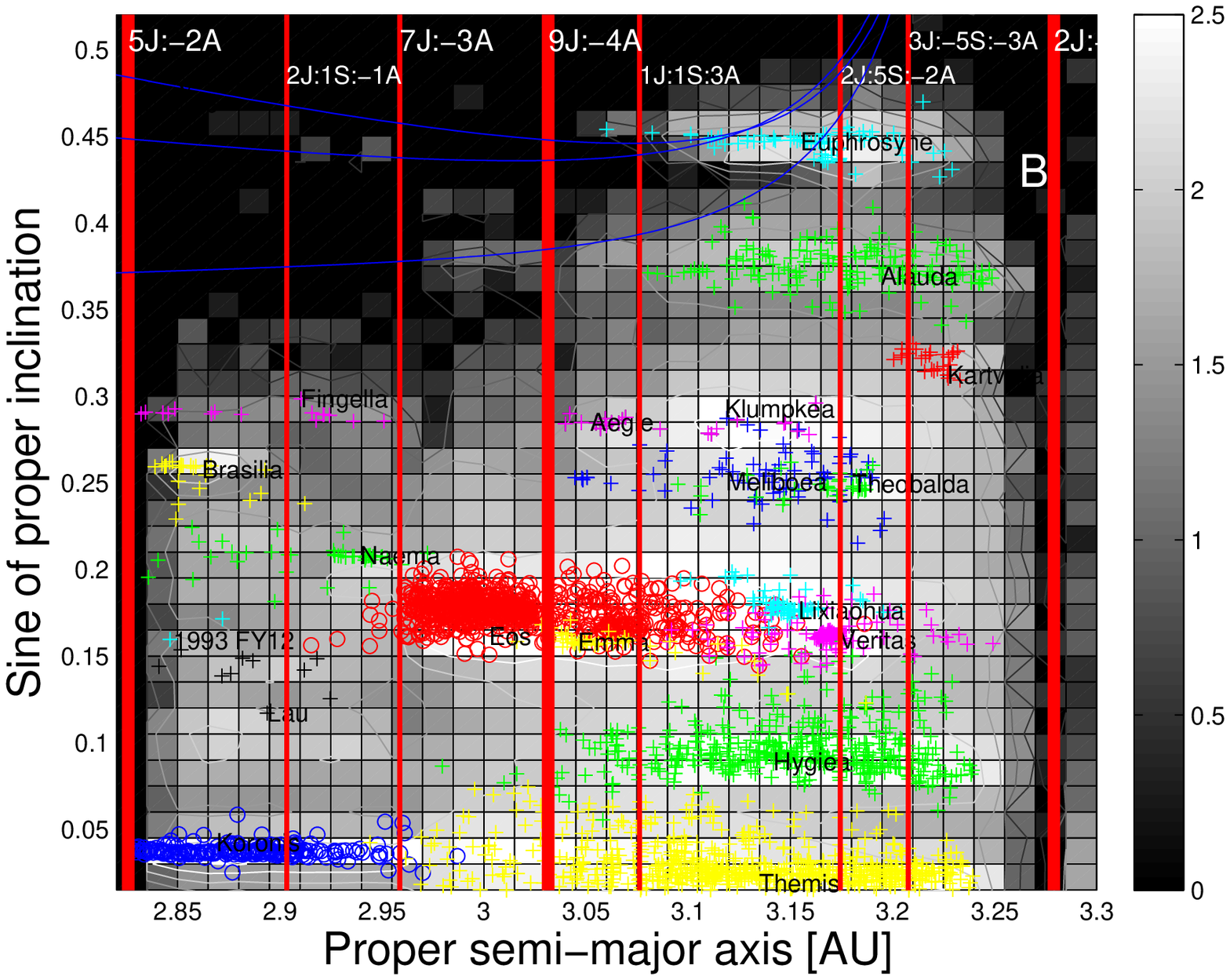}
  \end{minipage}

\caption{Panel A: An $(a,sin(i))$ projection of outer main belt 
asteroids in our multi-variate sample.  Panel B: contour plot
of the number density of asteroids in the proper element sample.  
Superimposed, we display the orbital location of asteroids
of families in the CX-complex (plus signs) or in the S-complex 
(circles).}
\label{fig: outer_el}
\end{figure*}

Fig.~\ref{fig: outer_el}, panel A, displays an $(a,sin(i))$ projection of
asteroids in our multi-variate sample in the outer main belt, with the same
symbols used for analogous figure in the inner and central main belt.
Here, however,  blue lines show the location of the main linear secular 
resonances, using the second order and fourth-degree secular perturbation
theory of Milani and Kne\v{z}evi\'{c} (1994)
to compute the proper frequencies $g$ and $s$ for the grid of $(a,e)$
and $(a,\sin(i))$ values shown in Fig.~\ref{fig: outer_el}, panel A, and the
values of angles and eccentricity of (31) Euphrosyne, the highly inclined
asteroid associated with the largest family in the region (Machuca and 
Carruba 2011).    In panel B 
of the same figure we display a density map of the outer 
main belt.  To quantitatively determine the local density of asteroids,  we
computed the $log_{10}$ of the number of all asteroids with proper elements 
per unit square in a 67 by 67 grid in $a$ (starting
at $a$~= 2.805 AU, with a step of 0.015 AU) and $sin(i)$ (starting at 
0, with a step of 0.015).  The other
symbols are the same as in Fig.~\ref{fig: inner_el}, panel A.
The reader may notice that regions with higher number density of asteroids
are associated with the families halo found in this work. Families halos 
are indeed more extended in proper elements domain than the core families 
found with the standard HCM.

\begin{figure*}

  \begin{minipage}[c]{0.5\textwidth}
    \centering \includegraphics[width=3.0in]{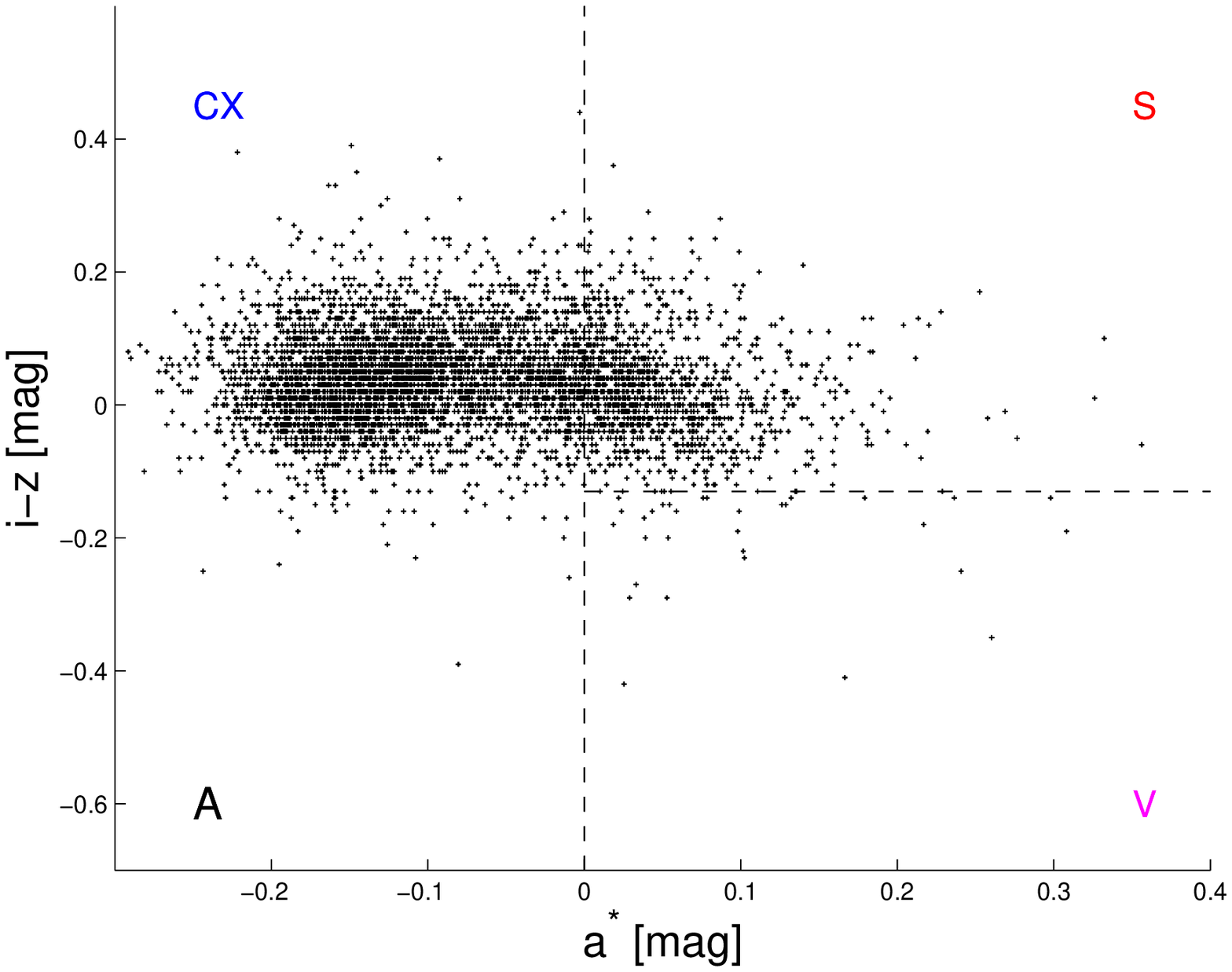}
  \end{minipage}%
  \begin{minipage}[c]{0.5\textwidth}
    \centering \includegraphics[width=3.0in]{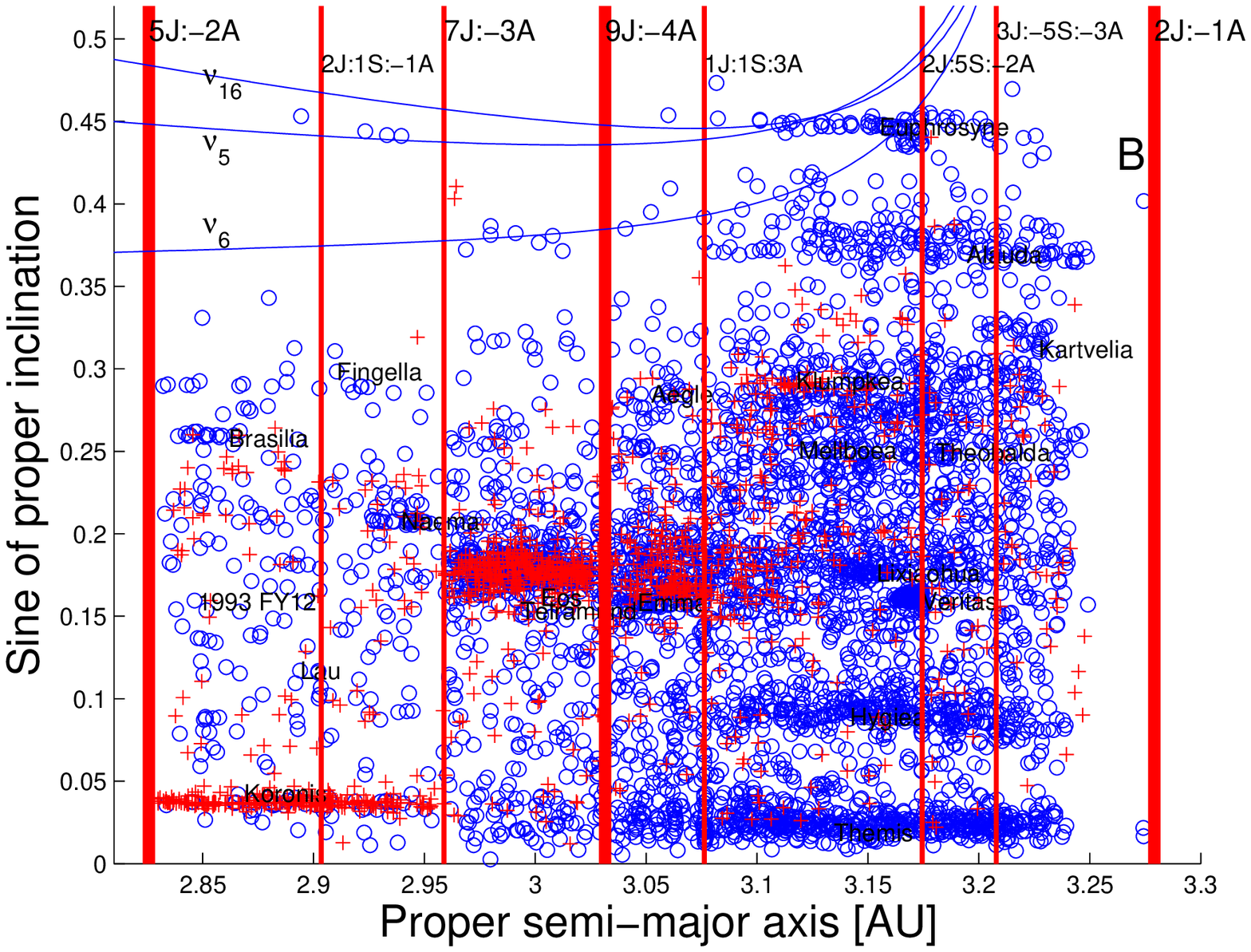}
  \end{minipage}

\caption{Panel A: an $(a^{*},i-z)$ projection of 
outer main belt asteroids in our multi-domain sample.  Panel B: a 
$(a,sin(i))$ projection of the same asteroids,
where objects in the CX complex are shown as blue circles, and
asteroids in the S-complex are identified as red plus signs.}
\label{fig: outer_PC}
\end{figure*}

To study how the family halos found in this work are related to the local
taxonomy, we also plotted in Fig.~\ref{fig: outer_PC} a projection 
in the $(a^{*},i-z)$ plane of all asteroids in our 
multi-domain sample (panel A), and an $(a,sin(i))$ projection of 
the same asteroids, (panel B).  
The great majority of asteroids in the region belong 
to the CX-complex, but there is a sizable minority of bodies belonging 
to the S-complex, that, as shown in Fig.~\ref{fig: outer_PC}, panel B,
are mostly associated with the Eos and Koronis families.


\begin{figure*}

  \begin{minipage}[c]{0.5\textwidth}
    \centering \includegraphics[width=3.0in]{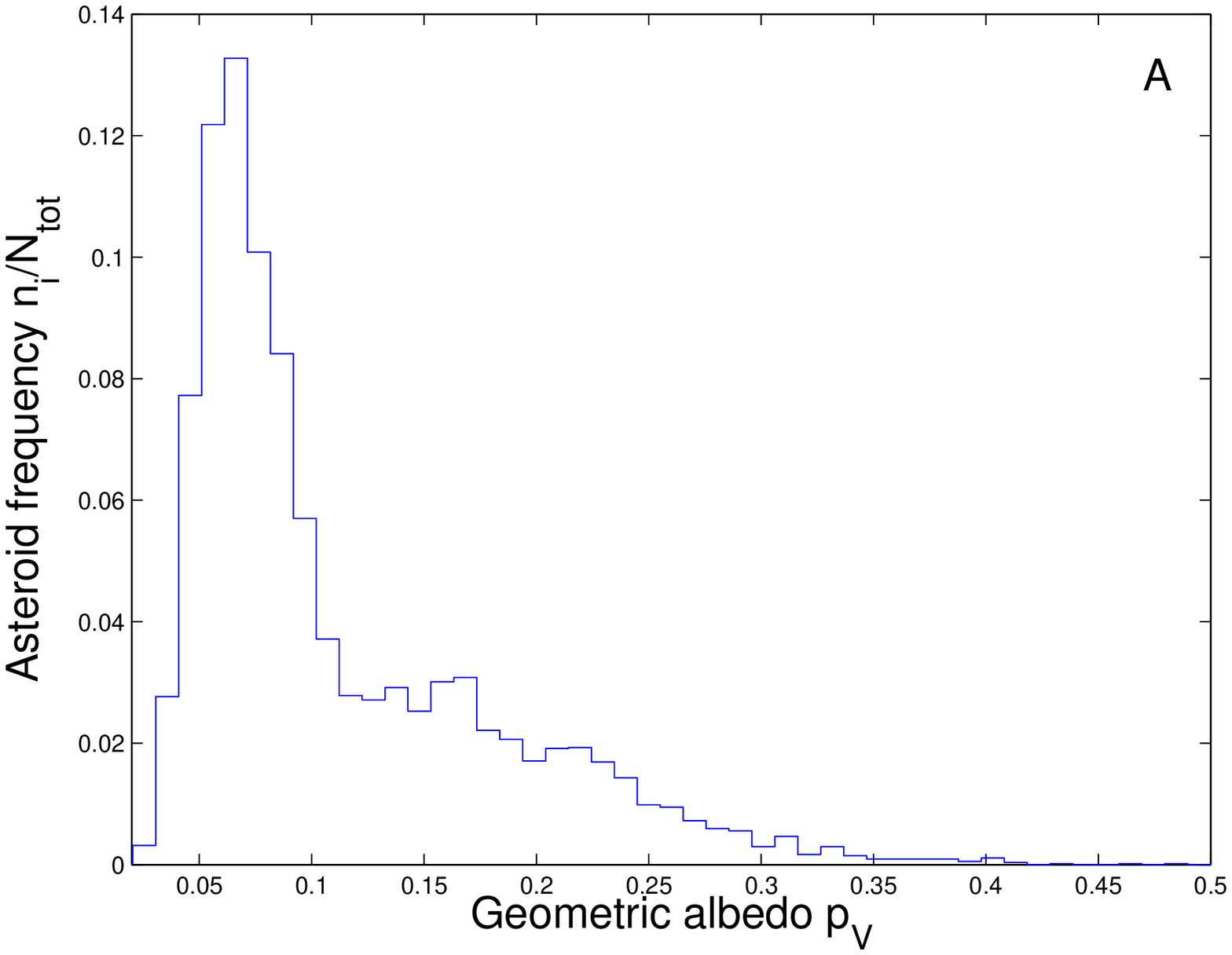}
  \end{minipage}%
  \begin{minipage}[c]{0.5\textwidth}
    \centering \includegraphics[width=3.0in]{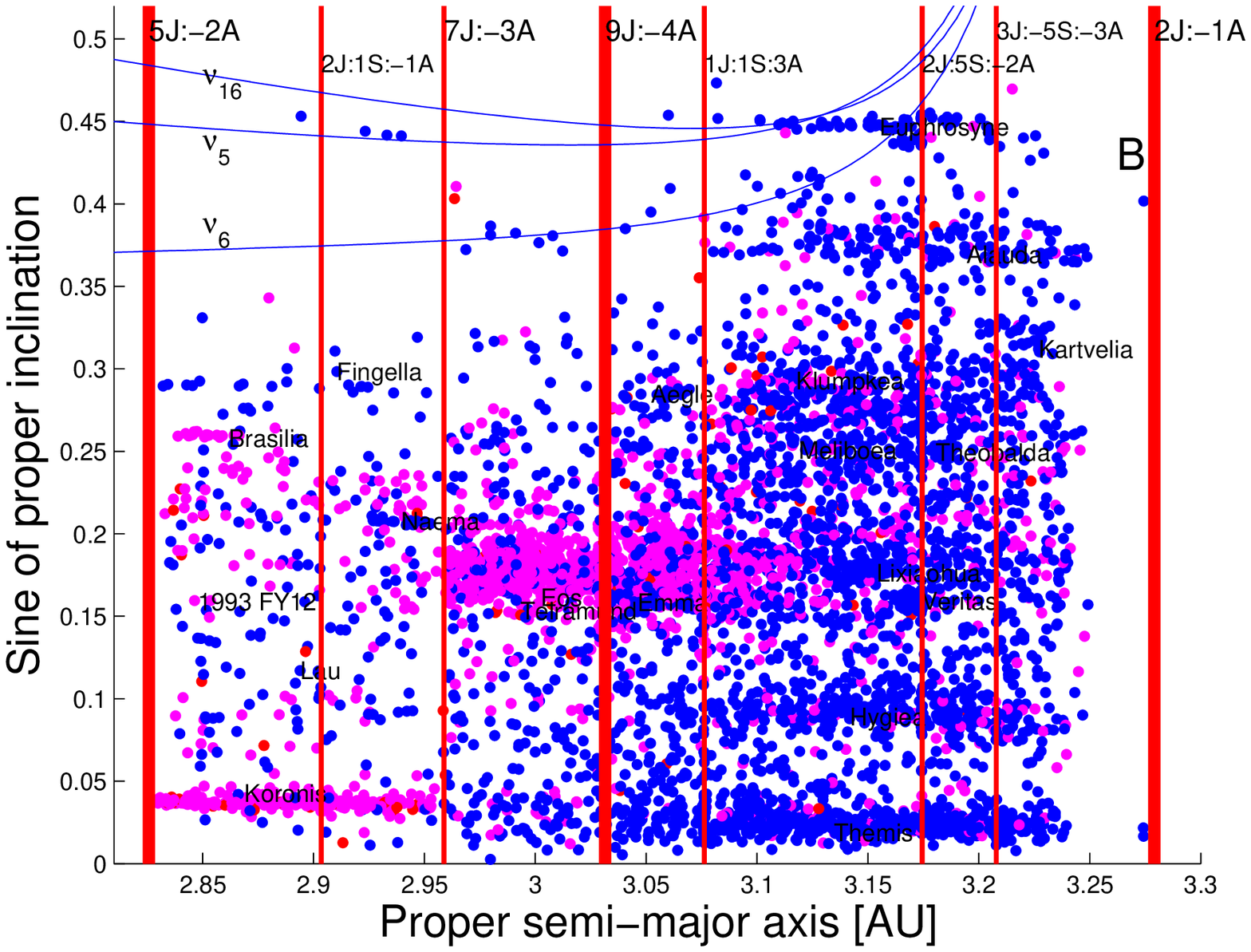}
  \end{minipage}

\caption{Panel A: a histogram of number frequency values 
$n_i/N_{Tot}$ as a function of geometric albedo $p_V$ for 
outer main belt asteroids in our multi-domain sample.  Panel B: an
$(a,sin(i))$ projection of the same asteroids,
where blue full dots are associated with asteroids with $p_V < 0.1$,
red full dots display asteroids with $0.1 < p_V < 0.3$, and magenta
full dots show asteroids with $p_V > 0.3$.}
\label{fig: outer_pv}
\end{figure*}

This is confirmed by WISE
$p_V$ geometrical albedo data, a histogram of which is presented
in Fig.~\ref{fig: outer_pv}, panel A.  Fig.~\ref{fig: outer_pv}, panel B,
displays an $(a,sin(i))$ projection of the same asteroids,
with the color code used in Fig.~\ref{fig: inner_pv}, panel B.
The great majority of objects
have low albedos, characteristics of the dark C-type objects that 
predominates in the outer main belt, but there is a fraction of asteroids,
that are associated with the Eos and Koronis families, with medium and
high values of geometric albedo.

Overall, we confirmed the results of the taxonomical analysis of 
Moth\'{e}-Diniz et al. (2005):  the outer main belt is dominated
by CX-complex dark asteroids, with the two notable exceptions of
the Eos and Koronis families.  S-complex asteroids in the local background
may be escapers from these two large families.   Dynamical studies
on the orbital evolution of members of these families are however 
needed to confirm this conclusion.

\section{Cybele group}
\label{sec: cybele}

The Cybele group, usually not considered part of the main belt, 
is located beyond the 2J:-1A mean motion resonance in semi-major
axis and its orbital region is usually defined to lie between
3.27 and 3.70 AU in proper $a$ and to $i < 30^{\circ}$.  Currently, there
are 1111 asteroid in the Cybele orbital region.
The largest collisional family in the region is associated with (87) Sylvia,
a triple asteroid (Vokrouhlick\'{y} et al. 2010).  The same
authors investigated the orbital region of two other large 
binary asteroids in the region, (107) Camila and (121) Hermione, that 
are currently not part of any recognizable family.  They 
concluded that, while it is possible that Yarkovsky/YORP driven
mobility in the orbital region of these asteroids may have depleted
possible local collisional families in timescales of 4 Byr, other
mechanisms, such as resonance sweeping or other perturbing effects
associated with the late Jupiter's inward migration may have been 
at play in the region in order to justify the current lack of
dynamical groups.  The AstDyS report three more families
in the region of the Cybele group: Huberta, Ulla, and (2000) EK76.
No halo group was found for the (2000) EK76 AstDyS cluster, that will not
therefore be treated in this section.  There were 128 asteroids in our 
multi-domain sample in this region, and we will start our analysis by 
investigating the Sylvia family halo.

\subsection{The Sylvia family}
\label{sec: sylvia}

The Sylvia family was first identified in Nesvorn\'{y} et al. (2006) 
and was the first dynamical group found in the Cybele region.
Its dynamical evolution was studied in detail in Vokrouhlick\'{y} et al.
(2010).   In this work we identified a CX-complex halo of 13 members
at a cutoff $d_{md}$ of 395~$m/s$.  There were no SDSS-MOC4 interlopers, 
and just one object (7.7\% of the total) was identified as 
a possible albedo interloper.

\subsection{The Huberta family}
\label{sec: huberta}

The Huberta family is a group reported at the AstDyS site.  We identified
a CX-complex halo of 4 members at a cutoff of 495~$m/s$.  One object
(25.0\% of the total) was a possible SDSS-MOC4 and albedo interloper.

\subsection{The Ulla family}
\label{sec: ulla}

The Ulla family is a very isolated group at relatively high inclination
of $\sin{(i)}\simeq 0.3$ and sligthly lower than 
the center of the ${\nu}_6$ secular resonance.
It is listed as a dynamical family at the AstDys
site, and we identified a small CX-complex halo of four members
for cutoffs larger than 220~$m/s$.  No interlopers
were identified in this halo.

\subsection{Cybele group: an overview}
\label{sec: cybele_overview}

We summarize our results for the Cybele group in 
Table~\ref{table: halo_cybele}, that has
the same format as similar tables used for the inner, central, and 
outer main belt.

\begin{table*}
\begin{center}
\caption{{\bf Asteroid families halos in the Cybele group.}}
\label{table: halo_cybele}
\vspace{0.5cm}
\begin{tabular}{|c|c|c|c|c|c|}
\hline
                 &                   &        &              &         & \\
First halo & $d_{md}$ cutoff value & Number of & Spectral & Number of SDSS-MOC4  & Number of $p_V$ \\ 
 member    &   [$m/s$] & members &    likely interlopers &  Complex & likely interlopers \\
                 &                   &         &             &         &  \\
\hline
                 &                   &         &             &         &  \\
(87) Sylvia: (18959)  &  395 & 13 & CX & 0  &  1\\
(260) Huberta: (260)   &  495 &  4 & CX & 0  &  1\\
(909) Ulla: (85036)    & $>$220 &  4 & CX & 0  &  0\\
                 &                   &         &             &         & \\
\hline
\end{tabular}
\end{center}
\end{table*}

\begin{figure*}

  \begin{minipage}[c]{0.5\textwidth}
    \centering \includegraphics[width=3.0in]{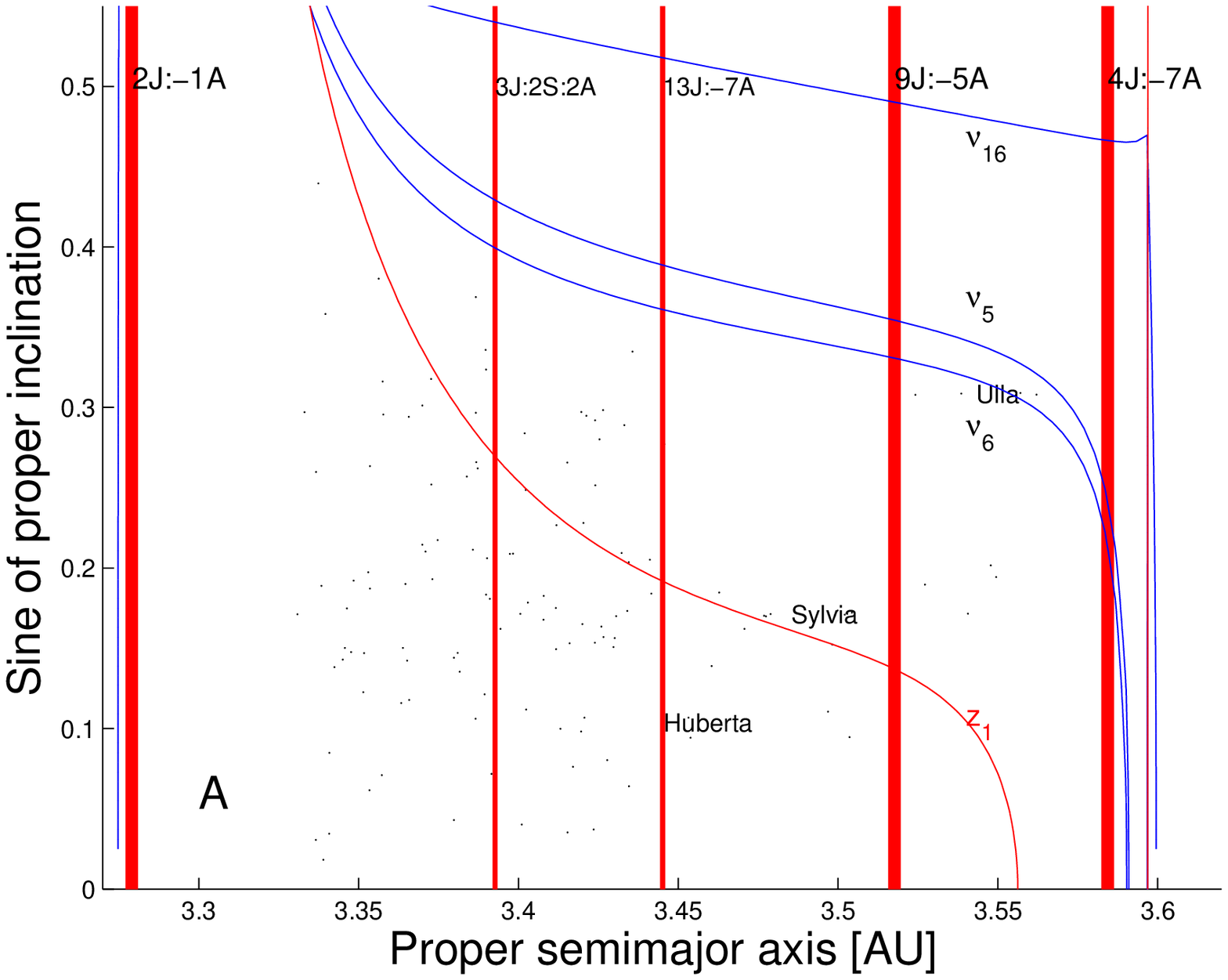}
  \end{minipage}%
  \begin{minipage}[c]{0.5\textwidth}
    \centering \includegraphics[width=3.0in]{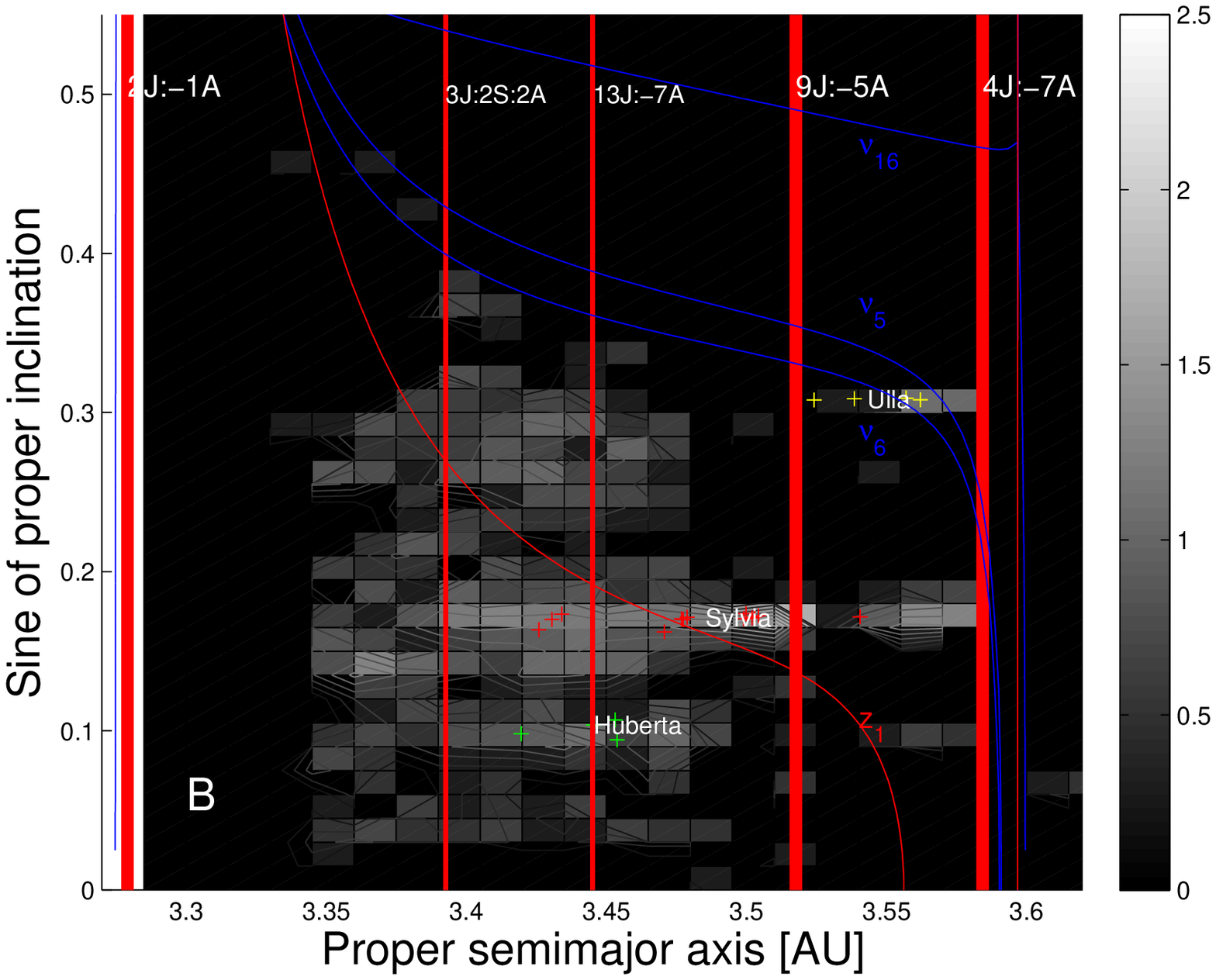}
  \end{minipage}

\caption{Panel A: An $(a,sin(i))$ projection of Cybele-group 
asteroids in our multi-variate sample.  Panel B: contour plot
of the number density of asteroids in the proper element sample.  
Superimposed, we display the orbital location of asteroids
of families in the CX-complex (plus signs).}
\label{fig: cybele_el}
\end{figure*}

Fig.~\ref{fig: cybele_el}, panel A, displays an $(a,sin(i))$ projection of
asteroids in our multi-variate sample in the outer main belt, with the same
symbols used for analogous figure for the inner, central, and 
outer main belt.
Here, however,  blue lines show the location of the main linear secular 
resonances, using the second order and fourth-degree secular perturbation
theory of Milani and Kne\v{z}evi\'{c} (1994)
to compute the proper frequencies $g$ and $s$ for the grid of $(a,e)$
and $(a,\sin(i))$ values shown in Fig.~\ref{fig: cybele_el}, panel A, and the
values of angles and eccentricity of (87) Sylvia, the asteroid 
associated with the largest family in the region (Vokrouhlick\'{y} et al. 
2010).   We also display the location of the $z_1$ secular
resonance as a red line, since this resonance is important in 
the dynamical evolution of the Sylvia group.  In panel B 
of the same figure we display a density map of the outer 
main belt, according to the approach described in Carruba and Michtchenko 
(2009). 
To quantitatively determine the local density of asteroids,  we
computed the $log_{10}$ of the number of all asteroids with proper elements 
per unit square in a 67 by 67 grid in $a$ (starting
at $a$~= 3.27 AU, with a step of 0.015 AU) and $sin(i)$ (starting at 
0, with a step of 0.015).    The other
symbols are the same as in Fig.~\ref{fig: inner_el}, panel B.

\begin{figure*}

  \begin{minipage}[c]{0.5\textwidth}
    \centering \includegraphics[width=3.0in]{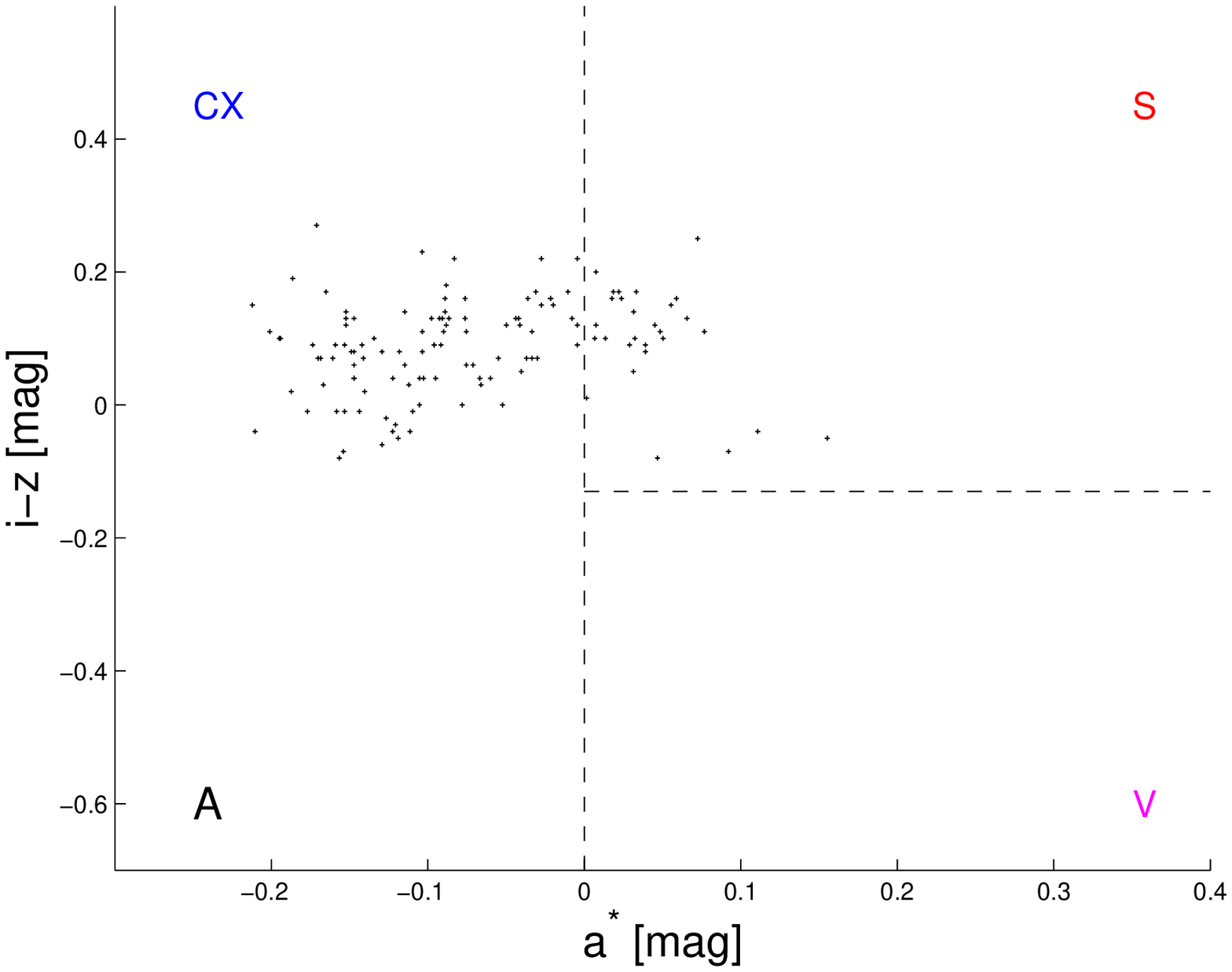}
  \end{minipage}%
  \begin{minipage}[c]{0.5\textwidth}
    \centering \includegraphics[width=3.0in]{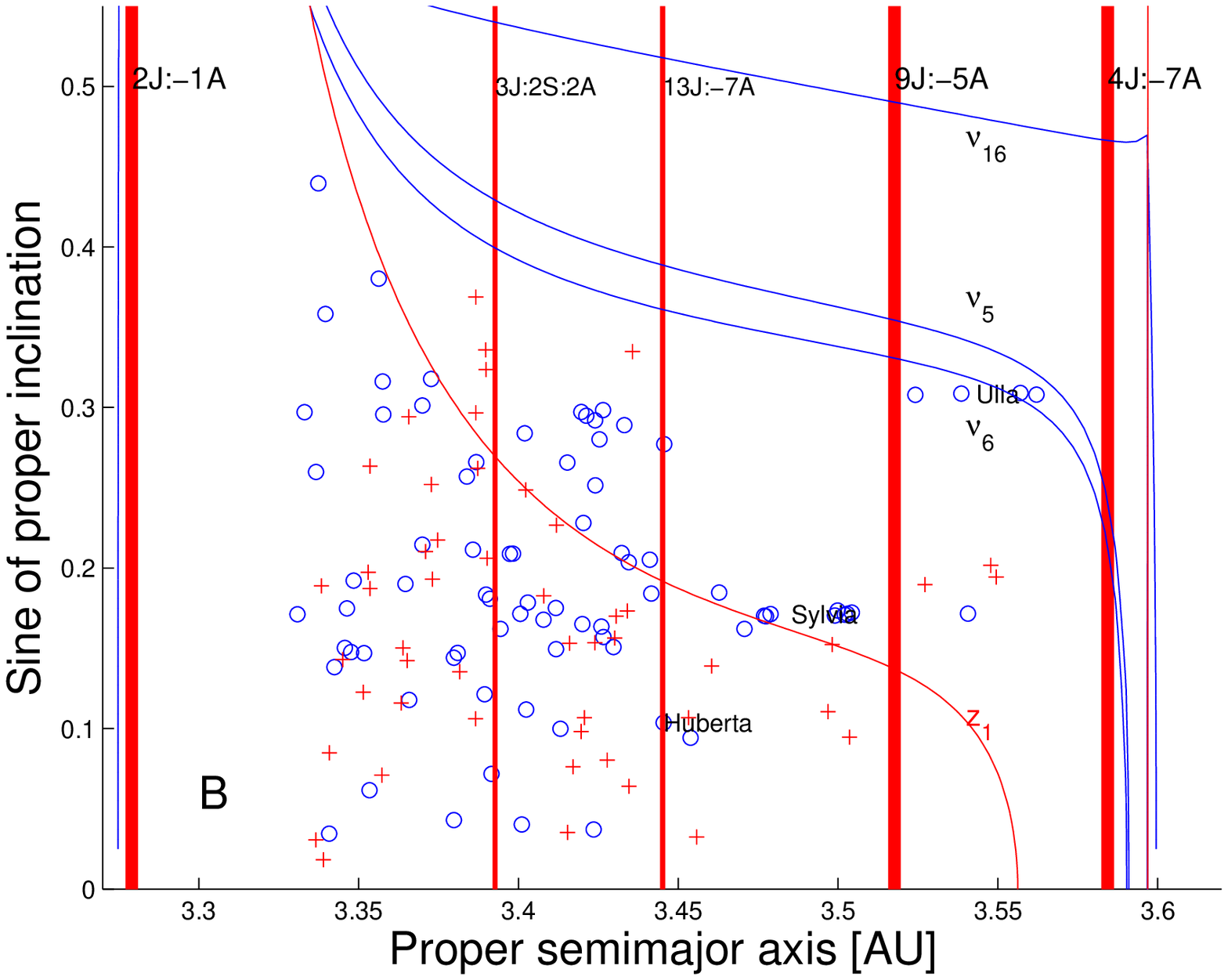}
  \end{minipage}

\caption{Panel A: an $(a^{*},i-z)$ projection of Cybele-group 
asteroids in our multi-domain sample.  Panel B: an
$(a,sin(i))$ projection of the same asteroids,
where objects in the CX complex are shown as blue circles, and
asteroids in the S-complex are identified as red plus signs.}
\label{fig: cybele_aiz}
\end{figure*}

\begin{figure*}

  \begin{minipage}[c]{0.5\textwidth}
    \centering \includegraphics[width=3.0in]{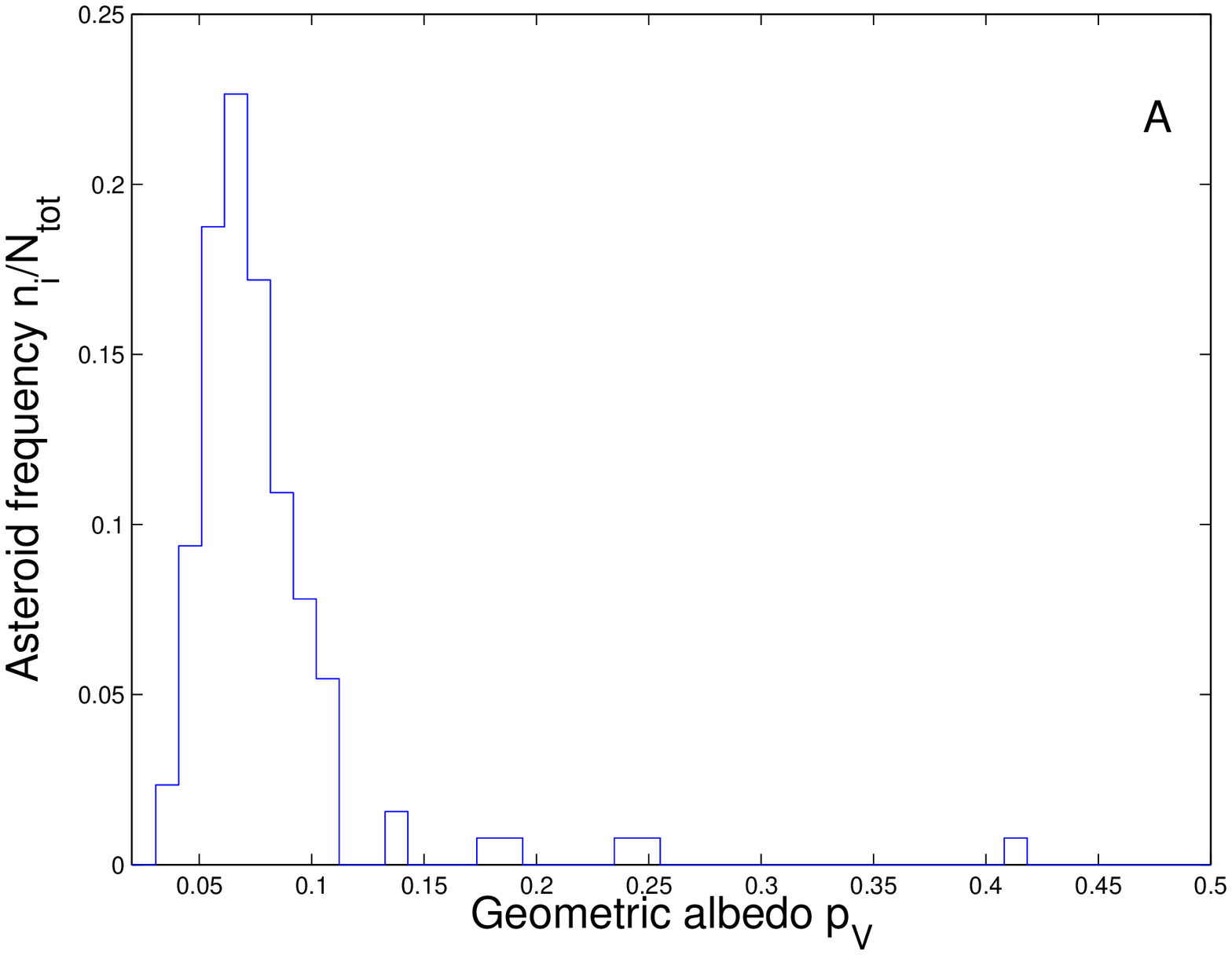}
  \end{minipage}%
  \begin{minipage}[c]{0.5\textwidth}
    \centering \includegraphics[width=3.0in]{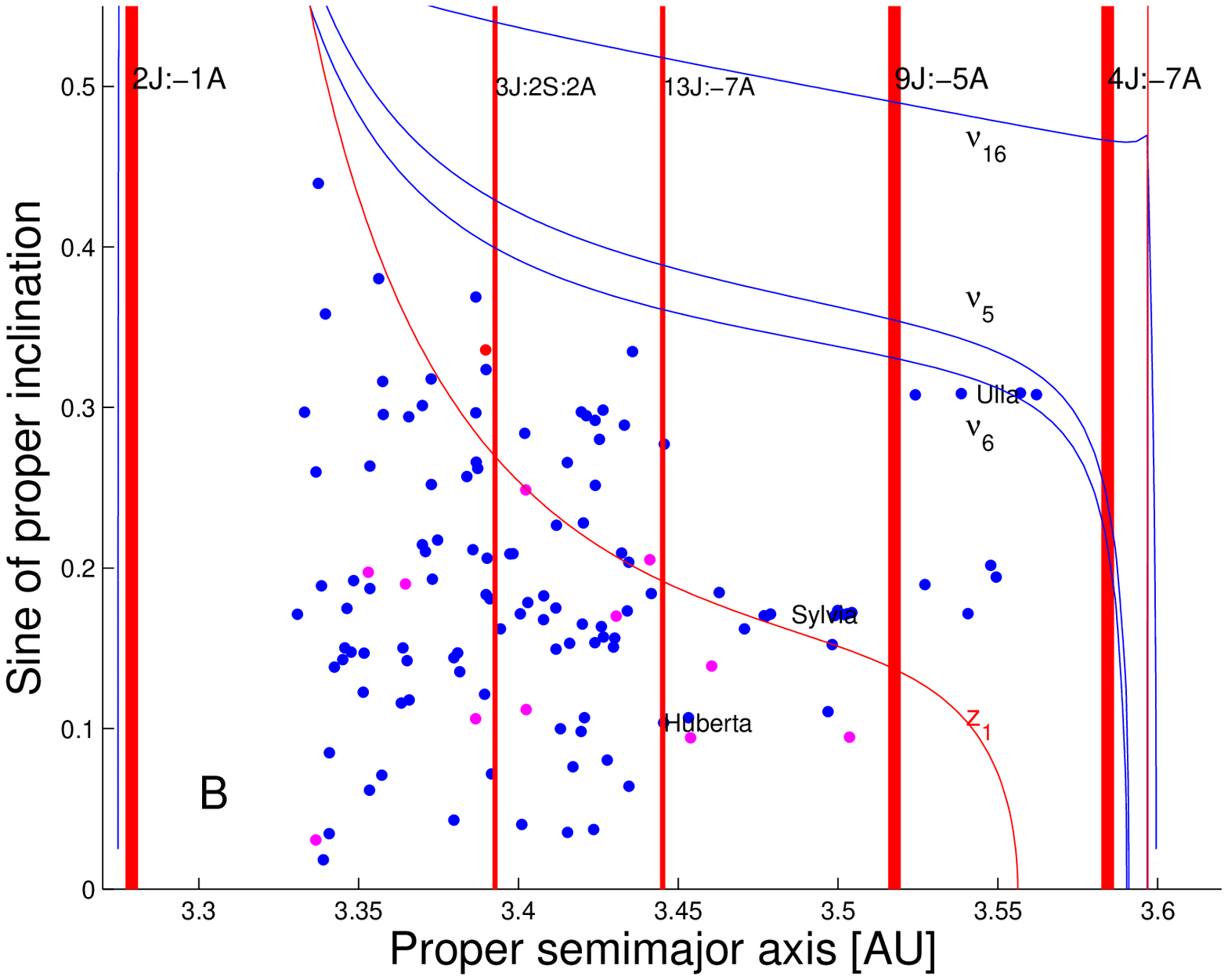}
  \end{minipage}

\caption{Panel A: a histogram of number frequency values 
$n_i/N_{Tot}$ as a function of geometric albedo $p_V$ for 
Cybele-group asteroids in our multi-domain sample.  Panel B: an
$(a,sin(i))$ projection of the same asteroids,
where blue full dots are associated with asteroids with $p_V < 0.1$,
red full dots display asteroids with $0.1 < p_V < 0.3$, and magenta
full dots show asteroids with $p_V > 0.3$.}
\label{fig: cybele_pv}
\end{figure*}

To study how the family halos found in this work are related to the local
taxonomy, we also plotted in Fig.~\ref{fig: cybele_aiz} a projection 
in the $(a^{*},i-z)$ plane of all asteroids in our 
multi-domain sample (panel A), and an $(a,sin(i))$ projection of 
the same asteroids, (panel B).  The majority of asteroids in the Cybele region
belong to the CX-complex, but there is a sizable minority of
S-complex bodies.  This is confirmed by WISE
$p_V$ geometrical albedo data, a histogram of which is presented
in Fig.~\ref{fig: cybele_pv}, panel A.  Fig.~\ref{fig: cybele_pv}, panel B,
displays an $(a,sin(i))$ projection of the same asteroids, with the 
same color code used in similar figures for the inner, central, and
outer main belt.  The vast majority of asteroids in the Cybele region
are dark objects, typical of CX-complex taxonomy. 
Overall, the predominance of dark, CX-complex asteroids in the Cybele 
group, confirms the taxonomical analysis performed by Vokrouhlick\'{y} et 
al. (2010).  The last region to be analyzed in this work will be that 
of the Hungaria asteroid family.

\section{The Hungaria region}
\label{sec: hungaria}

The Hungaria region is located at the inner edge of the asteroid main 
belt (at semi-major axis $a < 2$~AU), and it is located at high inclinations
and low to moderate eccentricities.  The limitations in eccentricity 
allow for a perihelion large enough to avoid strong interactions
with Mars, even considering secular changes in the Mars eccentricity.
(Milani et al. 2010).  The ${\nu}_3$ and ${\nu}_5$
secular resonances fix the dynamical limits of the Hungaria region
in inclination.
Only one family has been so far positively identified in the
Hungaria orbital region, the namesake (434) Hungaria group by 
Milani et al. (2010).  Other authors (Ca\~{n}ada-Assandri et al. 2013)
pointed out that the highly inclined Hungaria population is
dominated by S-type objects, and fairly distinguished from the 
C-complex population observed in the Hungaria dynamical family.
But no family in proper elements domain has yet been observed
in this highly inclined region.  We identified only 37 objects
in our multi-domain sample, with reasonable errors, and 
we will start our analysis 
of the Hungaria region by studying the Hungaria family halo.

\subsection{The Hungaria family}
\label{sec: hungaria_fam}

The most recent identification of the Hungaria family was obtained by
Milani et al. (2010), that also found no evidence for other possible
dynamical groups in the region (but identified possible sub-structures
inside the Hungaria family).  Here we identified a 2 CX-complex
halo at a cutoff of 590$~m/s$.  One object, 50\% of the total,
was however more compatible with S-complex taxonomy and could
therefore be a possible interlopers.  All asteroids in the halo
had $p_V > 0.1$, that is usually incompatible with a CX-complex
taxonomy, but that seems typical of Hungaria
objects, as discussed by Warner et al. (2009).

\subsection{The highly inclined Hungaria population}
\label{sec: hungaria_hi}

No family has been currently identified in the highly inclined
$(\sin(i) > 0.4)$ Hungaria region in the domain of proper elements.  
Ca\~{n}ada-Assandri et al. (2013) pointed out that 
the highly inclined Hungaria region is dominated by S-complex objects,
and taxonomically fairly different than the members of the CX-complex
Hungaria group.   While no family is yet identifiable in the proper
element domain in this region, we checked for the presence of 
a halo, possibly associated with hypothetical local 
frequency families.  We identified 
a S-complex halo of 7 members at a cutoff $d_{md} = 655~m/s$, around 
(2049) Grietje.  One object, 14.3\% of the total, is a possible SDSS-MOC4
interlopers, and there were no albedo interlopers.

\subsection{The Hungaria region: an overview}
\label{sec: hungaria_sum}

We summarize our results for the Hungaria region in 
Table~\ref{table: halo_hungaria}, that has
the same format as similar tables used for the inner, central, 
outer main belt, and Cybele group.

\begin{table*}
\begin{center}
\caption{{\bf Asteroid families halos in the Hungaria region.}}
\label{table: halo_hungaria}
\vspace{0.5cm}
\begin{tabular}{|c|c|c|c|c|c|}
\hline
                 &                   &        &              &         & \\
First halo & $d_{md}$ cutoff value & Number of & Spectral & Number of SDSS-MOC4  & Number of $p_V$ \\ 
 member    &   [$m/s$] & members &    likely interlopers &  Complex & likely interlopers \\
                 &                   &         &             &         &  \\
\hline
                 &                   &         &             &         &  \\
(434) Hungaria: (5968) &  590 & 2  & CX & 1 &  2\\
(2049) Grietje: (3043)& 655 &  7 &  S & 1 &  0\\
                 &                   &         &             &         & \\
\hline
\end{tabular}
\end{center}
\end{table*}

\begin{figure*}

  \begin{minipage}[c]{0.5\textwidth}
    \centering \includegraphics[width=3.0in]{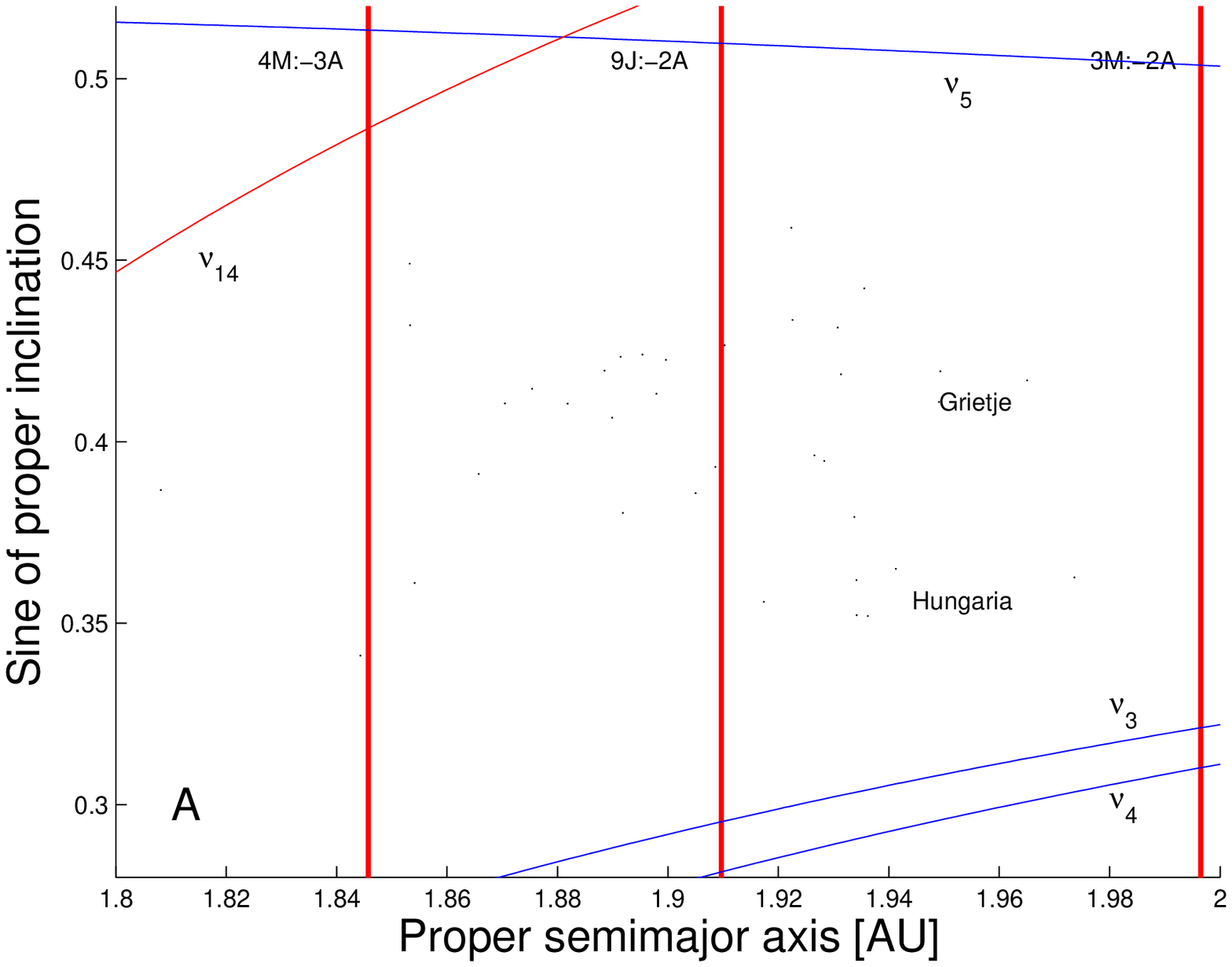}
  \end{minipage}%
  \begin{minipage}[c]{0.5\textwidth}
    \centering \includegraphics[width=3.0in]{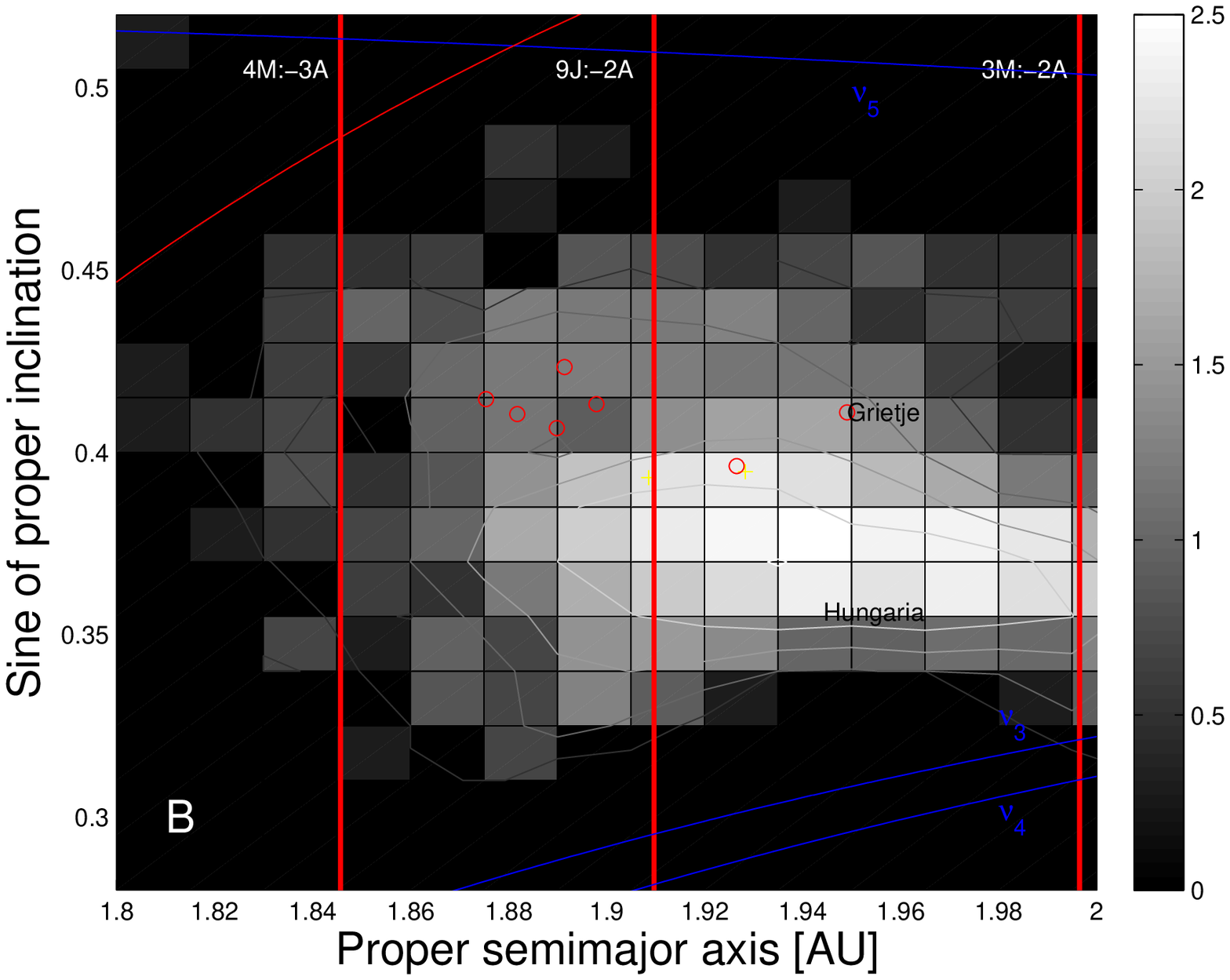}
  \end{minipage}

\caption{Panel A: An $(a,sin(i))$ projection of Hungaria-region 
asteroids in our multi-variate sample.  Panel B: contour plot
of the number density of asteroids in the proper element sample.  
Superimposed, we display the orbital location of asteroids
of families in the CX-complex (plus signs) and S-complex (circles).}
\label{fig: hungaria_el}
\end{figure*}

Fig.~\ref{fig: hungaria_el}, panel A, displays an $(a,sin(i))$ projection of
asteroids in our multi-variate sample in the outer main belt, with the same
symbols used for analogous figures in the inner, central, and outer main belt, 
and the Cybele region.
Here, however,  blue lines show the location of the ${\nu}_5$ linear secular 
resonances, using the second order and fourth-degree secular perturbation
theory of Milani and Kne\v{z}evi\'{c} (1994)
to compute the proper frequencies $g$ and $s$ for the grid of $(a,e)$
and $(a,\sin(i))$ values shown in Fig.~\ref{fig: cybele_el}, panel A, and the
values of angles and eccentricity of (434) Hungaria, the asteroid 
associated with the largest family in the region.   
We also display the location of the ${\nu}_4$, ${\nu}_3$ (blue lines), 
and ${\nu}_{14}$ (red line) secular
resonances, since this resonance are important in 
setting dynamical boundaries in the region;
The orbital position in the $(a,\sin(i))$ plane
of the first numbered asteroid in all the Hungaria
groups is also identified in Fig.~\ref{fig: cybele_el}, panel A.  In panel B 
of the same figure we display a density map of the outer 
main belt, according to the approach described in Carruba and Michtchenko 
(2009). 
To quantitatively determine the local density of asteroids,  we
computed the $log_{10}$ of the number of all asteroids with proper elements 
per unit square in a 15 by 24 grid in $a$ (starting
at $a$~= 1.8 AU, with a step of 0.015 AU) and $sin(i)$ (starting at 
0.25, with a step of 0.015).  
The other symbols are the same as in Fig.~\ref{fig: inner_el}, panel B.

\begin{figure*}

  \begin{minipage}[c]{0.5\textwidth}
    \centering \includegraphics[width=3.0in]{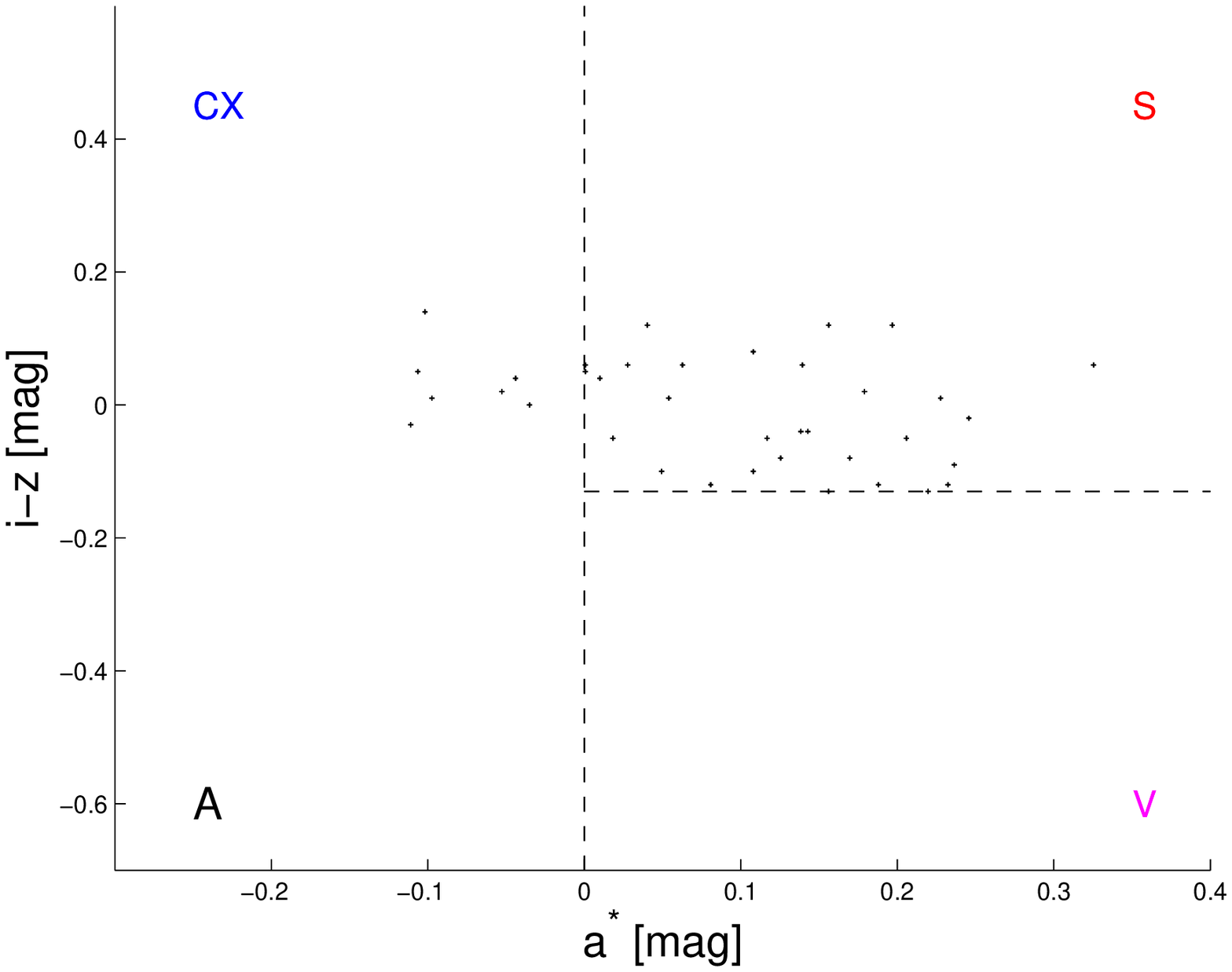}
  \end{minipage}%
  \begin{minipage}[c]{0.5\textwidth}
    \centering \includegraphics[width=3.0in]{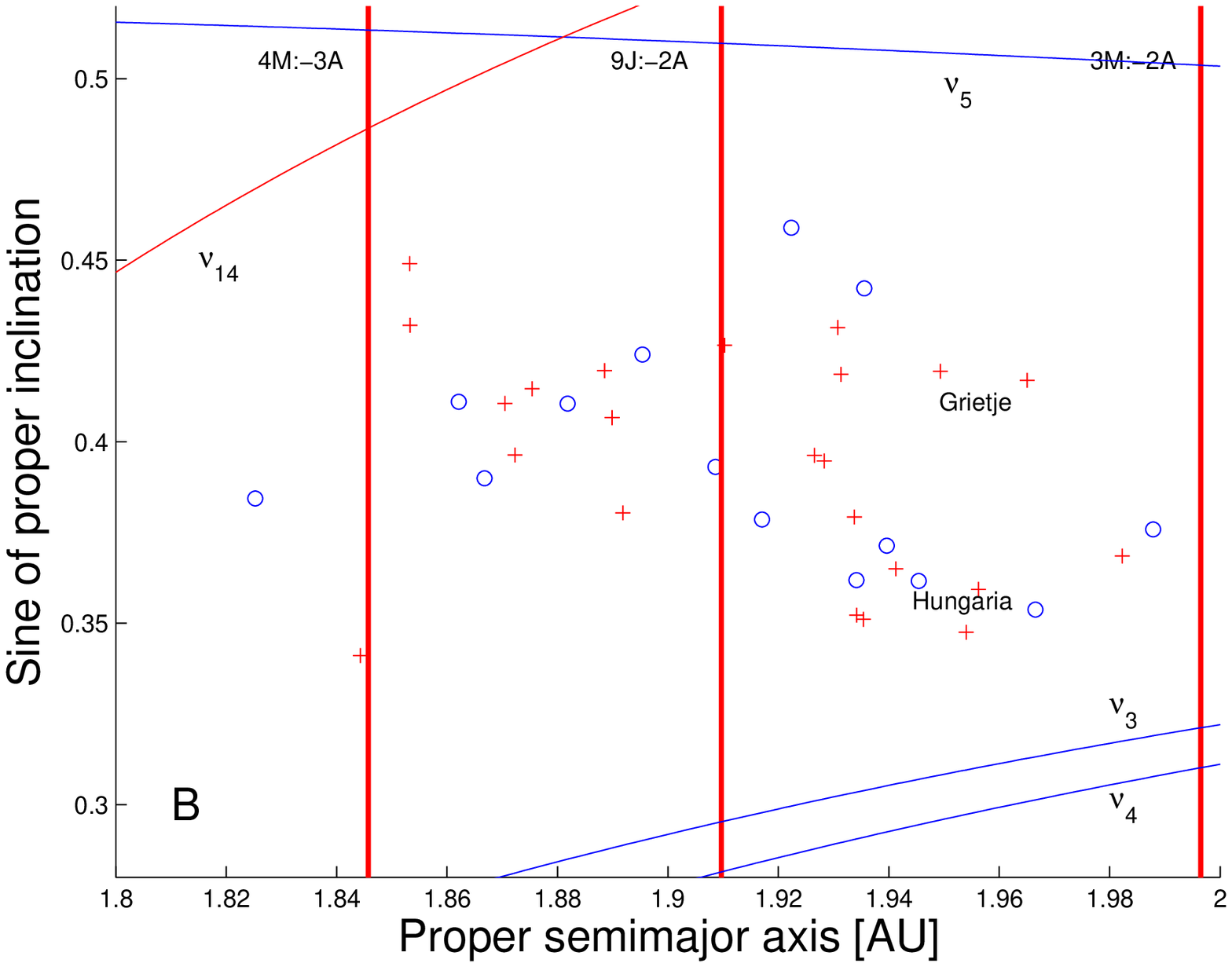}
  \end{minipage}

\caption{Panel A: an $(a^{*},i-z)$ projection of Hungaria-region 
asteroids in the in our multi-domain sample.  Panel B: an
$(a,sin(i))$ projection of the same asteroids,
where objects in the CX complex are shown as blue circles, and
asteroids in the S-complex are identified as red plus signs.}
\label{fig: hungaria_aiz}
\end{figure*}

To study how the family halos found in this work are related to the local
taxonomy, we also plotted in Fig.~\ref{fig: hungaria_aiz} a projection 
in the $(a^{*},i-z)$ plane of all asteroids in our 
multi-domain sample (panel A), and an $(a,sin(i))$ projection of 
the same asteroids, (panel B).  The majority of asteroids in the Hungaria region
in our sample belong to the S-complex, but there is a sizeable minority of
CX-complex bodies.

\begin{figure*}

  \begin{minipage}[c]{0.5\textwidth}
    \centering \includegraphics[width=3.0in]{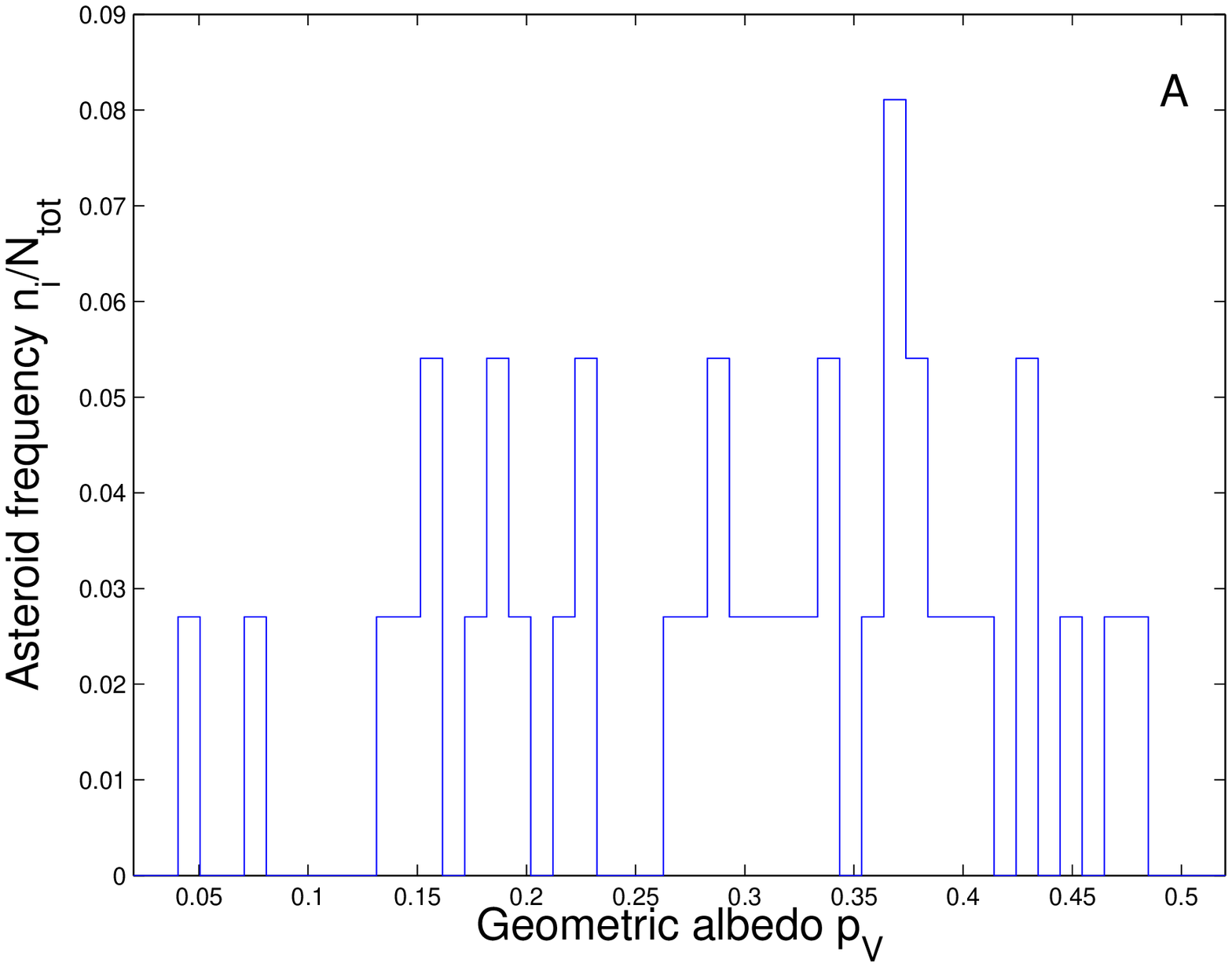}
  \end{minipage}%
  \begin{minipage}[c]{0.5\textwidth}
    \centering \includegraphics[width=3.0in]{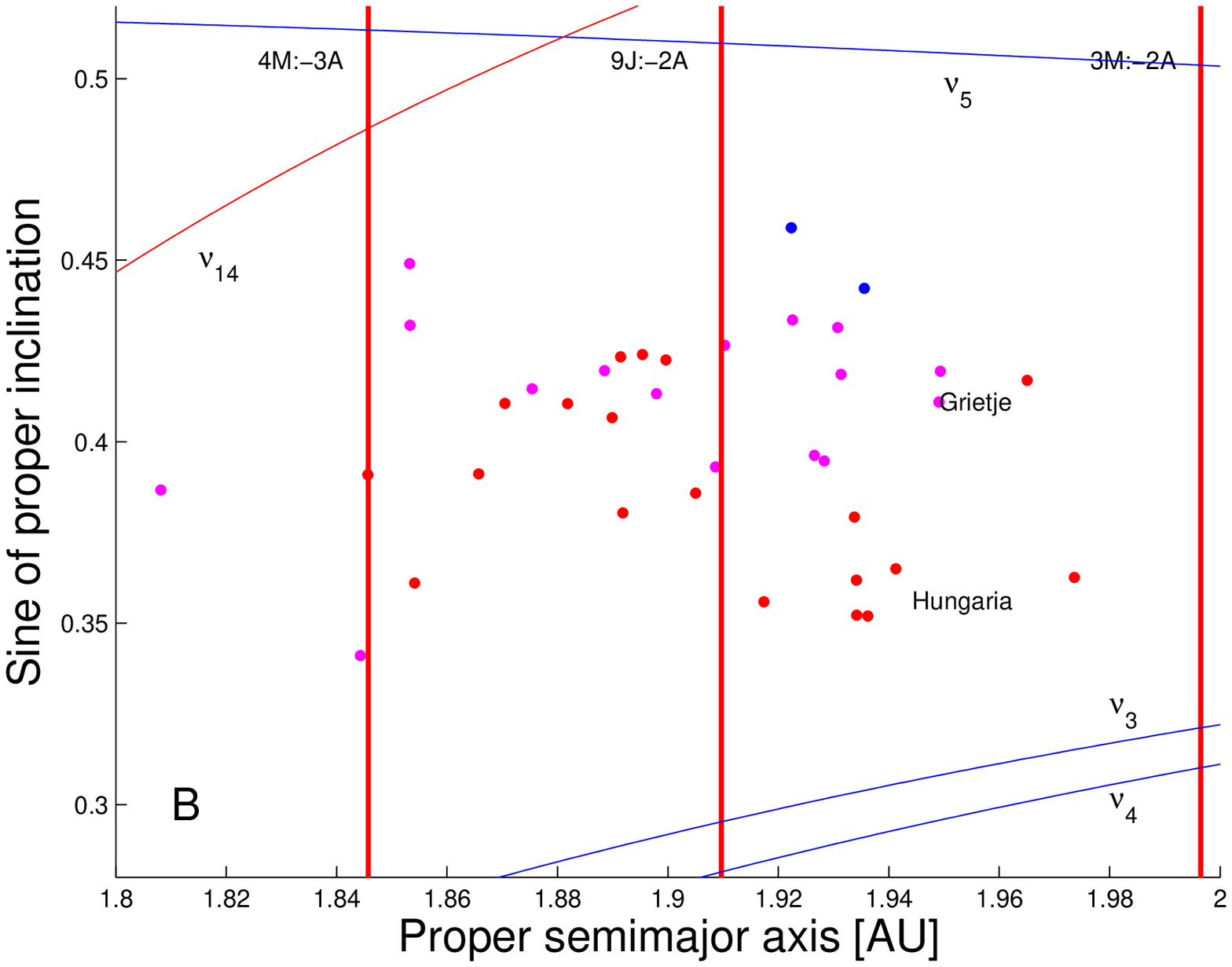}
  \end{minipage}

\caption{Panel A: a histogram of number frequency values 
$n_i/N_{Tot}$ as a function of geometric albedo $p_V$ for 
Hungaria-region asteroids in our multi-domain sample.  Panel B: an 
$(a,sin(i))$ projection of the same asteroids,
where blue full dots are associated with asteroids with $p_V < 0.1$,
red full dots display asteroids with $0.1 < p_V < 0.3$, and magenta
full dots show asteroids with $p_V > 0.3$.}
\label{fig: hungaria_pv}
\end{figure*}

The analysis of WISE
$p_V$ geometrical albedo data, a histogram of which is presented
in Fig.~\ref{fig: hungaria_pv}, panel A, 
show some peculiarities.  Fig.~\ref{fig: hungaria_pv}, panel B,
displays an $(a,sin(i))$ projection of the same asteroids, with the 
same color code used in similar figures for the inner, central, and
outer main belt.  The vast majority of asteroids in the Hungaria region
and the Hungaria family are very bright objects, 
typical of S-complex taxonomy.  The fact that many asteroids in
the Hungaria family show high albedo and CX-taxonomy remains yet 
to be explained.

\section{Conclusions}
\label{sec: concl}

In this work we:

\begin{itemize}
\item Introduced a new method to obtain asteroid families and asteroid
family halos based on a distance metric in a multi-domain composed
of proper elements, SDSS-MOC4 $(a^{*},i-z)$ colors, and WISE geometrical
albedo $p_V$.

\item Compared this new distance metric with other distance metrics
in domain of proper elements, proper elements and SDSS-MOC4 colors,
and proper elements and geometric albedo.  The method is at best a factor 
of two more efficient in eliminating interlopers than other methods, and at
worst it provides comparable results to groups found in domains of
proper elements and SDSS-MOC4 colors only.

\item Applied this method to all the major known families
in the asteroids' main belt, and in the Cybele and Hungaria orbital regions.
Overall, we identified sixty-two asteroid families halos, of which 
seven were in the inner main belt, thirty-one in the central main belt, 
nineteen in the outer main belt, three in the Cybele group, and two in the 
Hungaria region.  We confirm the taxonomical analysis performed 
by Moth\'{e}-Diniz et al. (2005), 
Nesvorn\'{y} et al. (2006), Carruba (2009a,b), (2010a,b) and other authors,
with some small discrepancies for a few minor families in the central main
belt.

\end{itemize}

Overall, apart from a few problematic cases such as the Eos family,
our method appears to provide robust results in terms of asteroid
family identification and in efficiency in eliminating interlopers
from the clusters.  While the sample of objects with data in all three
domains is still limited, we believe that such an approach may be certainly
more reliable than traditional HCM in identifying possible collisional
groups.  The possible future increase in the number of asteroids 
for which data in all three domains will be available, for instance
because of the GAIA mission, may provide in the future data-bases
for asteroid family identification much larger than the one used in 
this work.

Many other applications of this new approach are possible
with current data-bases.  An analysis of asteroid families in
domains of proper frequencies such as $(n,g,g+s)$ (Carruba and
Michtchenko 2007, 2009), where $g$ is the 
precession frequency of the longitude of pericenter, and $s$ the precession
frequency of the longitude of the node, SDSS-MOC4 colors, and WISE
albedo, may provide useful insights on the secular evolution of
asteroid families.  Many exciting years of discoveries are still
open, in our opinion, in the field of asteroid dynamics.

\section*{Acknowledgments}
We are grateful to the reviewer of this article, Ricardo Gil-Hutton, for 
suggestions and comments that significantly increased the quality of
this paper.   We would also
like to thank the S\~{a}o Paulo State Science Foundation 
(FAPESP) that supported this work via the grant 11/19863-3, and the
Brazilian National Research Council (CNPq, grant 305453/2011-4).
This publication makes use of data products from the Wide-field 
Infrared Survey Explorer, which is a joint project of the University 
of California, Los Angeles, and the Jet Propulsion Laboratory/California 
Institute of Technology, funded by the National Aeronautics and Space 
Administration.  This publication also makes use of data products 
from NEOWISE, which is a project of the Jet Propulsion 
Laboratory/California Institute of Technology, funded by the Planetary 
Science Division of the National Aeronautics and Space Administration.
Our data on asteroid families identification will be available at this site:  
http://www.feg.unesp.br/~vcarruba/Halos.html

\bsp

\label{lastpage}

\end{document}